\documentclass[12pt]{article}
\usepackage{amsmath,amssymb,amsfonts,amsthm,amscd}
\usepackage[english]{babel}
\usepackage[toc]{appendix}
\usepackage{float}
\newcommand{\be}{\begin{equation}}
\newcommand{\ee}{\end{equation}}
\newcommand{\bea}{\begin{eqnarray}}
\newcommand{\eea}{\end{eqnarray}}
\newcommand{\nn}{\nonumber \\}
\newcommand{\p}[1]{(\ref{#1})}
\newcommand{\lb}{\label}

\numberwithin{equation}{section}
\begin{document}
\begin{titlepage}
\begin{flushright}
JINR E2-2014-100\\
CBPF-NF-013/14
\end{flushright}

\vspace*{1cm}

\begin{center}
\renewcommand{\thefootnote}{$\star$}

{\Large\bf Superconformal mechanics in $SU(2|1)$
 superspace}

\vspace{2cm}
\renewcommand{\thefootnote}{$\times$}

{\large\bf
E.~Ivanov}${\,}^1$,\,\,\, {\large\bf S.~Sidorov}${\,}^1$,\,\,\,
{\large\bf F.~Toppan}${\,}^2$ \vspace{1cm}

${}^1${\it Bogoliubov  Laboratory of Theoretical Physics, JINR,}\\
{\it 141980 Dubna, Moscow region, Russia} \\
\vspace{0.1cm}

{\tt eivanov@theor.jinr.ru, sidorovstepan88@gmail.com}\\
\vspace{0.5cm}

${}^2${\it CBPF, Rua Dr. Xavier Sigaud 150, Urca,\\
cep 22290-180, Rio de Janeiro (RJ), Brazil} \\
\vspace{0.1cm}

{\tt toppan@cbpf.br}\\
\vspace{0.3cm} \setcounter{footnote}{0}

\end{center}
\vspace{0.2cm} \vskip 0.6truecm  \nopagebreak

\begin{abstract}
\noindent
Using the worldline $SU(2|1)$ superfield approach, we construct ${\cal N}=4$ superconformally invariant actions for the $d=1$ multiplets
$({\bf 1, 4, 3})$ and $({\bf 2, 4, 2})$. The $SU(2|1)$ superfield framework automatically implies the trigonometric realization
of the superconformal symmetry and the harmonic oscillator term in the corresponding component actions.
%in parallel with the standard inverse-square conformal potentials in some particular cases.
We deal with the general ${\cal N}=4$ superconformal
algebra $D(2,1;\alpha)$ and its central-extended $\alpha=0$ and $\alpha=-1$ $psu(1,1|2)\oplus su(2)$ descendants.
We capitalize on the observation that $D(2,1;\alpha)$ at $\alpha\neq 0$ can be treated as a closure of its two $su(2|1)$ subalgebras,
one of which defines the superisometry of the $SU(2|1)$ superspace, while the other is related to the first one through
the reflection of $\mu$, the parameter of contraction to the flat ${\cal N}=4, d=1$ superspace.
This closure property and its $\alpha=0$ analog suggest a simple criterion
for the $SU(2|1)$ invariant actions to be superconformal: they should be even functions of $\mu$. We find that the superconformal actions of
the multiplet $({\bf 2, 4, 2})$ exist only at $\alpha=-1, 0$ and are reduced to a sum of the free sigma-model type action and the conformal
superpotential yielding, respectively, the oscillator potential $\sim \mu^2$ and the standard conformal inverse-square potential in the bosonic sector. The
sigma-model action in this case can be constructed only on account of non-zero central charge in the superalgebra $su(1,1|2)$.
\end{abstract}

\vspace{2.5cm}

{\small

}

\end{titlepage}

\section{Introduction}

Recently, there was an essential progress in constructing and understanding  the rigid supersymmetric theories in curved superspace
which attract attention in connection with the general ``gauge/gravity'' correspondence (see, e.g., \cite{FS,DFS,SamSor} and references therein).
In \cite{DSQM,SKO}, two of us elaborated on the simplest $d=1$ analogs of such theories,
the $SU(2|1)$ supersymmetric quantum mechanics (SQM) models, proceeding from the $SU(2|1)$ covariant worldline superfield approach.
Two types of the worldline $SU(2|1)$ superspace as the proper supercosets of the supergroup $SU(2|1)$ were constructed.
Both superspaces are deformations of the standard ${\cal N}=4$, $d=1$ superspace (see \cite{superc} and references therein) by a mass parameter $m$.
The off- and on-shell deformed versions of the ${\cal N}=4, d=1$ multiplets $({\bf 1, 4, 3})$ and $({\bf 2, 4, 2})$ were studied
and proved to possess a number of interesting peculiarities as compared with their ``flat'' $m=0$ cousins. One of such new features is the necessary presence
of the harmonic oscillator terms $\sim m^2$ in the bosonic sectors of the corresponding invariant Lagrangians.
The ``weak supersymmetry'' model of ref. \cite{WS} and
the ``super K\"ahler oscillator'' models of refs. \cite{BN3,BN4} were recovered as the particular cases
of generic $SU(2|1)$ SQM associated, respectively, with the single multiplet $({\bf 1, 4, 3})$ and a few multiplets $({\bf 2, 4, 2})$. It is interesting to
inspect the superconformal subclass of the $SU(2|1)$ SQM models. This is the main subject of the present paper.

As was argued in \cite{PP}, conformal mechanics \cite{DFF} can be divided into three classes characterized by the {\it parabolic}, {\it trigonometric}
and {\it hyperbolic} realizations of the
$d=1$ conformal group $SO(2,1)\sim SL(2,\mathbb{R})$.
Earlier, supersymmetric extensions of conformal mechanics corresponding only to the parabolic transformations were mainly studied \cite{SCM,SCM1,superc}.
Motivated by \cite{PP}, the classification of superconformal ${\cal N}=4$ SQM models was recently extended by the {\it trigonometric/hyperbolic} type \cite{HT}.
The basic difference of the trigonometric/hyperbolic superconformal actions from the parabolic ones is the presence of oscillator potentials. The standard $d=1$
Poincar\'e supercharges present in the superconformal algebras are not squared to the canonical Hamiltonian in such models.
The actions of trigonometric/hyperbolic superconformal mechanics cannot be obtained from the standard ${\cal N}=4, d=1$ superfield approach,
while the parabolic actions are well described just within the latter\footnote{The possibility of adding an oscillator term to the DFF action \cite{DFF}
without breaking conformal symmetry was firstly noticed in \cite{Pash}. The ${\cal N}=2$ superconformal extensions of such actions were considered in \cite{DI,BK}.}.
It turns out that it is the $SU(2|1)$ superfield approach that is ideally suited for the comprehensive description of the trigonometric ${\cal N}=4$ superconformal
actions. The hyperbolic actions can be obtained from the trigonometric ones by a simple substitution.

Our construction is based on the appropriate two-parameter embedding of the superspace supergroup $SU(2|1)$ into the most general ${\cal N}=4, d=1$
superconformal group $D\left(2,1;\alpha\right)$, with the contraction parameter $m$ being redefined as $m \rightarrow -\alpha \mu$ and $\mu$
also appearing in the basic anticommutator on its own. At any
$\alpha \neq 0$ the whole conformal superalgebra $D\left(2,1;\alpha\right)$ can be obtained as a closure of the original superalgebra  $su(2|1)$
and its $-\mu$ counterpart, which suggests a simple selection rule for the superconformal $SU(2|1)$ SQM Lagrangians as those depending only on $\mu^2$. At $\alpha =0$,
the basic  $su(2|1)$ contracts into some flat ${\cal N}=4, d=1$ superalgebra which is still different from
the standard  ${\cal N}=4, d=1$ ``Poincar\'e'' superalgebra and involves the parameter $\mu$ in such a way that
$D\left(2,1;\alpha {=}0\right) \sim psu(1,1|2)\oplus su(2)$ (and its central extension) can be obtained as a closure of this flat superalgebra and its $-\mu$ counterpart
as subalgebras of $D\left(2,1;\alpha {=}0\right)$.  This important property makes it possible
to sort out the superconformal actions in the special $\alpha=0$ case too. Exploiting the closure property just mentioned, we find
the universal two-parameter family of the realizations of the conformal supergroup $D\left(2,1;\alpha\right)$ on the coordinates of the
 $SU(2|1)$ superspace, as well as on the superfields representing the off-shell multiplets  $({\bf 1, 4, 3})$ and $({\bf 2, 4, 2})$,
 at all admissible values of the parameter $\alpha$ (for the second multiplet, only $\alpha=-1$ and $\alpha=0$ are allowed). These realizations automatically
 prove to be trigonometric while the corresponding superconformal actions
 necessarily involve the oscillator-type terms $\sim \mu^2$. The parabolic realizations of $D\left(2,1;\alpha\right)$ and the corresponding actions
 are recovered in the limit $\mu =0$, in which  both $su(2|1)$ and its $\alpha =0$ analog go over into the standard $\mu$-independent
 ${\cal N}=4, d=1$ Poincar\'e superalgebra.

The paper is organized as follows. The salient features of the $SU(2|1)$ superspace approach are sketched in Section \ref{Sec. 2}.
In Section \ref{Sec. 3}, the embedding of $su(2|1)$ in $D\left(2,1;\alpha\right)$  is discussed along the lines outlined above and
the relevant $SU(2|1)$ superspace realizations  of $D\left(2,1;\alpha\right)$ are explicitly presented.
The study of the trigonometric models of superconformal mechanics associated with the multiplets $({\bf 1, 4, 3})$ and $({\bf 2, 4, 2})$
is the  subject of Sections \ref{Sec. 4} - \ref{Sec. 7}. We construct the superfield and component off- and on-shell actions for
various cases, distinguishing those which admit additional conformal inverse-square potentials in the bosonic sector.
The alternative (albeit equivalent) construction of the component superconformally invariant actions, based
on the $D$-module representation techniques, is briefly  outlined in Section \ref{Sec. 8} on the example of the multiplet $({\bf 2, 4, 2})$.
Section \ref{Sec. 9} is a summary of the basic results of the paper.
In Appendices, we collect some details concerning the central extensions of the superalgebra $D\left(2,1;\alpha\right)$ with $\alpha=-1$ (or $\alpha=0$),
the generalized chiral $SU(2|1)$ multiplets $({\bf 2, 4, 2})$, as well as the hyperbolic superconformal mechanics.

\setcounter{equation}{0}
\section{$SU(2|1)$ superspace}\label{Sec. 2}
First of all, we need to define the superalgebra $su(2|1)$. Its standard form is given by the following non-vanishing (anti)commutators:
\bea
    &&\lbrace Q^{i}, \bar{Q}_{j}\rbrace = 2m I^i_j +2\delta^i_j \tilde H,\qquad
    \left[I^i_j,  I^k_l\right]
    = \delta^k_j I^i_l - \delta^i_l I^k_j\,,\nn
    &&\left[I^i_j, \bar{Q}_{l}\right] = \frac{1}{2}\,\delta^i_j\bar{Q}_{l}-\delta^i_l\bar{Q}_{j}\, ,\qquad \left[I^i_j, Q^{k}\right]
    = \delta^k_j Q^{i} - \frac{1}{2}\,\delta^i_j Q^{k},\nn
    &&\left[\tilde H, \bar{Q}_{l}\right]=\frac{m}{2}\,\bar{Q}_{l}\,,\qquad \left[\tilde H, Q^{k}\right]=-\frac{m}{2}\,Q^{k}. \label{alg1}
\eea
The generators satisfy the following rules of the Hermitian conjugation:
\bea
    \left(Q^{k}\right)^{\dagger} = \bar{Q}_{k}\,,\qquad \left(\bar{Q}_{k}\right)^{\dagger} = Q^{k},\qquad
    \left(I_i^k\right)^{\dagger} = I^i_k\,,\qquad   \tilde{H}^{\dagger} = \tilde{H}.\label{SU21-conj}
\eea
The generators $I^i_j$ are the $SU(2)$ symmetry generators, while the mass-dimension generator $\tilde H$ corresponds to $U(1)$ symmetry.
The superalgebra \eqref{alg1} can be regarded as a deformation of the flat ${\cal N}=4$, $d=1$ ``Poincar\'e'' superalgebra by a real mass parameter $m$.
In the limit $m = 0$, $\tilde{H}$ becomes the Hamiltonian (alias the time-translation generator) and the generators $I^i_j$ define the outer $SU(2)$ automorphisms.

One can extend \eqref{alg1} by an external $U(1)$ automorphism symmetry ($R$-symmetry) generator $F$ which has non-zero commutation relations only
with the supercharges \cite{FS}:
\bea
    \left[F, \bar{Q}_{l}\right]=-\frac{1}{2}\,\bar{Q}_{l}\,,\qquad \left[F, Q^{k}\right]=\frac{1}{2}\,Q^{k}, \qquad \left(F\right)^\dagger = F .
\eea
After redefining $\tilde H \equiv H - m F $, the extended superalgebra $su(2|1)\oplus u(1)_{\rm ext}$ acquires
the form of a centrally extended superalgebra $\hat{su}(2|1)$:
\bea
    &&\lbrace Q^{i}, \bar{Q}_{j}\rbrace = 2m I^i_j +2\delta^i_j (H -2 m F) ,\qquad\left[I^i_j,  I^k_l\right]
    = \delta^k_j I^i_l - \delta^i_l I^k_j\,,\nn
    &&\left[I^i_j, \bar{Q}_{l}\right] = \frac{1}{2}\delta^i_j\bar{Q}_{l}-\delta^i_l\bar{Q}_{j}\, ,\qquad \left[I^i_j, Q^{k}\right]
    = \delta^k_j Q^{i} - \frac{1}{2}\delta^i_j Q^{k},\nn
    &&\left[F, \bar{Q}_{l}\right]=-\frac{1}{2}\,\bar{Q}_{l}\,,\qquad \left[F, Q^{k}\right]=\frac{1}{2}\,Q^{k}.\label{alg}
\eea
All other (anti)commutators are vanishing. The generator $H$ is the relevant central charge. This extended superalgebra is also
a deformation of the ${\cal N}=4$, $d=1$ Poincar\'e superalgebra. In the limit $m = 0$, $H$ becomes the Hamiltonian
and $I^i_j$, $F$ turn into the outer $U(2)$ automorphism generators.

In the present paper, we start from the framework of the $SU(2|1)$ superspace constructed in \cite{DSQM}.
The  $SU(2|1)$, $d=1$ superspace is identified with the following coset of the extended superalgebra \eqref{alg}:
\bea
    \frac{SU(2|1)\rtimes U(1)_{\rm ext}}{SU(2)\times U(1)_{\rm int}}\, \sim \,
    \frac{\{Q^{i},\bar{Q}_{j}, H, F, I^i_j \}}{\{I^i_j, F\}}\,.\label{coset}
\eea
It is convenient to deal with the superspace coordinates $\zeta := \{t,\theta_i,\bar{\theta}^k\}$ as in \cite{DSQM,SKO}.
They are related to those in the exponential parametrization of the supercoset \eqref{coset} as
\bea
    g = \exp\left\{\left(1-\frac{2m}{3}\,\bar{\theta}^k\theta_k\right) \left(\theta_{i}Q^{i} + \bar{\theta}^{j}\bar{Q}_{j}\right) \right\}\exp{\lbrace i t H\rbrace},\qquad
    \overline{\left(\theta_i\right)}=\bar{\theta}^i.
    \label{element}
\eea

The extended supergroup $\hat{SU}(2|1)$ acts as left shifts of the supercoset element \p{element}. The
corresponding supercharges are realized as
\bea
    &&Q^i=\frac{\partial}{\partial\theta_i}-2m\,\bar{\theta}^i\bar{\theta}^k\frac{\partial}{\partial\bar{\theta}^k}
    +i\bar{\theta}^i\partial_t -m\,\bar{\theta}^i\tilde{F} + m\,\bar{\theta}^k\left(1 -m\,\bar{\theta}^k\theta_k \right)\tilde{I}^i_k\,,\nn
    &&\bar{Q}_j=\frac{\partial}{\partial\bar{\theta}^j}
    +2 m\,\theta_j\theta_k\frac{\partial}{\partial\theta_k}+i\theta_j \partial_t - m\,\theta_j\tilde{F}
    + m\,\theta_k\left(1 -m\,\bar{\theta}^k\theta_k \right)\tilde{I}^k_j\,,
     \label{Q0}
\eea
and the bosonic generators as
\bea
    &&I^i_j=\left(\bar{\theta}^i\frac{\partial}{\partial\bar{\theta}^j}-\theta_j\frac{\partial}{\partial\theta_i}\right)
    -\frac12\,{\delta^i_j}\left(\bar{\theta}^k\frac{\partial}{\partial\bar{\theta}^k}-\theta_k\frac{\partial}{\partial\theta_k}\right),\nn
    &&H= i\partial_{t}\,,\qquad F=\frac{1}{2}\left(\bar{\theta}^k\frac{\partial}{\partial\bar{\theta}^k}
    -\theta_k\frac{\partial}{\partial\theta_k}\right).\label{U2}
\eea
Here, $\tilde{I}^k_j$ and $\tilde{F}$ are
matrix generators of the $U(2)$ representation by which the given superfield is rotated with respect to its external indices.
According to \eqref{Q0}, the supersymmetric transformations $\epsilon_i$\,, $\bar{\epsilon}^i=\overline{\left(\epsilon_i\right)}$
of the superspace coordinates are given by
\bea
    \delta \theta_{i}=\epsilon_{i} +
    2m\,\bar{\epsilon}^k\theta_k\theta_{i}\,,\qquad
    \delta \bar{\theta}^{i}=\bar{\epsilon}^{i} -
    2m\,\epsilon_k\bar{\theta}^k\bar{\theta}^{i}\,,\qquad
    \delta t=i\left(\bar{\epsilon}^k\theta_k + \epsilon_k\bar{\theta}^k\right). \label{transfBas}
\eea
The $SU(2|1)$ invariant integration measure is defined as
\bea
    d\zeta = dt\,d^2\theta\, d^2\bar{\theta}\left(1+2m\,\bar{\theta}^k\theta_k\right), \quad \delta  d\zeta = 0\,.\label{measure}
\eea

The covariant derivatives ${\cal D}^i$, $\bar{\cal D}_j$, ${\cal D}_{(t)}$ are defined by
the expressions\footnote{For Grassmann coordinates and variables we use the following conventions:
$\left(\chi\right)^2 =  \chi_i\chi^i\,, \;\left(\bar\chi\,\right)^2 = \bar{\chi}^i\bar{\chi}_i$\,.}
\bea
    {\cal D}^i &=& \left[1+{m}\,\bar{\theta}^k\theta_k
    -\frac{3m^2}{8} \left(\theta\right)^2\left(\bar{\theta}\,\right)^2\right]\frac{\partial}{\partial\theta_i}
    - {m}\,\bar{\theta}^i\theta_j\frac{\partial}{\partial\theta_j}-i\bar{\theta}^i \partial_t\nn
    &&+\,m\,\bar{\theta}^i \tilde{F}- {m}\,\bar{\theta}^j\left(1 -m\,\bar{\theta}^k\theta_k \right)\tilde{I}^i_j\,,\nn
    \bar{{\cal D}}_j &=& -\left[1+ {m}\,\bar{\theta}^k\theta_k
    -\frac{3m^2}{8} \left(\theta\right)^2\left(\bar{\theta}\,\right)^2\right]\frac{\partial}{\partial\bar{\theta}^j}
    + {m}\,\bar{\theta}^k\theta_j\frac{\partial}{\partial\bar{\theta}^k}+i\theta_j\partial_t\nn
    &&-\,m\,\theta_j\tilde{F}+ {m}\,\theta_k\left(1 -m\,\bar{\theta}^k\theta_k \right)\tilde{I}^k_j\,,\nn
    {\cal D}_{(t)} &=& \partial_t \label{cov}
\eea
and satisfy, together with $\tilde{I}^k_j,  \tilde{F}\,$, the superalgebra which mimics \eqref{alg}. Under the left $\hat{SU}(2|1)$ shifts of the coset element \p{coset}
the spinor covariant derivatives undergo the induced $SU(2)$ transformations in
their doublet indices and an induced $F$ transformation with respect to which ${\cal D}^i$ and $\bar{\cal D}_i$ possess opposite charges.
In the limit $m=0\,$, the formulas of the standard flat ${\cal N}=4$, $d=1$ superspace are recovered. The superfields given on the $SU(2|1)$
superspace \p{coset} can have external $SU(2)$ indices and $U(1)$ charges on which the proper matrix realizations
of the relevant generators act.

There exists an alternative definition of the $SU(2|1)$ superspace, in which the time coordinate is associated
as a coset parameter with the total internal
$U(1)$ generator $\tilde H = H - mF$, while $F$ is still placed into the stability subgroup \cite{SKO}. As was already mentioned, in the basis $(\tilde H, F)$
the generator $F$  is split from other generators, becoming the purely external $U(1)$ automorphism. The relevant supercoset
is schematically related to \p{coset} just
by replacing $H \rightarrow \tilde H$:
\bea
    \frac{SU(2|1)\rtimes U(1)_{\rm ext}}{SU(2)\times U(1)_{\rm ext}}\, \sim \,
    \frac{\{Q^{i},\bar{Q}_{j}, \tilde H, F, I^i_j \}}{\{I^i_j, F\}} \sim \frac{\{Q^{i},\bar{Q}_{j}, \tilde H, I^i_j \}}{\{I^i_j\}}\,.\label{coset22}
\eea
The same replacement $H \rightarrow \tilde H$ should be made in the coset element \p{coset}, giving rise to the coset element $\tilde{g}$. Due to the relation $\tilde{H} = H -mF$, these two coset elements are related as
\bea
\tilde{g} = g \exp\{-imtF\}. \lb{coscos}
\eea
Under the left shifts by the fermionic generators the
coordinates $\zeta = \{t, \theta_i, \bar\theta^k \}$ are transformed according to the same formulas \p{transfBas},
so they can also be treated as the parameters of the new supercoset. The difference from the first type of the $SU(2|1)$ superspace is
the absence of independent constant shift of the time coordinate, which can still be realized under the choice \p{coset}. The left $\tilde H$ shift
gives rise to a shift of $t$ accompanied by the proper $U(1)$ rotation of the Grassmann coordinates. The corresponding covariant spinor derivatives
differ from \p{cov} by the absence of the part $\sim \tilde{F}$ and by some overall phase factor ensuring them to transform only under
induced $SU(2)$ transformations. These modifications can be easily established
from the precise relation \p{coscos}. The corresponding superfields can carry only external $SU(2)$ indices.

\setcounter{equation}{0}
\section{Embedding of $su(2|1)$ into $D\left(2,1;\alpha\right)$}\label{Sec. 3}
The most general $d=1$, ${\cal N}=4$ superconformal algebra is $D\left(2,1;\alpha\right)$ \cite{superc,Sorba}. It is spanned by 8 fermionic
and 9 bosonic generators with the following non-vanishing (anti)commutators:
\begin{eqnarray}
    &&\{ {Q}_{\alpha i i^\prime},  {Q}_{\beta j j^\prime}\}=
2\,\Big[\epsilon_{ij}\epsilon_{i^\prime j^\prime} {T}_{\alpha\beta}+
\alpha \,\epsilon_{\alpha\beta}\epsilon_{i^\prime j^\prime} {J}_{ij}-
(1{+}\alpha)\,\epsilon_{\alpha\beta}\epsilon_{ij} {L}_{i^\prime j^\prime}\Big]\,,\label{basD} \\
    &&\left[{T}_{\alpha\beta}, {Q}_{\gamma i i^\prime}\right] =
-i\,\epsilon_{\gamma(\alpha}{Q}_{\beta) i i^\prime}\,,\qquad \left[{T}_{\alpha\beta}, {T}_{\gamma\delta}\right] =
i\left(\epsilon_{\alpha\gamma}{T}_{\beta\delta} +\epsilon_{\beta\delta}{T}_{\alpha\gamma}\right),\nn
    &&\left[{J}_{ij}, {Q}_{\alpha k i^\prime}\right] =
-i\,\epsilon_{k(i}{Q}_{\alpha j) i^\prime}\,,\qquad\left[{J}_{ij}, {J}_{kl}\right] =
i\left(\epsilon_{ik}{J}_{jl} +\epsilon_{jl}{J}_{ik}\right),\nn
    &&\left[{L}_{i^\prime j^\prime}, {Q}_{\alpha i k^\prime}\right] =
-i\,\epsilon_{k^\prime (i^\prime}{Q}_{\alpha i j^\prime)}\,,\qquad
    \left[{L}_{i^\prime j^\prime }, {L}_{k^\prime l^\prime }\right] =
i\left(\epsilon_{i^\prime k^\prime }{L}_{j^\prime l^\prime } +
\epsilon_{j^\prime l^\prime }{L}_{i^\prime k^\prime }\right).\label{D12}
\end{eqnarray}
The bosonic subalgebra is $su(2)\oplus su'(2)\oplus so(2,1)$ with the generators ${J}_{ik}$, ${L}_{i^\prime k^\prime }$ and ${T}_{\alpha\beta}$,
respectively. Switching $\alpha$ as $\alpha \leftrightarrow -(1 +\alpha)$ amounts to switching $SU(2)$ generators
as ${J}_{ik} \leftrightarrow {L}_{i^\prime k^\prime }\,$\footnote{More generally, the equivalent superalgebras are related through the substitutions $\alpha \rightarrow -(1 + \alpha), \; \alpha^{-1}\,$.}. The Hermitian conjugation rules are:
\begin{equation}\label{sca-conj}
({Q}_{\alpha i i^\prime}){}^\dag = \epsilon^{ij}\epsilon^{i^\prime j^\prime}{Q}_{\alpha jj^\prime}\,,\quad
({T}_{\alpha\beta})^\dag = {T}_{\alpha\beta}\,,\quad
({J}_{ij})^\dag = \epsilon^{ik}\epsilon^{jl}{J}_{kl}\,,\quad
({L}_{i^\prime j^\prime}){}^\dag = \epsilon^{i^\prime k^\prime}\epsilon^{j^\prime l^\prime}{L}_{k^\prime l^\prime}\,.
\end{equation}

The ${\cal N}=4$, $d=1$ Poincar\'e superalgebra can be defined as the following subalgebra of $D\left(2,1;\alpha\right)$:
\begin{eqnarray}
    &&\{ {Q}_{1 i i^\prime},  {Q}_{1 j j^\prime}\}=
2\epsilon_{ij}\epsilon_{i^\prime j^\prime} \hat{H},\label{N4}
\end{eqnarray}
where $\hat H$ is one of the generators of the conformal algebra $so(2,1)$ represented in \p{basD} and \p{D12} by the generators
${T}_{\alpha\beta}\,$. The standard conformal $so(2,1)$ generators are identified as
\bea
&&   \hat H := {T}_{11}\,,\qquad \hat K := {T}_{22}\,,\qquad \hat D := {T}_{12}\,, \label{T} \\
&&  \left[\hat{D},\hat{H}\right]=-i\hat{H},\qquad \left[\hat{D},\hat{K}\right]=i\hat{K},\qquad\left[\hat{H},\hat{K}\right]=2i\hat{D}.\label{Genso21}
\eea

In the degenerate case $\alpha=-1$ one may retain all eight fermionic generators ${Q}_{\alpha i i^\prime}$
and only six bosonic generators ${T}_{\alpha\beta}$, ${J}_{ij}$ forming together the superalgebra $psu(1,1|2)$ without central charge.
The second $SU(2)$ generators ${L}_{i^\prime j^\prime}$ drop out from the basic anticommutation relation \p{basD}. Yet, they can
be treated as the generators of some extra $SU'(2)$ automorphisms. Taking $\alpha=0$, one can suppress, in the same way,
the generators ${J}_{ij}$ in \eqref{basD}, ending up with $SU'(2)$ as the internal group and the first $SU(2)$ as the external automorphism group.
Thus in the cases $\alpha = -1$ and $\alpha = 0$ the supergroup $D\left(2,1;\alpha\right)$ is reduced to a semi-direct product:
\bea
    \alpha = -1,0,\qquad D\left(2,1;\alpha\right) \cong PSU(1,1|2)\rtimes SU(2)_{\rm ext}, \label{psu112}
\eea
with $SU(2)_{\rm ext}$ being generated, respectively, by $L_{i^\prime, j^\prime}$ or $J_{ij}$.
Note that in these exceptional cases one can extend the $psu(1,1|2)$ superalgebra by the proper $SU(2)_{\rm ext}$ triplets
of central charges \cite{SCM1}.  If these central charges are constant, the triplet can be reduced
to one central charge, which enlarges $psu(1,1|2)$ to $su(1,1|2)$ and simultaneously breaks $SU(2)_{\rm ext}$ to $U(1)_{\rm ext}$
(see Appendix \ref{App-A}).

We will be interested in the most general embedding of the superalgebra $su(2|1)$ into $D\left(2,1;\alpha\right)$. To this end, we pass to the new
basis in $D\left(2,1;\alpha\right)$ through the following linear relations:
\bea
    && \varepsilon^{ik}Q_{1k1'}=:-\frac{1}{2}\left(S^i + Q^i\right),\qquad Q_{1j2'}=:-\frac{1}{2}\left(\bar{S}_j +  \bar{Q}_j\right),\nn
    && \varepsilon^{ik}Q_{2k1'}=:\frac{i}{\mu}\left(Q^i - S^i\right),\qquad Q_{2j2'}=:-\frac{i}{\mu}\left(\bar{Q}_j - \bar{S}_j \right),\nn
    && T_{22}=:\frac{2}{\mu^2}\left[{\cal H} - \frac{1}{2}\left(T+\bar{T}\right)\right],\qquad T_{11}=:
    \frac{1}{2}\left[{\cal H} + \frac{1}{2}\left(T+\bar{T}\right)\right],\nn
    && T_{12}=T_{21}=:\frac{i}{2\mu}\left(T-\bar{T}\right),\qquad \mu \neq 0\,,\nn
    && L_{1'1'}=:-i C,\quad L_{2'2'}=: i \bar{C},\quad L_{1'2'}=L_{2'1'}=: -i F,\qquad J^i_j=: -i I^i_j\,.\label{sca}
\eea
Here $\mu$ is a real parameter of the mass dimension. In the new basis, the (anti)commutators \p{basD}, \p{D12} are rewritten as
\bea
    &&\lbrace Q^{i}, \bar{Q}_{j}\rbrace = -2\alpha\mu\, I^i_j +2\delta^i_j\left[{\cal H} + \left(1+\alpha\right)\mu\,F\right] ,\nn
    &&\{S^i, \bar{S}_j\}=2\alpha\mu\, I^i_j + 2\delta^i_j \left[{\cal H} -\left(1+\alpha\right) \mu\,F\right] ,\nn
    &&\{S^i, \bar{Q}_j\}=2\delta^i_j T,\qquad \{Q^i, \bar{S}_j\}=2\delta^i_j \bar{T},\nn
    &&\{Q^i,S^k\} = -2\left(1+\alpha\right)\mu\,\varepsilon^{ik}C,\qquad\{\bar{Q}_j, \bar{S}_k\}=2\left(1+\alpha\right)\mu\, \varepsilon_{jk}\bar{C}\,,
    \label{conf-anticomm}
\eea
\bea
    &&\left[I^i_j,  I^k_l\right] = \delta^k_j I^i_l - \delta^i_l I^k_j\,,\nn
    &&\left[I^i_j, \bar{Q}_{l}\right] = \frac{1}{2}\,\delta^i_j\bar{Q}_{l}-\delta^i_l\bar{Q}_{j}\, ,\qquad \left[I^i_j, Q^{k}\right]
    = \delta^k_j Q^{i} - \frac{1}{2}\,\delta^i_j Q^{k},\nn
    &&\left[I^i_j, \bar{S}_{l}\right] = \frac{1}{2}\,\delta^i_j\bar{S}_{l}-\delta^i_l\bar{S}_{j}\, ,\qquad \left[I^i_j, S^{k}\right]
    = \delta^k_j S^{i} - \frac{1}{2}\,\delta^i_j S^{k}\,,
\eea
\bea
    &&\left[C , \bar{C}\right]=2F,\qquad \left[F , C\right]=C,\qquad \left[F , \bar{C}\right]=-\bar{C},\nn
    &&\left[C,\bar{Q}_j\right]=-\varepsilon_{jl}S^l ,\quad \left[C,\bar{S}_j\right]=-\varepsilon_{jl}Q^l ,\quad \left[\bar{C},Q^i\right]=-\varepsilon^{ik}\bar{S}_k \,,
    \quad \left[\bar{C},S^i\right]=-\varepsilon^{ik}\bar{Q}_k \,, \nn
    &&\left[F, \bar{Q}_{l}\right]=-\frac{1}{2}\,\bar{Q}_{l}\,,\quad \left[F, Q^{k}\right]=\frac{1}{2}\,Q^{k}\,,\quad
    \left[F, \bar{S}_{l}\right]=-\frac{1}{2}\,\bar{S}_{l}\,,\quad \left[F, S^{k}\right]=\frac{1}{2}\,S^{k}\,,
\eea
\bea
    &&\left[T , \bar{T}\right] = - 2\mu\,{\cal H} , \qquad \left[{\cal H}, T\right]= \mu \,T,\qquad \left[{\cal H}, \bar{T}\right]=-\mu\,\bar{T}\,,\nn
    &&\left[T , Q^i\right]=- \mu \,S^i,\quad \left[T , \bar{S}_j\right]= - \mu \,\bar{Q}_j\,,\quad
    \left[\bar{T} , \bar{Q}_j\right]= \mu\,\bar{S}_j\,,\quad\left[\bar{T} , S^i\right]= \mu\, Q^i\,,\nn
    &&\left[ {\cal H}, \bar{S}_{l}\right]= -\frac{\mu}{2}\,\bar{S}_{l}\,,\quad \left[ {\cal H}, S^{k}\right]= \frac{\mu}{2}\, S^{k}\,,
    \quad \left[ {\cal H}, \bar{Q}_{l}\right]= \frac{\mu}{2}\,\bar{Q}_{l}\,,\quad \left[ {\cal H}, Q^{k}\right]= -\frac{\mu}{2}\, Q^{k}\,.\label{conf-alg}
\eea
The bosonic sector consisting of  the three mutually commuting algebras is now given by the following sets of the generators
\bea
  su(2)\oplus su'(2)\oplus so(2,1) \equiv  \{I^i_k\}\oplus\{F, C, \bar{C}\}\oplus\{{\cal H}, T, \bar{T}\}\,.\label{subalg}
\eea
According to \eqref{sca-conj} and \eqref{sca}, the conjugation rules are as follows
\bea
    &&\left(Q^{k}\right)^{\dagger} = \bar{Q}_{k}\,,\qquad \left(S^{k}\right)^{\dagger} = \bar{S}_{k}\,,\nn
    &&\left(F\right)^\dagger = F,\qquad\left(C\right)^{\dagger} = \bar{C}\,,\qquad
    \left(I_i^k\right)^{\dagger} = I^i_k\,,\qquad   {\cal H}^{\dagger} = {\cal H}\,,\qquad
    \left(T\right)^{\dagger} = \bar{T}\,.\label{Herm-conj}
\eea
Note the relation
\bea
{\cal H} = \hat H + \frac{\mu^2}{4}\,\hat K\,. \label{tildeHhatH}
\eea

In the contraction limit $\mu = 0$, the algebra \eqref{conf-anticomm} -- \eqref{conf-alg} becomes a kind of ${\cal N}=8, d=1$ Poincar\'e
superalgebra (with the common Hamiltonian ${\cal H}$) extended by the central charges $T, \bar T$ originating from the $so(2,1)$ generators.
The remaining two $su(2)$
subalgebras become outer automorphism algebras which form a semi-direct product with this ${\cal N}=8, d=1$
superalgebra\footnote{The full automorphism group $SO(8)$ of the ${\cal N}=8, d=1$ superalgebra is broken down to $SO(4) \sim SU(2)\times SU'(2)$ due to the presence
of central charges $T, \bar T$.}. At any $\mu \neq 0$, the relations \eqref{sca} defining the new basis contain no
singularities, and so eqs. \eqref{conf-anticomm} -- \eqref{conf-alg} yield an equivalent form
of the original superalgebra $D\left(2,1;\alpha\right)$. After coming back to the original superconformal generators any dependence
of the (anti)commutation relations on $\mu$ disappears while it still retains in the realizations of $D\left(2,1;\alpha\right)$
on the coordinates of the $SU(2|1)$ superspaces (see below). Taking the $\mu=0$ limit in this basis gives rise
to the standard parabolic realizations of  $D\left(2,1;\alpha\right)$ in the flat ${\cal N}=4, d=1$ superspaces.

The $su(2|1)$ basis in $D\left(2,1;\alpha\right)$ makes manifest some remarkable properties of this superalgebra which are implicit
in the ``standard'' basis.

\paragraph*{i.}It is straightforward to see that the superconformal algebra \p{conf-anticomm}-\p{conf-alg}
includes as a subalgebra  the following superalgebra $su(2|1)$:
\bea
    &&\lbrace Q^{i}, \bar{Q}_{j}\rbrace = -2\alpha\mu\, I^i_j +2\delta^i_j\left[{\cal H} + \left(1+\alpha\right)\mu\,F\right] ,\nn
    &&\left[I^i_j, \bar{Q}_{l}\right] = \frac{1}{2}\delta^i_j\bar{Q}_{l}-\delta^i_l\bar{Q}_{j}\, ,\qquad \left[I^i_j, Q^{k}\right]
    = \delta^k_j Q^{i} - \frac{1}{2}\delta^i_j Q^{k},\nn
    &&\left[F, \bar{Q}_{l}\right]=-\frac{1}{2}\,\bar{Q}_{l}\,,\qquad \left[F, Q^{k}\right]=\frac{1}{2}\,Q^{k},\nn
    &&\left[{\cal H}, \bar{Q}_{l}\right]= \frac{\mu}{2}\,\bar{Q}_{l}\,,\qquad \left[{\cal H}, Q^{k}\right]= -\frac{\mu}{2}\, Q^{k}.  \label{alpha}
\eea
These relations coincide with \p{alg} under the following identifications
\bea
&& m(\mu) = -\alpha\mu\,,\label{mu} \\
&& H(\mu) ={\cal H} + \mu\,F.\label{H}
\eea
We observe that the closure of the $SU(2|1)$ supercharges depends on the parameter $\alpha$, because the $SU(2)$ and $SU'(2)$ generators
${J}_{ij}= -iI_{ij}$ and $L_{i^\prime j^\prime} \sim \left\{F, C, \bar C\right\} $ appear in the basic anticommutator \eqref{basD} with
the factors $\alpha$, $1{+}\alpha$, respectively. The $U(1)$ generator $F$ in \p{alpha} comes from $su'(2)$, while the first
$su(2)$ with the generators $I_{ij}$ is just $su(2) \subset su(2|1)\,$.

\paragraph*{ii.}We see from \eqref{conf-anticomm} -- \eqref{conf-alg} that there exists another $su(2|1) \subset D\left(2,1;\alpha\right)$
generated by the generators $S^i, \bar S_j$ and corresponding to the identification
\bea
&& m(-\mu) = \alpha\mu\,,\label{mu2} \\
&& H(-\mu) ={\cal H} - \mu\,F\label{H2}
\eea
in \p{alg}. Hence, its (anti)commutation relations are obtained from \p{alpha} via the substitution $\mu \rightarrow -\mu$ and passing to
the new independent supercharges $(S_i, \bar S^j)$. As follows from \eqref{conf-anticomm} -- \eqref{conf-alg},
all the remaining generators of $D\left(2,1;\alpha\right)$ (i.e. $T, \bar T, C, \bar C$) appear in the cross-anticommutators of the supercharges
$(Q^i, \bar Q_j)$ with $(S_i, \bar S^j)$. Thus the superalgebra $D\left(2,1;\alpha\right)$ can be represented as a closure
of its two $su(2|1)$ supersubalgebras: $su(2|1)$ given by the relations \p{alpha} and another independent  $su(2|1)$
with the (anti)commutation relations obtained from those of the former $su(2|1)$ through the replacement $\mu \rightarrow -\mu$.
We were not able to find  such a  statement about the structure of
$D\left(2,1;\alpha\right)$ in the literature. This property is similar to the property that the ${\cal N}=1, d=4$ superconformal group $SU(2,2|1)$
can be viewed as a closure of its two different $OSp(1,4)$ subgroups related to each other through the analogous ``reflection'' of the anti-De-Sitter
radius as a parameter of contraction to the flat ${\cal N}=1, d=4$ Poincar\'e supersymmetry \cite{IvSo}. In what follows, this observation
will be useful for constructing $D\left(2,1;\alpha\right)$ invariant subclasses of the $SU(2|1)$ invariant actions.

\paragraph*{iii.}In the cases $\alpha = -1$ and $\alpha = 0$ the supergroup $D\left(2,1;\alpha\right)$ is reduced
to the semi-direct product \p{psu112}, with $SU(2)_{\rm ext}$ being generated, respectively, by $L_{i^\prime, j^\prime}$ or $J_{ij} = -i I_{ij}$.
The remaining $SU(2)$ subgroups
enter the relevant $PSU(1,1|2)$ factors. Each of the corresponding superalgebras $psu(1,1|2)$ can still be interpreted as a closure of its two  $su(2|1)$ subalgebras,
like in the
case of $\alpha \in \mathbb{R}\backslash \{0\}, \mathbb{R}\backslash \{-1\}$. In particular, the superalgebra \p{alpha} at $\alpha = -1$
is identical to \eqref{alg1} with $m = \mu$ and ${\cal H}$ as the $U(1)$ generator. The generator $F$ splits off as an external automorphism.

\paragraph*{iv.} One more peculiarity is associated with the presence of the ``composite'' deformation parameter $m = -\alpha \mu$ in \p{alpha}. It vanishes
not only in the standard contraction limit $\mu =0$, but also at $\alpha =0$ with $\mu \neq 0\,.$  For $\alpha = 0$, the superalgebra \eqref{alpha}
is reduced to the flat ${\cal N}=4$ superalgebra
\bea
    &&\lbrace Q^{i}, \bar{Q}_{j}\rbrace = 2\delta^i_j\left({\cal H} + \mu\,F\right) ,\nn
    &&\left[F, \bar{Q}_{l}\right]=-\frac{1}{2}\,\bar{Q}_{l}\,,\qquad \left[F, Q^{k}\right]=\frac{1}{2}\,Q^{k},\nn
    &&\left[{\cal H}, \bar{Q}_{l}\right]= \frac{\mu}{2}\,\bar{Q}_{l}\,,\qquad \left[{\cal H}, Q^{k}\right]= -\frac{\mu}{2}\, Q^{k}.  \label{alpha0}
\eea
This algebra is still a subalgebra of $D(2,1;\alpha{=}0)$. However, it does not coincide with the standard flat ${\cal N}=4, d=1$ Poincar\'e superalgebra
corresponding to the limit $\mu = 0$,
because the r.h.s. of the anticommutator in \p{alpha0} still involves $\mu$ and is a sum of ${\cal H}$ and the internal $U(1)$ charge $F$.
The $SU(2)$ generators ${I}_{i}^{j}$ now define automorphisms of both the superalgebra \eqref{alpha0} and the $\alpha=0$ superalgebra $psu(1,1|2)$,
while $F$ is an internal $U(1)$ generator. The whole $D(2,1;\alpha{=}0)$ superalgebra (including the $so(2,1)$ generators and those of the $su'(2) \sim \{F, C, \bar C \}$)
can now be treated as a closure of the superalgebra \p{alpha0} and its $\mu \rightarrow -\mu$ counterpart\footnote{In the $\alpha=0$ case one can still define
$su(2|1) \subset D(2,1;\alpha{=}0)$ which involves $su'(2)\sim \{F, C, \bar C \}$ as the internal subalgebra, as well as the proper analog of the $U(1)$ generator ${\cal H}$.}.

To avoid a confusion, let us point out that both \eqref{alpha0} and \eqref{N4} can of course be regarded as the Poincar\'e ${\cal N}=4, d=1$ superalgebras.
However, in contrast to \eqref{N4}, the superalgebra \eqref{alpha0} is embedded in the superconformal algebra in a different way,
with the Hamiltonian $H = {\cal H} + \mu F$ defined in \p{H}, instead of the standard $\hat H$ in \eqref{N4} (recall eq. \p{tildeHhatH}). In the limit $\mu \to 0$,
any difference between ${\cal H}$, $H$ and $\hat H$ disappears.

\paragraph*{v.} It is worth noting that the parameter $\alpha$ characterizes only the superconformal mechanics models, while the generic
$SU(2|1)$ models lack any dependence on it. So in the case of superconformal models we deal with the pair of parameters, $\alpha$ and $\mu$.
In the particular case $\alpha=-1$, we have $m=\mu$.

\paragraph*{vi.} Besides the $SU(2|1)$ superspaces \p{coset}, \p{coset22}, we can now consider another type of the $SU(2|1)$ superspace defined as the supercoset
\bea
    \frac{SU(2|1)\rtimes U(1)_{\rm ext}}{SU(2)\times U(1)_{\rm int}}\, \sim \,
    \frac{\{Q^{i},\bar{Q}_{j}, {\cal H}, F, I^i_j \}}{\{I^i_j, F\}}\,.\label{coset1}
\eea
According to \eqref{subalg}, this definition of superspace matches to the proper embedding of $SU(2|1)$
in $D\left(2,1;\alpha\right)$ for $\alpha \in \mathbb{R}\backslash\{0\}$:
\begin{equation}
\begin{CD}
    D(2,1;\alpha) @>{\rm bos}>> \quad SU(2) @. \;\times \;@. SU(2) @. \;\times \;@. SO(2,1), @.\\
    @VVV @VV{I}_{i}^{j}V @. @VV{F}V   @.    @VV{\cal H}V\\
     SU(2|1)\rtimes U(1)_{\rm ext} @>{\rm bos}>>  \quad  SU(2) @. \times @. U(1)_{\rm ext} @. \times @. U(1). @.
\end{CD} \label{emb}
\end{equation}
In the case $\alpha = -1$ corresponding to the second line in \p{emb},  one can omit the generator $F$ in \eqref{coset1} since it becomes an
external automorphism. So \p{coset1} is reduced to \p{coset22} in this case. For generic $\alpha \in \mathbb{R}\backslash\{-1\}$,
the coset \p{coset1} ``interpolates'' between \p{coset} and \p{coset22} since $F$ appears in the r.h.s. of the anticommutator in \p{alpha}
along with the generator ${\cal H}$, and so cannot be decoupled.

\paragraph*{vii.} In the limit $\alpha = 0$, the relevant coset is
\bea
    \frac{({\cal N}=4, d=1)\rtimes U(1)_{\rm ext}}{U(1)_{\rm int}}\, \sim \,
    \frac{\{Q^{i},\bar{Q}_{j}, {\cal H}, F\}}{\{F\}}\,,\label{coset2}
\eea
where $({\cal N}=4, d=1)\rtimes U(1)_{\rm ext}$ stands for the semi-direct product
of the supergroup with the algebra \eqref{alpha0} and the external $U(1)$ automorphism generated by $F \in \{F, C, \bar C\}$.
 We can deal with the coset superspace \p{coset2} in the standard manner, just substituting $\alpha =0$ into all the relations
 of the $SU(2|1)$ superspace formalism pertinent to the choice \p{coset1}.

\subsection{Superconformal generators}
Superconformal generators of \eqref{conf-anticomm} -- \eqref{conf-alg} can be naturally realized on the $SU(2|1)$ superspace  \eqref{coset1}.
An element of this supercoset is defined as
\bea
    g_1 = \exp\left\{\left(1+\frac{2\alpha \mu}{3}\,\bar{\theta}^k\theta_k \right) \left(\theta_{i}Q^{i}
    + \bar{\theta}^{j}\bar{Q}_{j}\right)\right\} \exp{\lbrace i t {\cal H}\rbrace},
    \label{element1}
\eea
where the superspace coordinates $\{t,\theta_i,\bar{\theta}^k\}$ coincide with those defined in \eqref{element}.
Because of the relation \p{H}, the coset elements \p{element1} and \p{element} are related as
\bea
g_1 = g\exp\{-i\mu t F\}\,. \label{coscos1}
\eea
In the particular case $\alpha=0$, the relevant superspace coset \eqref{coset2} is parametrized
by the flat superspace coordinates $\zeta_{(\alpha=0)} = \{t,\theta_i,\bar{\theta}^k\}$.
An element of this coset is obtained by setting $\alpha = 0$ in \p{element1}.

Dropping matrix parts of generators, one can obtain the $SU(2|1)$ supercharges for generic $\alpha$ just
through the substitution $m = -\alpha\mu$ in \eqref{Q0}:
\bea
    Q^i=\frac{\partial}{\partial\theta_i}+2\alpha \mu\,\bar{\theta}^i\bar{\theta}^k\frac{\partial}{\partial\bar{\theta}^k}
    +i\bar{\theta}^i\partial_t\,,\qquad
    \bar{Q}_j=\frac{\partial}{\partial\bar{\theta}^j}
    -2\alpha \mu\,\theta_j\theta_k\frac{\partial}{\partial\theta_k}+i\theta_j \partial_t\,.
     \label{Q}
\eea
They generate the $su(2|1)$ superalgebra \eqref{alpha} with the bosonic generators
\bea
    &&I^i_j=\left(\bar{\theta}^i\frac{\partial}{\partial\bar{\theta}^j}-\theta_j\frac{\partial}{\partial\theta_i}\right)
    -\frac12\,{\delta^i_j}\left(\bar{\theta}^k\frac{\partial}{\partial\bar{\theta}^k}-\theta_k\frac{\partial}{\partial\theta_k}\right),\nn
    &&{\cal H}= i\partial_{t}-\frac{\mu}{2}\left(\bar{\theta}^k\frac{\partial}{\partial\bar{\theta}^k}
    -\theta_k\frac{\partial}{\partial\theta_k}\right),\qquad F=\frac{1}{2}\left(\bar{\theta}^k\frac{\partial}{\partial\bar{\theta}^k}
    -\theta_k\frac{\partial}{\partial\theta_k}\right).\label{H3}
\eea
The extra supercharges of the superconformal algebra $D\left(2,1;\alpha\right)$ are defined as
\bea
    S^i &=& e^{-i\mu t}\bigg\{\left[1 - \left(1+2\alpha\right)\mu\,\bar{\theta}^k\theta_k
    - \frac{1}{4}\left(1+2\alpha\right)^2\mu^2\left(\theta\right)^2\left(\bar{\theta}\,\right)^2 \right]\frac{\partial}{\partial\theta_i}
    + 2\left(1+\alpha\right)\mu\,\bar{\theta}^i\theta_k\frac{\partial}{\partial\theta_k}\nn
    && + \, i\bar{\theta}^i\left[1+\left(1+2\alpha\right)\mu\,\bar{\theta}^k\theta_k\right]\partial_t \bigg\},\nn
    \bar{S}_j &=& e^{i\mu t} \bigg\{\left[1 - \left(1+2\alpha\right)\mu\,\bar{\theta}^k\theta_k
    - \frac{1}{4}\left(1+2\alpha\right)^2\mu^2\left(\theta\right)^2\left(\bar{\theta}\,\right)^2\right] \frac{\partial}{\partial\bar{\theta}^j}
     -2\left(1+\alpha\right)\mu\,\theta_j\bar{\theta}^k\frac{\partial}{\partial\bar{\theta}^k} \nn
     && +\, i\theta_j\left[1+\left(1+2\alpha\right)\mu\,\bar{\theta}^k\theta_k\right] \partial_t
    \bigg\}.\label{S}
\eea
The anticommutators of \eqref{Q} with \eqref{S} give the new bosonic generators
\bea
    &&T=e^{-i\mu t}\left\{i\left[1-\frac{1}{4}\left(1+2\alpha\right)\mu^2\left(\theta\right)^2\left(\bar{\theta}\,\right)^2\right]\partial_t
    +\mu\left[1 -\left(1+2\alpha\right)\mu\,\bar{\theta}^k\theta_k \right] \theta_i\frac{\partial}{\partial\theta_i}\right\},\nn
    &&\bar{T}=e^{i\mu t}\left\{i\left[1 - \frac{1}{4}\left(1+2\alpha\right)\mu^2\left(\theta\right)^2\left(\bar{\theta}\,\right)^2\right]\partial_t
     - \mu\left[1 - \left(1+2\alpha\right)\mu\,\bar{\theta}^k\theta_k \right] \bar{\theta}^i\frac{\partial}{\partial\bar{\theta}^i}\right\},\nn
    &&C = e^{-i\mu t}\varepsilon_{jl}\left[1+\left(1+2\alpha\right)\mu\,\bar{\theta}^k\theta_k\right]\bar{\theta}^j\frac{\partial}{\partial\theta_l}\,,\nn
    &&\bar{C}=e^{i\mu t}\varepsilon^{jl}\left[1+\left(1+2\alpha\right)\mu\,\bar{\theta}^k\theta_k\right]\theta_j\frac{\partial}{\partial\bar{\theta}^l}\,.\label{bos}
\eea
Under the $\varepsilon$, $\bar{\varepsilon}$ transformations generated by \eqref{S},
\bea
    \delta \theta_{i}&=&\left[1 - \left(1+2\alpha\right)\mu\,\bar{\theta}^k\theta_k
    - \frac{1}{4}\left(1+2\alpha\right)^2\mu^2\left(\theta\right)^2\left(\bar{\theta}\,\right)^2 \right]\varepsilon_i\,e^{-i\mu t}\nn
    &&+\, 2\left(1+\alpha\right)\mu\,\varepsilon_k\bar{\theta}^k\theta_i\,e^{-i\mu t},\nn
    \delta \bar{\theta}^{i}&=&\left[1 - \left(1+2\alpha\right)\mu\,\bar{\theta}^k\theta_k
    - \frac{1}{4}\left(1+2\alpha\right)^2\mu^2\left(\theta\right)^2\left(\bar{\theta}\,\right)^2\right] \bar{\varepsilon}^i\,e^{i\mu t}\nn
    &&-\,2\left(1+\alpha\right)\mu\,\bar{\varepsilon}^k\theta_k\bar{\theta}^i\, e^{i\mu t}, \nn
    \delta t &=& i\left(\bar{\varepsilon}^k\theta_k e^{i\mu t}
    + \varepsilon_k\bar{\theta}^k e^{-i\mu t}\right)\left[1+\left(1+2\alpha\right)\mu\,\bar{\theta}^k\theta_k\right],\label{trS}
\eea
the $SU(2|1)$ invariant measure \eqref{measure} is transformed as
\bea
    \delta_{\varepsilon}\,d\zeta = 2\mu\, d\zeta\left(1 - \mu \,\bar{\theta}^k\theta_k \right)
    \left( \bar{\varepsilon}^i \theta_i\, e^{i\mu t} - \varepsilon_i \bar{\theta}^i\, e^{-i\mu t} \right).\label{tr-measure}
\eea

Starting from the new coset given by \eqref{element1} and taking advantage of the relation \p{coscos1},
one can calculate the relevant covariant derivatives
\bea
    {\cal D}^i &=& e^{-\frac{i}{2}\mu t}\bigg\{\left[1-\alpha\mu\,\bar{\theta}^k\theta_k
    -\frac{3}{8}\, \alpha^2\mu^2\left(\theta\right)^2\left(\bar{\theta}\,\right)^2\right]\frac{\partial}{\partial\theta_i}
    +\alpha\mu\,\bar{\theta}^i\theta_j\frac{\partial}{\partial\theta_j}-i\bar{\theta}^i \partial_t\nn
    &&-\,\left(1+\alpha\right)\mu\,\bar{\theta}^i \tilde{F} + \alpha\mu\,\bar{\theta}^j\left(1 +\alpha\mu\,\bar{\theta}^k\theta_k \right)\tilde{I}^i_j\bigg\},\nn
    \bar{{\cal D}}_j &=& e^{\frac{i}{2}\mu t}\bigg\{-\left[1-\alpha\mu\,\bar{\theta}^k\theta_k
    -\frac{3}{8}\, \alpha^2\mu^2\left(\theta\right)^2\left(\bar{\theta}\,\right)^2\right]\frac{\partial}{\partial\bar{\theta}^j}
    -\alpha\mu\,\bar{\theta}^k\theta_j\frac{\partial}{\partial\bar{\theta}^k}+i\theta_j\partial_t\nn
    &&+\,\left(1+\alpha\right)\mu\,\theta_j\tilde{F} - \alpha\mu\,\theta_k\left(1 +\alpha\mu\,\bar{\theta}^k\theta_k \right)\tilde{I}^k_j\bigg\},\nn
    {\cal D}_{(t)} &=& \partial_t\,.\label{cov1}
\eea
Together with the matrix generators $\tilde{I}_k^i$, $\tilde{F}$ they mimic the superalgebra \eqref{alpha}.
In the particular case $\alpha=-1$, the matrix generator $\tilde{F}$ drops out from \eqref{cov1}, which is consistent
with the superalgebra \eqref{alpha} at $\alpha=-1$ \cite{SKO}. In this case the supercoset \p{coset1}, \p{element1} is reduced to the
supercoset \p{coset22} with $\tilde{H} = {\cal H}$ and $m = \mu$.
In the case $\alpha = 0$, the generators $\tilde{I}_k^i$ drop out (they become the outer automorphism ones). The $\alpha=0$ covariant
derivatives correspond to the degenerate supercoset \p{coset2}.

The redefinition \eqref{mu} allows one to avoid singularities at $\alpha = 0$.
Taking $\alpha = 0$ in the superconformal generators \eqref{Q} -- \eqref{bos}, one can naturally pass to the
generators corresponding to the coset space \eqref{coset2} with the relevant algebra \eqref{alpha0}.
Thus, within the $SU(2|1)$ superspace defined as the supercoset \p{coset1} with the elements \p{element1},
the superspace realization of superconformal generators has been written
in the universal form consistent with both choices $\alpha = 0$ and $\alpha \neq 0$, i.e., with any choice of $\alpha \in \mathbb{R}$.
This refers to the covariant derivatives \eqref{cov1} as well.

Any dependence of the superalgebra relations \eqref{conf-anticomm} -- \eqref{conf-alg} on the dimensionful parameter $\mu$ naturally disappears
after passing to the original basis \eqref{basD}, \eqref{D12}. However, in the realization of the generators \eqref{sca}
on the superspace coordinates the dependence on $\mu$ is still retained.
Thus the parameter $\mu$ is a deformation parameter of the particular {\it superspace realization} of \eqref{basD}, \eqref{D12}.
This new deformed realization corresponds to the trigonometric type of ${\cal N}=4$ superconformal mechanics \cite{HT}.
Sending $\mu \to 0$  in these realizations (and in the corresponding realizations on the $d=1$ fields) reduces
the deformed superconformal models to the standard superconformal mechanics models
of the parabolic type \cite{SCM,SCM1,superc}.

To be more precise, the trigonometric form of the conformal generators $\{{\cal H}, T, \bar{T}\}\,$,
\bea
    {\cal H}=i\partial_t\,,\qquad T = i e^{-i\mu t}\partial_t \,,\qquad
    \bar{T} = i e^{i\mu t}\partial_t\,,\label{trig}
\eea
is obtained as the bosonic truncations of the generators defined by eqs. \eqref{H3}, \eqref{bos}
(or an alternative realization of these generators given in the next subsection).
The standard $so(2,1)$ generators $\hat H, \hat K$ and $\hat D$ defined in \eqref{T} and \eqref{sca} are expressed, respectively, as
\bea
    \hat{H}=\frac{i}{2}\left(1 + \cos{\mu t}\,\right)\partial_t\,,\quad
    \hat{K}=\frac{2i}{\mu^2}\left(1 - \cos{\mu t}\,\right)\partial_t\,,\quad
    \hat{D}=\frac{i}{\mu}\sin{\mu t}\,\partial_t\,,\qquad \mu \neq 0\,. \label{trigonso21}
\eea
These generators satisfy the conventional relations of the $d=1$ conformal algebra:
\bea
    \left[\hat{D},\hat{H}\right]=-i\hat{H},\qquad \left[\hat{D},\hat{K}\right]=i\hat{K},\qquad\left[\hat{H},\hat{K}\right]=2i\hat{D}.
\eea
Thus, the definition of conformal superalgebra by eqs. \eqref{conf-anticomm} -- \eqref{conf-alg}
automatically provides the trigonometric form for the conformal algebra $so(2,1)$ \cite{PP}.

In the limit $\mu \to 0$ the generators \p{trigonso21} turn into the standard parabolic generators
\bea
    \hat{H}=i\partial_t\,,\qquad \hat{D}=it \partial_t\,,\qquad \hat{K}=i t^2\partial_t\,. \label{parab}
\eea
The same properties are inherent to the total set of the $D(2,1;\alpha)$ generators \eqref{sca} for $\mu \neq 0$. Thus, we treat
the superspace realization of the superconformal symmetry generators found in this paper as a trigonometric deformation of the
parabolic ${\cal N}=4, d=1$ superconformal generators constructed in \cite{SCM,SCM1,superc}.

The main reason for considering the basis \eqref{trig} is that the generator ${\cal H}= \hat{H} + \frac{1}{4}\,\mu^2\hat{K}$
is directly given by the time-derivative, ${\cal H}=i\partial_t$ \cite{PP}.
Another peculiarity of this basis concerns Cartan generator (diagonal generator) of conformal algebra \cite{DFF}.
In \eqref{trig} we have the Hamiltonian ${\cal H}$ as Cartan generator,
while in the parabolic basis \eqref{parab} the Cartan generator of $so(2,1)$ is associated with the dilatation generator $\hat{D}$.
Thus the relevant quantum mechanical system must be solved in terms of eigenvalues
and eigenstates of the quantum Hamiltonian ${\cal H}= \hat{H} + \frac{1}{4}\,\mu^2\hat{K}$ which just coincides with the ``improved''
Hamiltonian of the $d=1$ conformal mechanics \cite{DFF} ensuring the energy spectrum to be bounded from below\footnote{
The  orthogonal combination ${\cal H}_{\rm h}= \hat{H} - \frac{1}{4}\,\mu^2\hat{K}$ corresponds to the hyperbolic case discussed in Appendix \ref{App-C}.
It yields a non-unitary model.}.
In the next subsection we will demonstrate that there is a basis in the $SU(2|1)$ superspace in which the full generator ${\cal H}$ defined in \p{H3}
(not only its bosonic truncation)  becomes just $i\partial_t$.

\subsection{An alternative realization of superconformal generators}\label{subsect. 3.2}
According to \eqref{conf-anticomm}, the supercharges $Q$ form the $su(2|1)$ superalgebra
with the deformation parameter $\mu$, while the supercharges $S$ form the $su(2|1)$ superalgebra with $-\mu$. Analogously, in the case $\alpha=0$,
the relevant deformed superalgebras are \eqref{alpha0} and its $-\mu$ counterpart. In the limit $\mu=0$ both sets of supercharges reproduce the
same flat ${\cal N}=4, d=1$ supercharges.

Here we demonstrate that, after the appropriate redefinition of the $SU(2|1)$ superspace coordinates,  the whole set of the superconformal generators
can be constructed in terms of the pair of deformed supercharges $Q(\mu)$ and $S(\mu)\equiv Q(-\mu)\,$.
This explains why the $SU(2|1)$ and superconformal transformations of the component fields obtained below for the multiplets
$({\bf 1, 4, 3})$, $({\bf 2, 4, 2})$ can be represented as deformations of the standard ${\cal N}=4, d=1$ transformations of component fields,
with the deformation parameters $\mu$ and $-\mu$, respectively.

The new coordinates $\{t,\tilde{\theta}_j,\bar{\tilde\theta}^i\}$ represent the same supercoset \eqref{coset1}
and are related to the previously employed super-coordinates as
\bea
    \tilde{\theta}_j = e^{\frac{i}{2}\mu t}\theta_j\left[1+\frac{1}{2}\left(1+2\alpha\right)\mu\,\bar{\theta}^k\theta_k\right],
    \qquad \bar{\tilde\theta}^i =\overline{\Big(\tilde{\theta}_j\,\Big)} =
    e^{-\frac{i}{2}\mu t}\bar{\theta}^i\left[1+\frac{1}{2}\left(1+2\alpha\right)\mu\,\bar{\theta}^k\theta_k\right].\label{altpar}
\eea
The supercharges \eqref{Q} are rewritten as
\bea
    Q^i &=& e^{\frac{i}{2}\mu t}\bigg\{\left[1 + \frac{1}{2}\left(1+2\alpha\right)\mu\,\bar{\tilde\theta}^k\tilde{\theta}_k
    - \frac{1}{16}\left(1+2\alpha\right)\mu^2\,(\tilde{\theta})^2 \big(\bar{\tilde\theta}\,\big)^2\right]\frac{\partial}{\partial\tilde{\theta}_i}
    -\left(1+\alpha\right)\mu\,\bar{\tilde\theta}^i\tilde{\theta}_k\frac{\partial}{\partial\tilde{\theta}_k}\nn
    &&
    +\,\alpha\mu\,\bar{\tilde\theta}^i\bar{\tilde\theta}^k\frac{\partial}{\partial\bar{\tilde\theta}^k}
    +\, i\bar{\tilde\theta}^i \left[1 - \frac{1}{2}\left(1+2\alpha\right)\mu\,\bar{\tilde\theta}^k\tilde{\theta}_k\right]\partial_t\bigg\},
    \nn
    \bar{Q}_j &=& e^{-\frac{i}{2}\mu t} \bigg\{\left[1 + \frac{1}{2}\left(1+2\alpha\right)\mu\,\bar{\tilde\theta}^k\tilde{\theta}_k
    - \frac{1}{16}\left(1+2\alpha\right)\mu^2\,(\tilde{\theta})^2 \big(\bar{\tilde\theta}\,\big)^2\right] \frac{\partial}{\partial\bar{\tilde\theta}^j}
     +\left(1+\alpha\right)\mu\,\tilde{\theta}_j\bar{\tilde\theta}^k\frac{\partial}{\partial\bar{\tilde\theta}^k} \nn
     && -\,\alpha\mu\,\tilde{\theta}_j\tilde{\theta}_k\frac{\partial}{\partial\tilde{\theta}_k}
     + i\tilde{\theta}_j\left[1-\frac{1}{2}\left(1+2\alpha\right)\mu\,\bar{\tilde\theta}^k\tilde{\theta}_k\right] \partial_t\bigg\}.\label{Q2}
\eea
The new form of the supercharges \eqref{S} is given by
\bea
    S^i &=& e^{-\frac{i}{2}\mu t}\bigg\{\left[1 - \frac{1}{2}\left(1+2\alpha\right)\mu\,\bar{\tilde\theta}^k\tilde{\theta}_k
    - \frac{1}{16}\left(1+2\alpha\right) \mu^2\,(\tilde{\theta})^2 \big(\bar{\tilde\theta}\,\big)^2 \right]\frac{\partial}{\partial\tilde{\theta}_i}
    + \left(1+\alpha\right)\mu\,\bar{\tilde\theta}^i\tilde{\theta}_k\frac{\partial}{\partial\tilde{\theta}_k}
    \nn
    &&
    -\,\alpha\mu\,\bar{\tilde\theta}^i\bar{\tilde\theta}^k\frac{\partial}{\partial\bar{\tilde\theta}^k}
    +\, i\bar{\tilde\theta}^i \left[1 + \frac{1}{2}\left(1+2\alpha\right)\mu\,\bar{\tilde\theta}^k\tilde{\theta}_k\right]\partial_t\bigg\},
    \nn
    \bar{S}_j &=& e^{\frac{i}{2}\mu t} \bigg\{\left[1 - \frac{1}{2}\left(1+2\alpha\right)\mu\,\bar{\tilde\theta}^k\tilde{\theta}_k
    - \frac{1}{16}\left(1+2\alpha\right) \mu^2 \,(\tilde{\theta})^2 \big(\bar{\tilde\theta}\,\big)^2 \right] \frac{\partial}{\partial\bar{\tilde\theta}^j}
     -\left(1+\alpha\right)\mu\,\tilde{\theta}_j\bar{\tilde\theta}^k\frac{\partial}{\partial\bar{\tilde\theta}^k} \nn
     && +\, \alpha\mu\,\tilde{\theta}_j\tilde{\theta}_k\frac{\partial}{\partial\tilde{\theta}_k}
     + i\tilde{\theta}_j\left[1+ \frac{1}{2}\left(1+2\alpha\right)\mu\,\bar{\tilde\theta}^k\tilde{\theta}_k\right] \partial_t\bigg\}.\label{S2}
\eea
We observe that they are obtained from the supercharges \eqref{Q2} just through the change of the sign of $\mu$, $S(\mu) \equiv Q(-\mu)$.
The bosonic generators \eqref{H3} of $SU(2|1)$ are written as
\bea
    &&I^i_j=\left(\bar{\tilde\theta}^i\frac{\partial}{\partial\bar{\tilde\theta}^j}
    -\tilde{\theta}_j\frac{\partial}{\partial\tilde{\theta}_i}\right)
    -\frac12\,{\delta^i_j}\left(\bar{\tilde\theta}^k\frac{\partial}{\partial\bar{\tilde\theta}^k}
    -\tilde{\theta}_k\frac{\partial}{\partial\tilde{\theta}_k}\right),\nn
    &&F=\frac{1}{2}\left(\bar{\tilde\theta}^k\frac{\partial}{\partial\bar{\tilde\theta}^k}
    -\tilde{\theta}_k\frac{\partial}{\partial\tilde{\theta}_k}\right),\qquad
    {\cal H} = i\partial_{t}\,.\label{bosU2}
\eea
In this new realization the Hamiltonian ${\cal H}$ takes the correct form as the time translation generator.
The rest of the bosonic generators \eqref{bos} is rewritten as
\bea
    T &=& e^{-i\mu t}\bigg\{i\left[1 - \frac{1}{4}\left(1+2\alpha\right) \mu^2\,(\tilde{\theta})^2 \big(\bar{\tilde\theta}\,\big)^2\right]\partial_t
    + \frac{\mu}{2}\left(\bar{\tilde\theta}^k\frac{\partial}{\partial\bar{\tilde\theta}^k}
    +\tilde{\theta}_k\frac{\partial}{\partial\tilde{\theta}_k}\right)\nn
    &&+\,\frac{1}{2}\left(1+2\alpha\right)\mu^2\,\bar{\tilde\theta}^i\tilde{\theta}_i\left(\bar{\tilde\theta}^k\frac{\partial}{\partial\bar{\tilde\theta}^k}
    -\tilde{\theta}_k\frac{\partial}{\partial\tilde{\theta}_k}\right)\bigg\},
    \nn
    \bar{T} &=& e^{i\mu t}\bigg\{i\left[1-\frac{1}{4}\left(1+2\alpha\right) \mu^2\,(\tilde{\theta})^2 \big(\bar{\tilde\theta}\,\big)^2\right]\partial_t
    - \frac{\mu}{2}\left(\bar{\tilde\theta}^k\frac{\partial}{\partial\bar{\tilde\theta}^k}
    +\tilde{\theta}_k\frac{\partial}{\partial\tilde{\theta}_k}\right)\nn
    &&+\,\frac{1}{2}\left(1+2\alpha\right)\mu^2\,\bar{\tilde\theta}^i\tilde{\theta}_i\left(\bar{\tilde\theta}^k\frac{\partial}{\partial\bar{\tilde\theta}^k}
    -\tilde{\theta}_k\frac{\partial}{\partial\tilde{\theta}_k}\right)\bigg\},\nn
    C &=& \varepsilon_{jl}\bar{\tilde\theta}^j\frac{\partial}{\partial\tilde{\theta}_l}\,,\qquad
    \bar{C}\; = \;\varepsilon^{jl}\tilde{\theta}_j\frac{\partial}{\partial\bar{\tilde\theta}^l}\,.
\eea
Note that the supercharges \eqref{Q2}, \eqref{S2} acquired the exponential factors $\sim e^{\pm\frac{i}{2}\mu t}$ which
are needed for ensuring the correct commutation relations with ${\cal H} =i\partial_t$. Also note that
the $su(2)$ and $su'(2)$ generators now include no $\mu$ dependence at all, while the $so(2,1)$ generators
$T$ and $\bar T$ are just related by the reflection $\mu \leftrightarrow -\mu\,,\; T(-\mu) = \bar T(\mu)\,$.
So the property that the whole superalgebra $D(2,1;\alpha)$ is contained in the closure of $\left(Q_i(\mu)\,,\; \bar Q^j(\mu)\right)$ and
$\left(S_i(\mu) =Q_i(-\mu)\,, \; \bar S^j(\mu) = \bar Q^j(-\mu)\right)$ becomes manifest in the new parametrization
of the $SU(2|1)$ superspace.

For further use, we give the new basis form of the $SU(2|1)$ invariant measure \eqref{measure}:
\bea
    d\tilde\zeta = dt\, d^2\tilde\theta\, d^2\bar{\tilde{\theta}}\left(1+ \mu\,\bar{\tilde{\theta}}^k\tilde{\theta}_k\right).
    \label{tilde-measure}
\eea
Under the $\varepsilon$, $\bar{\varepsilon}$ transformations generated by \eqref{S2} it is transformed as
\bea
    \delta_{\varepsilon}\,d\tilde{\zeta} = 2\mu\,d\tilde{\zeta} \left[1 - \frac{1}{2}\left(3+2\alpha\right)\mu\,\bar{\tilde\theta}^k\tilde{\theta}_k \right]
    \left( \bar{\varepsilon}^i \tilde{\theta}_i e^{\frac{i}{2}\mu t} - \varepsilon_i \bar{\tilde\theta}^i e^{-\frac{i}{2}\mu t}\right).
\eea

\setcounter{equation}{0}
\section{The multiplet $({\bf 1, 4, 3})$}\label{Sec. 4}
\subsection{Constraints}
The multiplet $({\bf 1, 4, 3})$ was described in \cite{DSQM} in the framework of the $SU(2|1)$ superspace \eqref{coset}.
It is represented by the real neutral superfield $G$ satisfying the $SU(2|1)$ covariantization of
the standard $({\bf 1, 4, 3})$ multiplet constraints
\bea
    \varepsilon^{lj}{\cal \bar D}_l\, {\cal \bar D}_j \,G =\varepsilon_{lj}{\cal D}^l\, {\cal D}^j\, G =0\,,
    \qquad \left[{\cal D}^i , {\cal \bar D}_i\right] G = 4m\,G\,.
    \label{143constr0}
\eea
They are solved by
\bea
    G&=&\left[ 1 -m\,\bar{\theta}^k\theta_k + m^2\left(\theta\right)^2\left(\bar{\theta}\,\right)^2\right]x
    +\frac{\ddot{x}}{4}\left(\theta\right)^2\left(\bar{\theta}\,\right)^2 - i\,\bar{\theta}^k\theta_k \left(\theta_i \,\dot{\psi}^i
    +\bar{\theta}^j\,\dot{\bar{\psi}}_j\right)\nn
    &&+\left(1- 2m\,\bar{\theta}^k\theta_k\right)\left(\theta_i \,\psi^i-\bar{\theta}^j\,\bar{\psi}_j\right)
    + \bar{\theta}^j\theta_i \,B^i_j \,, \quad B^k_k = 0\,.\label{Gstruct}
\eea

For studying superconformal properties of this $SU(2|1)$ supermultiplet it will be more convenient to reformulate it
in the superspace \eqref{coset1}. By rewriting the constraints \eqref{143constr0} through the covariant derivatives \eqref{cov1} as
\bea
    \varepsilon^{lj}{\cal \bar D}_l\, {\cal \bar D}_j \,G =\varepsilon_{lj}{\cal D}^l\, {\cal D}^j\, G =0\,,
    \qquad \left[{\cal D}^i , {\cal \bar D}_i\right] G = -4\alpha\mu\, G\,,
    \label{143constr}
\eea
we obtain
\bea
    G&=& x\left[ 1 + \alpha \mu\,\bar{\theta}^k\theta_k + \alpha^2\mu^2\left(\theta\right)^2\left(\bar{\theta}\,\right)^2\right]
    +\frac{\ddot{x}}{4}\left(\theta\right)^2\left(\bar{\theta}\,\right)^2 - i\bar{\theta}^k\theta_k\left(\theta_i \,\dot{\psi}^i \,e^{\frac{i}{2}\mu t}
    +\bar{\theta}^j\,\dot{\bar{\psi}}_j\, e^{-\frac{i}{2}\mu t}\right)\nn
    &&+\left[1 + \frac{1}{2}\left(1+4\alpha\right)\mu\,\bar{\theta}^k\theta_k\right]\left(\theta_i \,\psi^i \,e^{\frac{i}{2}\mu t}
    -\bar{\theta}^j\,\bar{\psi}_j\, e^{-\frac{i}{2}\mu t}\right)
    + \bar{\theta}^j\theta_i \,B^i_j \,,\label{G}
\eea
where we have redefined
\bea
    \psi^i \rightarrow \psi^i e^{\frac{i}{2}\mu t} ,\qquad \bar{\psi}_j \rightarrow \bar{\psi}_j e^{-\frac{i}{2}\mu t}.\label{psipsi}
\eea
This field redefinition makes the $U(1)$ generator $F$ act only on fermionic fields and ensures that the operator
${\cal H}$ is realized on the component fields as the pure time derivative $i\partial_t$ without additional $U(1)$ rotation terms.
We see that the irreducible set of the off-shell component fields is $x(t)$, $\psi^i(t)$, $\bar\psi_i(t)$, $B^i_j (t)$ $(B^k_k =0)$,
i.e., $G$ reveals just the $({\bf 1, 4, 3})\,$ content.
In the contraction limit $\mu = 0$, it is reduced to the ordinary $({\bf 1, 4, 3})$ superfield.

As the most important requirement, the constraints \eqref{143constr} (rewritten through the covariant derivatives \eqref{cov1}) must be covariant
under the superconformal symmetry $D\left(2,1;\alpha\right)$.
From this requirement, one can actually restore  the supercharges \eqref{S} and the bosonic generators \eqref{bos} as the differential operators acting on the superspace \p{coset1}.
Moreover, it implies that these extra generators for the multiplet $({\bf 1, 4, 3})$ should be extended by the proper weight terms.
The supercharges \eqref{S} are extended as
\bea
    S^i &=& e^{-i\mu t}\bigg\{\left[1 - \left(1+2\alpha\right)\mu\,\bar{\theta}^k\theta_k
    - \frac{1}{4}\left(1+2\alpha\right)^2\mu^2\left(\theta\right)^2\left(\bar{\theta}\,\right)^2 \right]\frac{\partial}{\partial\theta_i}
    + 2\left(1+\alpha\right)\mu\,\bar{\theta}^i\theta_k\frac{\partial}{\partial\theta_k}\nn
    && + \, i\bar{\theta}^i\left[1+\left(1+2\alpha\right)\mu\,\bar{\theta}^k\theta_k\right]\partial_t
    + 2 \alpha\mu\,\bar{\theta}^i\left(1 - \mu \,\bar{\theta}^k\theta_k \right) \bigg\},\nn
    \bar{S}_j &=& e^{i\mu t} \bigg\{\left[1 - \left(1+2\alpha\right)\mu\,\bar{\theta}^k\theta_k
    - \frac{1}{4}\left(1+2\alpha\right)^2\mu^2\left(\theta\right)^2\left(\bar{\theta}\,\right)^2\right] \frac{\partial}{\partial\bar{\theta}^j}
     -2\left(1+\alpha\right)\mu\,\theta_j\bar{\theta}^k\frac{\partial}{\partial\bar{\theta}^k} \nn
     && +\, i\theta_j\left[1+\left(1+2\alpha\right)\mu\,\bar{\theta}^k\theta_k\right] \partial_t  - 2\alpha \mu\,\theta_j
     \left(1 - \mu \,\bar{\theta}^k\theta_k \right) \bigg\}.\label{S1}
\eea
Respectively, the bosonic generators are modified as
\bea
    T&=&e^{-i\mu t}\left\{i\left[1-\frac{1}{4}\left(1+2\alpha\right)\mu^2\left(\theta\right)^2\left(\bar{\theta}\,\right)^2\right]\partial_t
    +\mu\left[1 -\left(1+2\alpha\right)\mu\,\bar{\theta}^k\theta_k \right] \theta_i\frac{\partial}{\partial\theta_i}\right\}\nn
    && +\,\alpha\mu\, e^{-i\mu t}\left[1 - \mu \,\bar{\theta}^k\theta_k + \frac{1}{4}\left(1-2\alpha\right) \mu^2\left(\theta\right)^2\left(\bar{\theta}\,\right)^2\right],\nn
    \bar{T}&=&e^{i\mu t}\left\{i\left[1 -\frac{1}{4}\left(1+2\alpha\right)\mu^2\left(\theta\right)^2\left(\bar{\theta}\,\right)^2\right]\partial_t
     - \mu\left[1 - \left(1+2\alpha\right)\mu\,\bar{\theta}^k\theta_k \right] \bar{\theta}^i\frac{\partial}{\partial\bar{\theta}^i}\right\}\nn
    && -\,\alpha\mu\, e^{i\mu t}\left[1 - \mu\,\bar{\theta}^k\theta_k + \frac{1}{4}\left(1-2\alpha\right) \mu^2\left(\theta\right)^2\left(\bar{\theta}\,\right)^2\right],\nn
    C &=& e^{-i\mu t}\varepsilon_{jl}\left[1+\left(1+2\alpha\right)\mu\,\bar{\theta}^k\theta_k\right]\bar{\theta}^j\frac{\partial}{\partial\theta_l} + \alpha\mu\left(\bar{\theta}\,\right)^2 e^{-i\mu t},\nn
    \bar{C}&=&e^{i\mu t}\varepsilon^{jl}\left[1+\left(1+2\alpha\right)\mu\,\bar{\theta}^k\theta_k\right]\theta_j\frac{\partial}{\partial\bar{\theta}^l} - \alpha\mu\left(\theta\right)^2 e^{i\mu t}.
\eea
These modifications of the additional $D(2,1;\alpha)$ generators imply the following ``passive'' transformation law
for the superfield $G$ under the $\varepsilon_i, \bar\varepsilon^j$ transformations
\bea
    \delta_{\varepsilon}G   = 2\alpha \mu \left(1 - \mu \,\bar{\theta}^k\theta_k \right)
    \left( \bar{\varepsilon}^i \theta_i\, e^{i\mu t} - \varepsilon_i \bar{\theta}^i\, e^{-i\mu t} \right) G\,.
    \label{trG}
\eea
All other transformations are produced by commuting \p{trG} with the odd $SU(2|1)$ transformations which are generated by the pure
differential operators \p{Q}. It is worth pointing out once more that all additional weight terms in the $D(2,1;\alpha)$ generators
are necessary for the $D(2,1;\alpha)$ covariance of the $({\bf 1, 4, 3})$ constraints \eqref{143constr} and, in fact, can be deduced from
requiring this covariance. Making the bosonic truncation of the conformal generators with the weight terms,
\bea
    {\cal H}=i\partial_t\,,\qquad T =  e^{-i\mu t}\left(i\partial_t + \alpha\mu\right) ,\qquad
    \bar{T} = e^{i\mu t}\left(i\partial_t -  \alpha\mu\right),
\eea
one observes that $\alpha$ can be identified with the scaling dimension parameter $\lambda_{D}$ for the multiplet $({\bf 1, 4, 3})$ \cite{HT}.

{\it Digression}. In Section 3.2, we showed that, after passing to the new superspace basis  $\{t,\tilde{\theta}_j,\bar{\tilde\theta}^i\}$, the
differential parts of the $D(2,1;\alpha)$ supercharges in the $\mu$-representation satisfy the relation $S(\mu) = Q(-\mu)$,
thus making manifest
the property that $D(2,1;\alpha)$ is the closure of two its $su(2|1)$ subalgebras, one defined at $\mu$ and the other at $-\mu\,$.
Due to the presence of the additional weight terms, the supercharges \eqref{S1} written in the new basis do not longer exhibit this nice correspondence.
To restore it, one needs to make the appropriate $\theta$-dependent  rescaling of the superfield $G$,
\bea
G = A\,G_0\,, \lb{AG0}
\eea
and to pick up the factor $A$ in such a way that the extra weight terms acquired by the supercharges $Q^i(\mu)$ and $S^i(\mu)$ when acting on $G_0$
ensure the needed relation. The factor $A$ is defined up to a freedom associated with a real parameter $\beta$:
\bea
A(\tilde{\theta})  &=& 1 + \alpha \mu\,\bar{\tilde\theta}^k\tilde{\theta}_k - \frac{1}{2}\,\beta\mu^2(\tilde\theta)^2\big(\bar{\tilde\theta}\,\big)^2\,, \lb{Afactor} \\
G_0(t,\tilde{\theta})  &=& \left[1 + \frac{1}{2}\left(\beta-\alpha\right)\mu^2(\tilde\theta)^2\big(\bar{\tilde\theta}\,\big)^2\right] x
+\frac{\ddot{x}}{4}\,(\tilde\theta)^2\big(\bar{\tilde{\theta}}\,\big)^2
    +\tilde{\theta}_i \,\psi^i
    -\bar{\tilde\theta}^j\,\bar{\psi}_j \nn
    && - i\bar{\tilde\theta}^k\tilde{\theta}_k\left(\tilde{\theta}_i \,\dot{\psi}^i +\bar{\tilde\theta}^j\,\dot{\bar{\psi}}_j \right)+
    \bar{\tilde\theta}^j\tilde{\theta}_i \,B^i_j\,.\lb{G0}
\eea
The $\epsilon$ and $\varepsilon$ variations of $G_0$ are related just through the substitution $\mu \rightarrow -\mu\,,$
\bea
    &&\delta_{\epsilon}G_0   = -\mu \left[\alpha -\frac12({4\beta}- {3}\alpha)\mu\,\bar{\tilde\theta}^k\tilde{\theta}_k \right]
    \left( \bar{\epsilon}^i \tilde{\theta}_i\, e^{-\frac{i}{2}\mu t} - \epsilon_i \bar{\tilde{\theta}}^i\, e^{\frac{i}{2}\mu t} \right) G_0\,,\nn
    &&\delta_{\varepsilon}G_0 = \mu \left[\alpha +\frac12({4\beta}- {3}\alpha)\mu\,\bar{\tilde\theta}^k\tilde{\theta}_k \right]
    \left( \bar{\varepsilon}^i \tilde{\theta}_i\, e^{\frac{i}{2}\mu t} - \varepsilon_i\bar{\tilde{\theta}}^i\, e^{-\frac{i}{2}\mu t} \right) G_0\,, \lb{NewVar}
\eea
and imply the following expressions for the total $D(2,1;\alpha)$ generators in the realization on $G_0$:
\bea
    Q^i &=& e^{\frac{i}{2}\mu t}\bigg\{\left[1 + \frac{1}{2}\left(1+2\alpha\right)\mu\,\bar{\tilde\theta}^k\tilde{\theta}_k
    - \frac{1}{16}\left(1+2\alpha\right)\mu^2\,(\tilde{\theta})^2 \big(\bar{\tilde\theta}\,\big)^2\right]\frac{\partial}{\partial\tilde{\theta}_i}
    -\left(1+\alpha\right)\mu\,\bar{\tilde\theta}^i\tilde{\theta}_k\frac{\partial}{\partial\tilde{\theta}_k}\nn
    &&
    +\,\alpha\mu\,\bar{\tilde\theta}^i\bar{\tilde\theta}^k\frac{\partial}{\partial\bar{\tilde\theta}^k}
    +\, i\bar{\tilde\theta}^i \left[1 - \frac{1}{2}\left(1+2\alpha\right)\mu\,\bar{\tilde\theta}^k\tilde{\theta}_k\right]\partial_t
    -\alpha \mu\,\bar{\tilde{\theta}}^i + \frac12(4\beta- {3} \alpha)\mu^2\,\bar{\tilde{\theta}}^i\bar{\tilde\theta}^k\tilde{\theta}_k
    \bigg\},
    \nn
    \bar{Q}_j &=& e^{-\frac{i}{2}\mu t} \bigg\{\left[1 + \frac{1}{2}\left(1+2\alpha\right)\mu\,\bar{\tilde\theta}^k\tilde{\theta}_k
    - \frac{1}{16}\left(1+2\alpha\right)\mu^2\,(\tilde{\theta})^2 \big(\bar{\tilde\theta}\,\big)^2\right] \frac{\partial}{\partial\bar{\tilde\theta}^j}
     +\left(1+\alpha\right)\mu\,\tilde{\theta}_j\bar{\tilde\theta}^k\frac{\partial}{\partial\bar{\tilde\theta}^k} \nn
     && -\,\alpha\mu\,\tilde{\theta}_j\tilde{\theta}_k\frac{\partial}{\partial\tilde{\theta}_k}
     + i\tilde{\theta}_j\left[1-\frac{1}{2}(1+2\alpha)\mu\,\bar{\tilde\theta}^k\tilde{\theta}_k\right] \partial_t
     +\alpha \mu\, \tilde{\theta}_i - \frac12(4\beta- {3}\alpha)\mu^2\,\tilde{\theta}_i\bar{\tilde\theta}^k\tilde{\theta}_k
     \bigg\}, \nn
S^i(\mu) &=& Q^i(-\mu)\,, \qquad \bar{S}_j(\mu) = \bar{Q}_j(-\mu)\,.\lb{NewSuperch}
\eea
One can directly check that their (anti)commutators form the superalgebra $D\left(2,1;\alpha\right)$. The $so(2,1)$ generators $T$, $\bar{T}$ are given by
\bea
    T &=& e^{-i\mu t}\bigg\{i\left[1 - \frac{1}{4}\left(1+2\alpha\right) \mu^2\,(\tilde{\theta})^2 \big(\bar{\tilde\theta}\,\big)^2\right]\partial_t
    + \frac{\mu}{2}\left(\bar{\tilde\theta}^k\frac{\partial}{\partial\bar{\tilde\theta}^k}
    +\tilde{\theta}_k\frac{\partial}{\partial\tilde{\theta}_k}\right)\nn
    &&+\,\frac{1}{2}\left(1+2\alpha\right)\mu^2\,\bar{\tilde\theta}^i\tilde{\theta}_i\left(\bar{\tilde\theta}^k\frac{\partial}{\partial\bar{\tilde\theta}^k}
    -\tilde{\theta}_k\frac{\partial}{\partial\tilde{\theta}_k}\right)+\alpha\mu\left[1 +\left(\frac{3}{4}-\frac{\beta}{\alpha}\right)
    \mu^2\,(\tilde{\theta})^2 \big(\bar{\tilde\theta}\,\big)^2\right]\bigg\},
    \nn
    \bar{T} &=& e^{i\mu t}\bigg\{i\left[1-\frac{1}{4}\left(1+2\alpha\right) \mu^2\,(\tilde{\theta})^2 \big(\bar{\tilde\theta}\,\big)^2\right]\partial_t
    - \frac{\mu}{2}\left(\bar{\tilde\theta}^k\frac{\partial}{\partial\bar{\tilde\theta}^k}
    +\tilde{\theta}_k\frac{\partial}{\partial\tilde{\theta}_k}\right)\nn
    &&+\,\frac{1}{2}\left(1+2\alpha\right)\mu^2\,\bar{\tilde\theta}^i\tilde{\theta}_i\left(\bar{\tilde\theta}^k\frac{\partial}{\partial\bar{\tilde\theta}^k}
    -\tilde{\theta}_k\frac{\partial}{\partial\tilde{\theta}_k}\right)-\alpha\mu\left[1 +\left(\frac{3}{4}
    -\frac{\beta}{\alpha}\right) \mu^2\,(\tilde{\theta})^2 \big(\bar{\tilde\theta}\,\big)^2\right]\bigg\}.\lb{NewT}
\eea
The rest of bosonic generators contain no weight terms.

The parameter $\beta$ appears neither in the structure constants of $D\left(2,1;\alpha\right)$ nor in the superconformal component actions (see next Subsections),
so it can be chosen at will. One choice is $\beta = \frac34 \alpha\,,$ which ensures the simplest structure of the weight terms in \p{NewSuperch}, \p{NewT}, \p{NewVar}.
Another possible choice is $\beta = \alpha$, under which the superfield $G_0$ in \p{G0} contains no $\mu$ dependence at all.
In this case, the $SU(2|1)$ constraints \eqref{143constr} are reduced to the linear combination of the flat constraints:
\bea
&&    \varepsilon_{ik} D^i D^k\, G_0 =  \varepsilon^{ik} \bar{D}_i\bar{D}_k \,G_0 = 0\,,\qquad \left[D^i , \bar{D}_i\right]G_0 = 0\,, \label{flat}\\
&& D^i = \frac{\partial}{\partial\tilde{\theta}_i}-i\,\bar{\tilde{\theta}}^i\partial_{t}\,,\qquad
    \bar{D}_{j} = -\frac{\partial}{\partial\bar{\tilde{\theta}}^j}+i\,\tilde{\theta}_j\partial_{t} \,.
\eea
These constraints are still covariant under the relevant trigonometric realization of $D\left(2,1;\alpha\right)$
(with $\beta =\alpha$ in \p{NewSuperch}, \p{NewT}, \p{NewVar}). The corresponding superconformal actions of $G_0$ written as integrals over
the $SU(2|1)$ superspace do not coincide with the standard ones constructed as integrals over flat ${\cal N}=4, d=1$ superspace.\\

As the final remark, we note that the constraints \eqref{143constr} can be generalized as
\bea
    \varepsilon^{lj}{\cal \bar D}_l\, {\cal \bar D}_j \,\tilde G =\varepsilon_{lj}{\cal D}^l\, {\cal D}^j\, \tilde G =0\,,
    \qquad\left[{\cal D}^i , {\cal \bar D}_i \right]\tilde G = -4 \alpha\mu\, \tilde G - 4 c \,.\label{143gen}
\eea
Their solution is
\bea
    \tilde G \left(x,\psi,\bar{\psi},B\right) = G \left(x,\psi,\bar{\psi},B\right)
    + c\,\bar{\theta}^j\theta_j\left(1+2\alpha\mu\,\bar{\theta}^k\theta_k\right),\label{G1}
\eea
where $G \left(x,\psi,\bar{\psi},B\right)$ was defined in \p{G}. Once again, this solution can be adapted to the supercoset \eqref{coset1}.
We observe that the superconformal covariance of the corresponding version of the constraints \eqref{143gen} implies the additional condition
\bea
    c\,{\cal D}_l\, {\cal D}^l\left(1 - \mu\,\bar{\theta}^k\theta_k \right)
    \left( \bar{\varepsilon}^i \theta_i\, e^{i\mu t} - \varepsilon_i\bar{\theta}^i \, e^{-i\mu t} \right)=0\,.\label{cConstr}
\eea
Substituting the explicit expressions \eqref{cov1} for the covariant derivatives, one can show that at $c\neq 0$ the condition \p{cConstr}
is satisfied only
for $\alpha=-1$. Then the superfield $\tilde{G}$ transform as
\bea
    \delta_{\varepsilon}\tilde{G}   = -2 \mu \left(1 - \mu \,\bar{\theta}^k\theta_k \right)
    \left( \bar{\varepsilon}^i \theta_i\, e^{i\mu t} - \varepsilon_i \bar{\theta}^i\, e^{-i\mu t} \right) \tilde{G}\,.
    \label{trG1}
\eea
Thus at $c\neq 0$ the relevant superconformal group is reduced to the supergroup $PSU(1,1|2) \rtimes U(1)$. At $c=0$, any
$\alpha \neq 0$ is admissible, including $\alpha =-1$\footnote{At $c=0, \, \alpha=-1$ the whole automorphism $SU(2)_{\rm ext}$ is
a symmetry of the superfield constraints. It is reduced to $U(1)_{\rm ext}$ only at $c\neq 0\,$.}. In what follows, the special case $\alpha =0$
will be considered separately.

\subsection{$SU(2|1)$ invariant Lagrangians}
One can construct the general Lagrangian and action for the $SU(2|1)$ multiplet $({\bf 1, 4, 3})$ as
\bea
     \quad S(\tilde{G}) = \int dt\, {\cal L}=-\int d\zeta\, f(\tilde{G})\,.
    \label{1}
\eea
We consider the invariant Lagrangians for the superfield $\tilde G$ satisfying the generalized constraints \eqref{143gen} with $c\neq 0$.
The action for the superfield $G$ subject to the constraints \p{143constr} can be then obtained by setting $c=0$.

Any action with an arbitrary Lagrangian function $f(\tilde{G})$ is $SU(2|1)$ invariant and provides a deformation of the standard $({\bf 1, 4, 3})\,$ models.
Substituting the expression \p{G1} for $\tilde{G}$ into \eqref{1} and doing there the Berezin integration, we obtain the component off-shell Lagrangian
\bea
    {\cal L}&=&\dot{x}^2 g(x)+i\left(\bar{\psi}_i\dot{\psi}^i -\dot{\bar{\psi}}_i\psi^i\right)g(x)+ \frac12 \,B^i_j B^j_i\,g(x)
    -B^i_j\left(\frac12\, {\delta^j_i}\,\bar{\psi}_k\psi^k -\bar{\psi}_i\psi^j \right)g'(x)\nn
    && -\frac{1}{4} \left(\psi\right)^2\left(\bar{\psi}\,\right)^2  g''(x) - \left[\left(1+2\alpha\right) g(x)
    + {\alpha x g'(x)}\right]\mu\,\bar{\psi}_i\psi^i - {\alpha^2\mu^2 x^2} g(x) - c\, g'(x)\,\bar{\psi}_i\psi^i  \nn
    && - \, 2c \alpha \mu\, x g(x) -c^2\, g(x)\,,
    \label{off-shell}
\eea
where $g := f''$ and primes mean differentiation in $x$\,, $f' = \partial_x f$\,, etc.
The parameter $c$ produces new additional potential-type terms in the Lagrangian.

The $\epsilon$\,, $\bar{\epsilon}$ transformation law of \eqref{G1}\,,
\bea
    \delta \tilde G =-\left[\epsilon_{i}Q^{i}+\bar{\epsilon}^{j}\bar{Q}_{j}\,, \tilde G\right],
\eea
implies the following $SU(2|1)$ transformation laws for the component fields:
\bea
\label{transf}
    \delta x &=& \bar{\epsilon}^k\bar{\psi}_k\, e^{-\frac{i}{2}\mu t} - \epsilon_k\psi^k\, e^{\frac{i}{2}\mu t} ,\qquad
    \delta \psi^i = e^{-\frac{i}{2}\mu t}\left( i\bar{\epsilon}^i \dot x + \alpha\mu\,\bar{\epsilon}^i x + c \,\bar{\epsilon}^i  + \bar{\epsilon}^k B^i_k\right) ,\nn
    \delta B^i_j &=& -\,2i\left[ \epsilon_j\dot{\psi}^i\, e^{\frac{i}{2}\mu t} +\bar{\epsilon}^i\dot{\bar{\psi}}_j\, e^{-\frac{i}{2}\mu t}
    -  \frac12\,{\delta^i_j}\left(\epsilon_k\dot{\psi}^k\, e^{\frac{i}{2}\mu t} +\bar{\epsilon}^k\dot{\bar{\psi}}_k\, e^{-\frac{i}{2}\mu t}\right)\right]\nn
    &&
    - \, \left(1+2\alpha\right)\mu\left[  \bar{\epsilon}^i\bar{\psi}_j\, e^{-\frac{i}{2}\mu t}- \epsilon_j\psi^i\, e^{\frac{i}{2}\mu t}
     - \frac12{\delta^i_j}\left(\bar{\epsilon}^k\bar{\psi}_k \, e^{-\frac{i}{2}\mu t} - \epsilon_k\psi^k \, e^{\frac{i}{2}\mu t}\right)\right].
\eea

We can simplify the Lagrangian \eqref{off-shell} by passing to the new bosonic field $y(x)$ with the free kinetic term. From the equality
\be
\dot{x}^2 g(x) = \frac12 \,\dot y^2 , \label{xy1}
\ee
we find the equation
\be
y'(x) = \sqrt{2 g(x)}\,,\qquad y'(x) = \frac{1}{x'(y)}\label{xy}
\ee
and define
\bea
    \chi^i = \psi^i y'(x),\qquad \tilde{B}_j^i =
\frac{1}{2}\,B_j^i\, y'(x),\qquad V(y)=\frac{x(y)}{x'(y)}\,. \label{xy2}
\eea
Solving last of eqs. \p{xy2} as
\bea
x(y) = \exp\Big\{\int^y\frac{d\tilde y}{V(\tilde y)}\Big\},
\eea
we can cast the Lagrangian \eqref{off-shell} in the form
\bea
{\cal L} &=& \frac{\dot{y}^2}{2}+\frac{i}{2}\left(\bar{\chi}_i\dot{\chi}^i-\dot{\bar{\chi}}_i\chi^i\right)+\tilde{B}^i_j\tilde{B}^j_i
    - \frac{V'(y)-1}{V(y)}\,\tilde{B}^j_i \left(\delta^i_j \bar{\chi}_k\chi^k - 2\bar{\chi}_j\chi^i\right) \nn
    && -\,\frac{V''(y)V(y) +\left[2V'(y)-3\right]\left[V'(y)-1\right]}{4V^2(y)}\left(\chi\right)^2\left(\bar{\chi}\,\right)^2
    \nn
    &&- \,\partial_y\left[\alpha\mu V(y) +c\,V(y)\exp{\bigg\lbrace-\int^y\frac{d\tilde y}{V(\tilde y)} \bigg\rbrace}\right] \bar{\chi}_i\chi^i
    - \frac{\mu}{2}\,\bar{\chi}_i\chi^i\nn
    &&- \,\frac{1}{2}\left[\alpha\mu V(y) +c\,V(y)\exp{\bigg\lbrace-\int^y\frac{d\tilde y}{V(\tilde y)} \bigg\rbrace}\right]^2 .\label{off-shell1}
\eea
Here, $V(y)$ can be regarded as an arbitrary function due to the arbitrariness of $g(x)$ in \eqref{xy}.
Thus we have finally obtained the $SU(2|1)$ Lagrangian involving an arbitrary function
and extended by additional terms which depend on the parameter $c\,$. In the new representation, the supersymmetry
transformations acquire the form
\bea
    \delta y &=&\bar{\epsilon}^k \bar{\chi}_k \,e^{-\frac{i}{2}\mu t}- \epsilon_k\,\chi^k e^{\frac{i}{2}\mu t},\nn
    \delta \chi^i &=& e^{-\frac{i}{2}\mu t}\bigg[ i\bar{\epsilon}^i \dot y  +  \alpha\mu\,\bar{\epsilon}^i V(y)
    +c\,\bar{\epsilon}^i V(y)\exp{\bigg\lbrace-\int^y\frac{d\tilde y}{V(\tilde y)} \bigg\rbrace}+2\bar{\epsilon}^k \tilde{B}^i_k
     \nn
    &&+\, \chi^i\left(\bar{\epsilon}^k \bar{\chi}_k - \epsilon_k\chi^k \,e^{i\mu t}\right)\frac{V'(y)-1}{V(y)}\bigg],\nn
    \delta \tilde{B}^i_j &=& -\,i\left[ \epsilon_j\dot{\chi}^i\, e^{\frac{i}{2}\mu t} +\bar{\epsilon}^i\dot{\bar{\chi}}_j \, e^{-\frac{i}{2}\mu t}
    - \frac12 {\delta^i_j}\left(\epsilon_k\dot{\chi}^k\, e^{\frac{i}{2}\mu t} +\bar{\epsilon}^k\dot{\bar{\chi}}_k\, e^{-\frac{i}{2}\mu t}\right)\right]\nn
    &&- \,\frac{\mu}{2}\left(1+2\alpha\right)\left[  \bar{\epsilon}^i\bar{\chi}_j\, e^{-\frac{i}{2}\mu t}- \epsilon_j\chi^i \, e^{\frac{i}{2}\mu t}
    - \frac12{\delta^i_j}\left(\bar{\epsilon}^k\bar{\chi}_k \, e^{-\frac{i}{2}\mu t}-\epsilon_k\chi^k \, e^{\frac{i}{2}\mu t}\right)\right]\nn
    && + \,\tilde{B}^i_j \left(\bar{\epsilon}^k{\bar{\chi}}_k\, e^{-\frac{i}{2}\mu t} -\epsilon_k{\chi}^k\, e^{\frac{i}{2}\mu t}\right)
    \frac{V'(y)-1}{V(y)}\nn
     &&+ \, i \dot{y}\left[ \epsilon_j{\chi}^i\, e^{\frac{i}{2}\mu t} +\bar{\epsilon}^i{\bar{\chi}}_j \, e^{-\frac{i}{2}\mu t}
    - \frac12 {\delta^i_j}\left(\epsilon_k{\chi}^k\, e^{\frac{i}{2}\mu t} +\bar{\epsilon}^k{\bar{\chi}}_k\, e^{-\frac{i}{2}\mu t}\right)\right]
    \frac{V'(y)-1}{V(y)}\,.
\eea
In the particular case $c=0$, the models described by these transformations and the Lagrangian \eqref{off-shell1} correspond
to the off-shell form of ``weak supersymmetry'' models \cite{WS}.

\subsection{Superconformal mechanics with $c=0$}
The superconformal $({\bf 1, 4, 3})$ action with $c=0$ can be written in the superfield formulation as
\bea
    S_{\rm sc}^{(\alpha)}({G})=-\int d\zeta\, f_{\rm sc}^{(\alpha)}(G)\,, \label{143conf}
\eea
where the corresponding superfield function $f(G)$ is given by
\bea
f_{\rm sc}^{(\alpha)}(G)=
\left\{
\begin{array}{l l}
    \frac{1}{8\left(\alpha + 1\right)}\, G^{-\frac{1}{\alpha}} & \quad \text{for $\alpha\neq -1,0$}\,,\\
    \frac{1}{8}\, G\ln G & \quad \text{for $\alpha=-1$}\,.\label{f}
\end{array} \right.
\qquad \Rightarrow  \quad g(x)=\frac{x^{-\frac{1}{\alpha}-2}}{8\alpha^2}\,.
\eea
Using \eqref{trG} and \eqref{tr-measure}, one can check that the action \eqref{143conf} is indeed invariant
with respect to the superconformal group $D\left(2,1;\alpha\right)$\,.

We can consider few special cases, e.g., $\alpha=-1$\,, $\alpha=-1/2$\,.
As we will see in the next subsection, in the case $\alpha=-1$ the action \eqref{143conf} can be generalized
to incorporate the non-zero parameter $c$ defined in \eqref{143gen}.
The case $\alpha=-1/2$ corresponds to the free action. We cannot treat the $\alpha = 0$ case as a particular case
of the $SU(2|1)$ models under consideration since we defined the $SU(2|1)$ superspace for $\alpha \neq 0$, while passing  to $\alpha =0$ amounts
to contraction of the original $SU(2|1)$ supergroup into the supergroup with the flat algebra \p{alpha0}. Nevertheless, as we will see soon,
the $\alpha =0$ superconformal action can still be constructed within the properly modified superfield approach based on the contracted supergroup.

Doing $\theta$-integral in the superfield action \eqref{143conf} and making the redefinition \p{mu},
we calculate the superconformal Lagrangian as\footnote{The term $\sim \bar{\psi}\psi$ vanishes
because of the identity $\left(-\frac{1}{\alpha}-2\right)g(x) = x g'(x)$
for \eqref{f}.}
\bea
    {\cal L}_{\rm sc}^{(\alpha)}&=&\dot{x}^2 g(x)+i\left(\bar{\psi}_i\dot{\psi}^i -\dot{\bar{\psi}}_i\psi^i\right)g(x)+ \frac12{B^i_j B^j_i}\,g(x)
    - B^i_j\left(\frac12 \,{\delta^j_i}\,\bar{\psi}_k\psi^k -\bar{\psi}_i\psi^j \right)g'(x)\nn
    && -\,\frac{1}{4} \left(\psi\right)^2\left(\bar{\psi}\,\right)^2  g''(x)- \alpha^2\mu^2 x^2 g(x)\,.                         \label{off-shell3}
\eea
We observe that it depends only on $\mu^2$, not on $\mu$. Taking advantage of the redefinitions just mentioned,
one can conveniently rewrite  the transformations \eqref{transf} as
\bea
\label{transf1}
    \delta x &=& \bar{\epsilon}^k\bar{\psi}_k\, e^{-\frac{i}{2}\mu t} - \epsilon_k\psi^k\, e^{\frac{i}{2}\mu t} ,\qquad
    \delta \psi^i = e^{-\frac{i}{2}\mu t}\left( i\bar{\epsilon}^i \dot x + \alpha \mu\,\bar{\epsilon}^i x  + \bar{\epsilon}^k B^i_k\right) ,\nn
    \delta B^i_j &=& -\,2i\left[ \epsilon_j\dot{\psi}^i\, e^{\frac{i}{2}\mu t} +\bar{\epsilon}^i\dot{\bar{\psi}}_j\, e^{-\frac{i}{2}\mu t}
    -  \frac12\,{\delta^i_j}\left(\epsilon_k\dot{\psi}^k\, e^{\frac{i}{2}\mu t} +\bar{\epsilon}^k\dot{\bar{\psi}}_k\, e^{-\frac{i}{2}\mu t}\right)
    \right]\nn
    &&
    - \, \left(1+2\alpha\right)\mu\left[  \bar{\epsilon}^i\bar{\psi}_j\, e^{-\frac{i}{2}\mu t}- \epsilon_j\psi^i\, e^{\frac{i}{2}\mu t}
     - \frac12 {\delta^i_j}\left(\bar{\epsilon}^k\bar{\psi}_k \, e^{-\frac{i}{2}\mu t} - \epsilon_k\psi^k \, e^{\frac{i}{2}\mu t}\right)\right].
\eea
The Lagrangian \p{off-shell3} is invariant under the second $SU(2|1)$ transformations with the parameters $\varepsilon$, $\bar{\varepsilon}$,
\bea
\label{transf2}
    \delta x &=& \bar{\varepsilon}^k\bar{\psi}_k\, e^{\frac{i}{2}\mu t} - \varepsilon_k\psi^k \,e^{-\frac{i}{2}\mu t} ,\qquad
    \delta \psi^i = e^{\frac{i}{2}\mu t}\left(i\bar{\varepsilon}^i \dot x - \alpha \mu\,\bar{\varepsilon}^i x + \bar{\varepsilon}^k B^i_k\right),\nn
    \delta B^i_j &=& -\,2i\left[ \varepsilon_j\dot{\psi}^i\, e^{-\frac{i}{2}\mu t} +\bar{\varepsilon}^i\dot{\bar{\psi}}_j\, e^{\frac{i}{2}\mu t}
    -  \frac12{\delta^i_j}\left(\varepsilon_k\dot{\psi}^k\, e^{-\frac{i}{2}\mu t} +\bar{\varepsilon}^k\dot{\bar{\psi}}_k\, e^{\frac{i}{2}\mu t}
    \right)\right]\nn
    &&
    + \, \left(1+2\alpha\right)\mu\left[  \bar{\varepsilon}^i\bar{\psi}_j\, e^{\frac{i}{2}\mu t}- \varepsilon_j\psi^i\, e^{-\frac{i}{2}\mu t}
     - \frac12{\delta^i_j}\left(\bar{\varepsilon}^k\bar{\psi}_k \, e^{\frac{i}{2}\mu t} - \varepsilon_k\psi^k \, e^{-\frac{i}{2}\mu t}\right)\right],
\eea
which correspond to the supercharges \eqref{S1}. We see that \p{transf1} and \p{transf2} are related by the replacement $\mu \rightarrow -\mu$
in accord with the structure of $D(2,1;\alpha)$ as the closure of these two $su(2|1)$ subalgebras.

The parabolic transformations of the $({\bf 1, 4, 3})$ component fields can be obtained from the trigonometric transformations \p{transf1}, \p{transf2} in two steps.
First, one passes to the new pair $\{\epsilon', \bar{\epsilon}'\}$, $\{\varepsilon', \bar{\varepsilon}'\}$ of infinitesimal parameters
with opposite dimensions by redefining the old parameters as
\bea
    \epsilon_i = \frac{1}{2}\,{\epsilon'_i} + \frac{i}{\mu}\,{\varepsilon'_i}\,,\quad
    \varepsilon_i = \frac{1}{2}\,{\epsilon'_i} - \frac{i}{\mu}\,{\varepsilon'_i}\,,\qquad {\rm and \;\; c.c.}.\label{parabolic}
\eea
This redefinition just corresponds to passing to the original basis for the $D(2,1;\alpha)$ supercharges, in which the super Poincar\'e
supercharges and those of the superconformal boosts have the opposite dimensions. Only after that we can send $\mu \rightarrow 0$
and obtain the parabolic transformations.
This procedure is universal  and can be performed for the transformations of superfields, component fields and superspace coordinates,
regardless of the type of the realization of $D(2,1;\alpha)$. In this way one can, e.g., deduce the parabolic transformations
of the integration measure \eqref{measure}, which becomes the standard flat measure $dt\,d^2\theta \,d^2\bar{\theta}$
in the limit $\mu = 0$.

Using \eqref{xy1} -- \eqref{xy2}, we can calculate the function $V(y)$ corresponding to \eqref{f}:
\bea
    V(y)=-\frac{y}{2\alpha}\,,\qquad V'(y)=-\frac{1}{2\alpha}\,,\qquad \frac{V'(y)-1}{V(y)}= \frac{1+2\alpha}{y}\,.\label{V}
\eea
As a result, we obtain the superconformal Lagrangian in the form
\bea
{\cal L}_{\rm sc}^{(\alpha)} &=& \frac{\dot{y}^2}{2}+\frac{i}{2}\left(\bar{\chi}_i\dot{\chi}^i-\dot{\bar{\chi}}_i\chi^i\right)
+\tilde{B}^i_j\tilde{B}^j_i - \frac{1+2\alpha}{y}\,\tilde{B}^j_i\left(\delta^i_j \bar{\chi}_k\chi^k - 2\bar{\chi}_j\chi^i\right)
    \nn
&&  - \,\frac{\left(1 + 3\alpha \right)\left(1+2\alpha\right)}{2 y^2}\left(\chi\right)^2\left(\bar{\chi}\,\right)^2
 - \frac{\mu^2}{8}\, y^2 .\label{off-shell2}
\eea
It is invariant (modulo a total derivative) under the following $SU(2|1)$ odd transformations
\bea
    \delta y &=&\bar{\epsilon}^k \bar{\chi}_k\, e^{-\frac{i}{2}\mu t} - \epsilon_k\chi^k \, e^{\frac{i}{2}\mu t},\nn
    \delta \chi^i &=& e^{-\frac{i}{2}\mu t} \left[i\bar{\epsilon}^i \dot y   -  \frac{\mu}{2}\,\bar{\epsilon}^i y
     + 2\,\bar{\epsilon}^k \tilde{B}^i_k  + \chi^i\left(\bar{\epsilon}^k \bar{\chi}_k
     - \epsilon_k\chi^k \, e^{i\mu t}\right)\frac{1+2\alpha}{y}\right] ,\nn
    \delta \tilde{B}^i_j &=& -\,i\left[ \epsilon_j\dot{\chi}^i\, e^{\frac{i}{2}\mu t} +\bar{\epsilon}^i\dot{\bar{\chi}}_j \, e^{-\frac{i}{2}\mu t}
    - \frac12{\delta^i_j}\left(\epsilon_k\dot{\chi}^k\, e^{\frac{i}{2}\mu t} +\bar{\epsilon}^k\dot{\bar{\chi}}_k\, e^{-\frac{i}{2}\mu t}\right)\right]\nn
    &&- \,\frac{\mu}{2}\left(1+2\alpha\right)\left[  \bar{\epsilon}^i\bar{\chi}_j\, e^{-\frac{i}{2}\mu t}- \epsilon_j\chi^i \, e^{\frac{i}{2}\mu t}
    - \frac12{\delta^i_j}\left(\bar{\epsilon}^k\bar{\chi}_k \, e^{-\frac{i}{2}\mu t}-\epsilon_k\chi^k \, e^{\frac{i}{2}\mu t}\right)\right]\nn
    && + \,\tilde{B}^i_j \left(\bar{\epsilon}^k{\bar{\chi}}_k\, e^{-\frac{i}{2}\mu t} -\epsilon_k{\chi}^k\, e^{\frac{i}{2}\mu t}\right)\frac{1+2\alpha}{y}\nn
     &&+ \, i \dot{y}\left[ \epsilon_j{\chi}^i\, e^{\frac{i}{2}\mu t} +\bar{\epsilon}^i{\bar{\chi}}_j \, e^{-\frac{i}{2}\mu t}
    - \frac12{\delta^i_j}\left(\epsilon_k{\chi}^k\, e^{\frac{i}{2}\mu t} +\bar{\epsilon}^k{\bar{\chi}}_k\, e^{-\frac{i}{2}\mu t}\right)\right]\frac{1+2\alpha}{y}\,.
\eea
Changing $\mu$ in these transformations as $\mu \to -\mu$, one obtains the transformations associated with the extra generators $S(\mu)=Q(-\mu)$.
Since the Lagrangian \eqref{off-shell2} depends only on $\mu^2$ like \p{off-shell3}, it is automatically invariant under these $S^i$ transformations
and, hence, under the full $D(2,1;\alpha)$.

Thus in the present case we deal with the superconformal mechanics corresponding to the trigonometric transformations \cite{HT}.
Another type of superconformal mechanics is that associated with the  parabolic transformations, and
its superfield description is based on the standard ${\cal N}=4$, $d=1$ superspace.
The only difference is that the trigonometric type action \eqref{off-shell2} has an additional oscillator term.
Thus by sending $\mu \to 0$, the parabolic type of superconformal mechanics can be restored.

The property that the component superconformal trigonometric actions are even functions of the parameter $\mu$  can be established
already at the superfield level. One should pass to the $SU(2|1)$ superspace basis $\{t,\tilde{\theta}_j,\bar{\tilde\theta}^i\}$,
in which the property $S^i(\mu) = Q^i(-\mu)$ is valid and the integration measure is defined by \p{tilde-measure},
and express the superfield $G$ through $G_0$ according to eqs. \p{AG0} - \p{G0}:
\bea
    S_{\rm sc}^{(\alpha)}(G) &=&-\int dt\, d^2\tilde\theta\, d^2\bar{\tilde{\theta}}\left(1+ \mu\,\bar{\tilde{\theta}}^k\tilde{\theta}_k\right) G^{-\frac{1}{\alpha}}\nn
    &=&-\int dt\, d^2\tilde\theta\, d^2\bar{\tilde{\theta}}
    \left[1 + \frac{1}{4}\left(\alpha - 1 + \frac{2\beta}{\alpha}\right)\mu^2\,(\tilde\theta)^2\big(\bar{\tilde\theta}\,\big)^2\right]
    (G_0)^{-\frac{1}{\alpha}},\; \alpha \neq -1, 0 \lb{Snewbas}
\eea
and
\bea
    S_{\rm sc}^{(\alpha=-1)}(G) &=&-\int dt\, d^2\tilde\theta\, d^2\bar{\tilde{\theta}}\left(1+ \mu\,\bar{\tilde{\theta}}^k\tilde{\theta}_k\right) G\ln{G} \nn
    &=& -\int dt\, d^2\tilde\theta\, d^2\bar{\tilde{\theta}}
    \Big\{\left[1 - \frac{1}{2}\left(1+\beta\right)\mu^2(\tilde\theta)^2\big(\bar{\tilde\theta}\,\big)^2\right]G_0\ln{G_0} \nn
    &&-\, \frac{\mu^2}{4}\left(1 + 2\beta\right)(\tilde\theta)^2\big(\bar{\tilde\theta}\,\big)^2\, G_0\Big\}, \lb{Snewbas1}
\eea
where one should take into account that
\bea
    \int dt\, d^2\tilde\theta\, d^2\bar{\tilde{\theta}}\left(\mu\,\bar{\tilde{\theta}}^k\tilde{\theta}_k\, G_0 \right)=0\,.\nonumber
\eea
All terms with the manifest  $\theta$s in \p{Snewbas}, \p{Snewbas1}, equally as the superfield $G_0$, depend only on $\mu^2$. Also,
it is easy to show that all $\beta$-dependent terms in these actions are canceled among themselves. For any
other trigonometric superconformal action treated below (e.g, in the  $\alpha=-1, c\neq 0$ case), it is possible to show in a similar way that,
in the appropriate superfield formulation, they depend only on $\mu^2$  like in the component field formulations.

\subsection{The model with  $\alpha=-1, c\neq 0$}
Let us consider the case of $c \neq 0$ for which the superconformal invariance requires that $\alpha=-1$ ($m=\mu$).
The corresponding supergroup is $D\left(2,1;\alpha{=}-1\right)=PSU(1,1|2)\rtimes SU(2)_{\rm ext}$,
but the constraints \eqref{143gen} are covariant only with respect to $PSU(1,1|2) \rtimes U(1)_{\rm ext}$.

The corresponding superfield action is
\bea
    S^{(\alpha = -1)}_{\rm sc}(\tilde{G}) = -\int d\zeta\,\tilde G\,\ln\tilde{G}\,.\lb{cneq0S}
\eea
Starting from the general $SU(2|1)$ invariant component Lagrangian \eqref{off-shell} with $c \neq 0$ and substituting there $f(\tilde G) \rightarrow
 \tilde{G}\ln \tilde G$, we obtain, up to an additive constant, the following $c \neq 0\,, \,\alpha =-1$
generalization of the superconformal Lagrangian \eqref{off-shell3}
\bea
    {\cal L}^{(\alpha = -1, \,c)}_{\rm sc} &=&\frac{\dot{x}^2}{x} +\frac{i}{x}\left(\bar{\psi}_i\dot{\psi}^i -\dot{\bar{\psi}}_i\psi^i\right)+ \frac{B^i_j B^j_i}{2x}
    +\frac{B^i_j}{x^2}\left(\frac12\,{\delta^j_i}\,\bar{\psi}_k\psi^k -\bar{\psi}_i\psi^j \right)\nn
    && -\frac{1}{2 x^3}\left(\psi\right)^2\left(\bar{\psi}\,\right)^2  - {\mu^2 x} + \frac{c\,\bar{\psi}_i\psi^i}{x^2}
      - \frac{c^2}{x} \,.
\eea
Here, the new term $\sim \bar{\psi}\psi$ is responsible for reducing superconformal symmetry to $PSU(1,1|2) \rtimes U(1)$.
This action is invariant under the supersymmetry transformations, with $\delta \psi^i$ being a generalization of the relevant transformations
in \eqref{transf1}, \eqref{transf2}:
\bea
    \delta \psi^i = e^{-\frac{i}{2}\mu t}\left( i\bar{\epsilon}^i \dot x  - {\mu}\,\bar{\epsilon}^i x
    + \bar{\epsilon}^k B^i_k + c \,\bar{\epsilon}^i x \right)
    + e^{\frac{i}{2}\mu t}\left( i\bar{\varepsilon}^i \dot x  + {\mu}\,\bar{\varepsilon}^i x
    + \bar{\varepsilon}^k B^i_k + c\,\bar{\varepsilon}^i x \right).
\eea
Transformations of the bosonic fields are the same as in \eqref{transf1}, \eqref{transf2}.

Passing to the action with free kinetic terms, we find the relevant function $V(y)$ to be
\bea
    V(y)=\frac{y}{2}\,,\qquad V'(y)=\frac{1}{2}\,,\qquad \frac{V'(y)-1}{V(y)}= -\frac{1}{y}\,.
\eea
In accordance with \eqref{off-shell1}, we also should take into account additional terms involving $c$\,.
Thus the superconformal Lagrangian \eqref{off-shell2} is generalized to this special case as
\bea
{\cal L}^{(\alpha = -1, \,c)}_{\rm sc} &=& \frac{\dot{y}^2}{2}+\frac{i}{2}\left(\bar{\chi}_i\dot{\chi}^i-\dot{\bar{\chi}}_i\chi^i\right)
+\tilde{B}^i_j\tilde{B}^j_i + \frac{\tilde{B}^j_i }{y}\left(\delta^i_j \bar{\chi}_k\chi^k - 2\bar{\chi}_j\chi^i\right)
- \frac{1}{y^2}\left(\chi\right)^2\left(\bar{\chi}\,\right)^2
    \nn
&&  +\, \frac{c}{2y^2}\,\bar{\chi}_i\chi^i - \frac{\mu^2 y^2}{8} - \frac{c^2}{8y^2}\,.\label{BKoffact}
\eea
The relevant on-shell Lagrangian,
\bea
{\cal L}^{(\alpha = -1, \,c)}_{\rm sc} = \frac{\dot{y}^2}{2}+\frac{i}{2}\left(\bar{\chi}_i\dot{\chi}^i-\dot{\bar{\chi}}_i\chi^i\right)
 - \frac{1}{4y^2}\left(\chi\right)^2\left(\bar{\chi}\,\right)^2
    + \frac{c}{2y^2}\,\bar{\chi}_i\chi^i  - \frac{\mu^2 y^2}{8} - \frac{c^2}{8y^2}\,,\label{BKact}
\eea
as a superconformal Lagrangian was previously found in \cite{BK}\footnote{One needs to perform a redefinition of fields
in order to show the coincidence of these two Lagrangians.}.
The $SU(2|1)$ superspace approach allowed us to find the off-shell superfield form of \p{BKact}.

\subsection{The $\alpha=0$ model}\label{143-0}
Inspecting the Lagrangian \eqref{off-shell3}, we observe that the limit $\alpha \rightarrow -0$ is divergent and
the opposite limit $\alpha \rightarrow +0$ yields ${\cal L}^{(\alpha = 0)}_{\rm sc}=0$.
Nevertheless, we can unambiguously define this limit for the Lagrangian \eqref{off-shell3} by introducing an inhomogeneity parameter $\rho$ \cite{HT}.

The limit $\alpha \rightarrow 0$ can be obtained, if we redefine the Lagrangian \eqref{off-shell3}
by shifting the field $x$ as
\bea
    x\to x + \frac{\rho}{\alpha}\,.
\eea
The homogeneous Lagrangian \eqref{off-shell3} is rewritten as
\bea
    {\cal L}_{\rm sc}^{(\alpha, \,\rho)} &=& \frac{\alpha^{\frac{1}{\alpha}}}{8}\left(\alpha x+\rho\right)^{-\frac{1}{\alpha}-2}\left[\dot{x}^2
    +i\left(\bar{\psi}_i\dot{\psi}^i -\dot{\bar{\psi}}_i\psi^i\right)+ \frac12\,{B^i_j B^j_i}\right]\nn
    &&+\,\frac{\alpha^{\frac{1}{\alpha}}\left(1+2\alpha\right)}{8}\,B^i_j\left(\frac12\,{\delta^j_i}\,
    \bar{\psi}_k\psi^k -\bar{\psi}_i\psi^j \right)\left(\alpha x+\rho\right)^{-\frac{1}{\alpha}-3}\nn
    && -\, \frac{\alpha^{\frac{1}{\alpha}}\left(1+2\alpha\right)\left(1+3\alpha\right)}{32}
    \left(\psi\right)^2\left(\bar{\psi}\,\right)^2\left(\alpha x+\rho\right)^{-\frac{1}{\alpha}-4}
    - \frac{\alpha^{\frac{1}{\alpha}}\mu^2}{8}\left(\alpha x+\rho\right)^{-\frac{1}{\alpha}}.\label{off-shellalpha0}
\eea
Detaching the divergent factor $\sim(\frac{\alpha}{\rho})^{\frac{1}{\alpha}}$ and sending $\alpha \rightarrow 0$ in the remainder,
we obtain the Lagrangian ${\cal L}^{(\alpha = 0, \,\rho)}_{\rm sc}$ as
\bea
    {\cal L}^{(\alpha = 0, \,\rho)}_{\rm sc} &=& e^{-\frac{x}{\rho}}\left[\dot{x}^2 +i\left(\bar{\psi}_i\dot{\psi}^i -\dot{\bar{\psi}}_i\psi^i\right)
    + \frac12\,{B^i_j B^j_i}\right]
    +\frac{B^i_j}{\rho}\left(\frac12\,{\delta^j_i}\, \bar{\psi}_k\psi^k -\bar{\psi}_i\psi^j \right)e^{-\frac{x}{\rho}}\nn
    && -\, \frac{1}{4\rho^2}\left(\psi\right)^2\left(\bar{\psi}\,\right)^2 e^{-\frac{x}{\rho}} - \mu^2\rho^2 e^{-\frac{x}{\rho}}.\label{143-00}
\eea
Following the same procedure as in \eqref{xy1} -- \eqref{xy2}, we can obtain the Lagrangian which coincides with \eqref{off-shell2} at $\alpha = 0$ \cite{HT}.
For ensuring the superconformal invariance in this case, one needs to extend the transformations \eqref{transf1}, \eqref{transf2}
for $\alpha = 0$ by the inhomogeneous parts
\bea
    \delta_{(\rho)}\psi^i = \rho \mu\left(\bar{\epsilon}^i e^{-\frac{i}{2}\mu t}-\bar{\varepsilon}^i e^{\frac{i}{2}\mu t}\right),\quad
    \delta_{(\rho)}x = \delta_{(\rho)}B^i_k=0\,.
\eea
This modification entails the appearance of inhomogeneous pieces in the conformal transformations of $x$,
\bea
    T x =  e^{-i\mu t}\left(i\dot{x} + \rho\mu\right) ,\qquad
    \bar{T}x = e^{i\mu t}\left(i\dot{x} -  \rho\mu\right).
\eea
The standard conformal $so(2,1)$ generators defined in \eqref{T}, \eqref{sca} act on $x$ as
\bea
    &&\hat{H}x=\frac{i}{2}\left(1 + \cos{\mu t}\,\right)\dot{x}- \frac{i}{2}\,\rho\mu\sin{\mu t}\,,\nn
    &&\hat{K}x=\frac{2i}{\mu^2}\left(1 - \cos{\mu t}\,\right)\dot{x}+\frac{2i}{\mu}\,\rho\sin{\mu t}\,,\nn
    &&\hat{D}x=\frac{i}{\mu}\sin{\mu t}\,\dot{x} + i\rho\cos{\mu t}\,.
\eea

The superconformal superfield action \eqref{143conf} is not defined at $\alpha=0$.
Nevertheless, the superfield description of \eqref{143-00} can be given in the framework of
the supercoset \eqref{coset2} associated with the $\alpha =0$ superalgebra \p{alpha0}. According to \eqref{G}, the superfield $G$ is written as
\bea
    G&=& x    +\frac{\ddot{x}}{4}\left(\theta\right)^2\left(\bar{\theta}\,\right)^2 - i\bar{\theta}^k\theta_k\left(\theta_i \,\dot{\psi}^i \,e^{\frac{i}{2}\mu t}
    +\bar{\theta}^j\,\dot{\bar{\psi}}_j\, e^{-\frac{i}{2}\mu t}\right)\nn
    &&+\left(1 + \frac{\mu}{2}\,\bar{\theta}^k\theta_k\right)\left(\theta_i \,\psi^i \,e^{\frac{i}{2}\mu t}-\bar{\theta}^j\,\bar{\psi}_j\, e^{-\frac{i}{2}\mu t}\right)
    + \bar{\theta}^j\theta_i \,B^i_j \label{alpha0constr}
\eea
and satisfies the standard ``flat'' $({\bf 1, 4, 3})$ constraints
\bea
    \varepsilon^{lj}{\cal \bar D}_l\, {\cal \bar D}_j \,G =\varepsilon_{lj}{\cal D}^l\, {\cal D}^j\, G =0\,,
    \qquad \left[{\cal D}^i , {\cal \bar D}_i\right] G = 0\,,\label{stand143}
\eea
where the covariant derivatives are\footnote{Though the superfield $G$ has no external $U(1)$ charge and the generator $\tilde{F}$ yields zero on
$G$, it is non-vanishing when acting on the covariant derivative itself. Nevertheless, it is direct to check that in the $\alpha =0$
constraints \p{stand143} such contributions are canceled against terms coming from the phase factors in the definition \p{cov2}.}
\bea
    &&{\cal D}^i = e^{-\frac{i}{2}\mu t}\left(\frac{\partial}{\partial\theta_i}
    -i\bar{\theta}^i \partial_t -\mu\,\bar{\theta}^i \tilde{F} \right),\qquad
    \bar{{\cal D}}_j = e^{\frac{i}{2}\mu t}\left(-\frac{\partial}{\partial\bar{\theta}^j}
    +i\theta_j\partial_t + \mu\,\theta_j\tilde{F} \right), \label{cov2}\\
    &&{\cal D}_{(t)} = \partial_t\,, \qquad \{{\cal D}^i,\bar{{\cal D}}_j\} = 2\delta^i_j \left(i \partial_{t} + \mu \tilde{F}\right). \label{alpha0antic}
\eea
Then the component Lagrangian \eqref{143-00} is reproduced from the superfield action
\bea
     \quad S^{(\alpha = 0, \,\rho)}_{\rm sc}(G) = \int dt\, {\cal L}^{(\alpha = 0, \,\rho)}_{\rm sc}=-\rho^2\int dt\,d^2\theta \,d^2\bar{\theta}\,
     e^{-\frac{G}{\rho}-\mu\,\bar{\theta}^k\theta_k}.
     \label{143conf0}
\eea
The ``passive'' superfield infinitesimal transformation of $G$ involves only the inhomogeneous piece
\bea
    \delta_{(\rho)} G = -\rho \mu \left(\bar{\epsilon}^k\theta_k - \epsilon_k \bar{\theta}^k\right)
    + \rho \mu \left(1+\mu\,\bar{\theta}^k\theta_k\right)\left(\bar{\varepsilon}^k\theta_k e^{i\mu t}
    - \varepsilon_k \bar{\theta}^k e^{-i\mu t}\right), \lb{alpha0transf}
\eea
since its standard homogeneous part \eqref{trG} vanishes at $\alpha = 0$.

Note that the superfield $G$ at $\alpha =0$, though being defined in fact on the flat ${\cal N}=4$ superspace, still possesses  an unusual
inhomogeneous transformation law \eqref{alpha0transf} under the ${\cal N}=4, d=1$ Poincar\'e supersymmetry to which, at $\alpha =0$,
the $\epsilon, \bar\epsilon$ transformations are reduced. We  can reformulate this model in terms of the superfield $u$ having the standard homogeneous
transformation law under  the ${\cal N}=4, d=1$ Poincar\'e supersymmetry
\bea
&& u = G+\rho\mu\,\bar{\theta}^k\theta_k\,, \quad \delta_{\epsilon}u =0\,, \label{uG} \\
&&  \varepsilon^{lj}{\cal \bar D}_l\, {\cal \bar D}_j \,u =\varepsilon_{lj}{\cal D}^l\, {\cal D}^j\, u =0\,,
    \qquad \left[{\cal D}^i , {\cal \bar D}_i\right] u = -4\rho\mu\,. \label{uConstr}
\eea
The inhomogeneity of the full odd superconformal transformation law of $u$ is retained only in the part $\sim \varepsilon_i,
\bar\varepsilon^k$ associated with the generators $S^i, \bar S_k$:
\bea
    \delta_{(\rho)} u = 2 \rho \mu \left(1-\mu\,\bar{\theta}^k\theta_k\right)\left(\bar{\varepsilon}^k\theta_k e^{i\mu t}
    - \varepsilon_k \bar{\theta}^k e^{-i\mu t}\right). \label{confu}
\eea
The action \p{143conf0} is rewritten in the form in which it does not involve explicit $\theta$:
\bea
     \quad S^{(\alpha = 0, \,\rho)}_{\rm sc}(u) = -\rho^2\int dt\,d^2\theta \,d^2\bar{\theta}\,
     e^{-\frac{u}{\rho}}\,.\lb{143conf0u}
\eea

We also note that the $\mu$ dependence in the solution \p{alpha0constr} is fake because it can be removed by the inverse
phase transformation of fermionic fields as $\psi^i \rightarrow \psi^i e^{-\frac{i}{2}\mu t}$. Then the whole $\mu$ dependence
in the component actions \p{off-shellalpha0}, \p{143-00} is generated by the $\theta$ dependent term in \p{143conf0} or
the $\theta$-dependent additional term in $u$ defined in \p{uG} (if  one prefers the $u$-representation \p{143conf0u} for the
superconformal action). The definition of the fermionic fields as in \p{alpha0constr} is convenient since it ensures the absence
of the fermionic ``mass terms'' $\sim \mu\psi^i\bar\psi_i$ in  \p{off-shellalpha0}, \p{143-00}.  Despite
the fact that at $\alpha =0$ we deal with the standard flat ${\cal N}=4$ superfield $u$, the superconformal transformations
\p{confu} still correspond to the trigonometric realization of the conformal subgroup $SO(2,1)$, as well as of the full $PSU(1,1|2)$.
The parabolic realization  is achieved by redefining the fermionic parameters as in \p{parabolic} and then sending $\mu \rightarrow 0$
in the resulting transformations, like in other cases.

As the final remark, we notice that the $\alpha=0$ analog of the superconformal action \eqref{cneq0S} with $c\neq 0$ and $\alpha=-1$ can be obtained \cite{ikrpash}
by considering the superfield action dual to \eqref{cneq0S}:
\bea
     \quad S^{(\alpha = 0, \,\rho,\, \tilde{c})}_{\rm sc}(G)  &=& \int dt\,d^2\theta \,d^2\bar{\theta}\,
     \left[ -\rho^2 e^{-\frac{G}{\rho}-\mu\,\bar{\theta}^k\theta_k} + \tilde{c}\left(\bar{\theta}^1\theta_1 - \bar{\theta}^2\theta_2\right)G\right] \nn
     &=&
     \int dt\,d^2\theta \,d^2\bar{\theta}\,
     \left[ -\rho^2 e^{-u} + \tilde{c}\left(\bar{\theta}^1\theta_1 - \bar{\theta}^2\theta_2\right)u\right].
\eea
It can be checked that the relevant component Lagrangian coincides with the off-shell Lagrangian \eqref{BKoffact},
modulo the replacements of all $SU(2)$ indices by the $SU'(2)$ indices (on which the generators $\{F,C,\bar{C}\}$ act) and the substitution $c \rightarrow \tilde{c}\,$.

\setcounter{equation}{0}
\section{The multiplet $({\bf 2, 4, 2})$}\label{Sec. 5}
\subsection{Chiral $SU(2|1)$ superfields}\label{51}
In this Section, we will consider the multiplet $({\bf 2, 4, 2})$, proceeding from the superspace \eqref{coset}.
Also, in \cite{SKO} the multiplet $({\bf 2, 4, 2})$ was generalized by exploiting the superspace coset \eqref{coset22}.
Such a generalization will be addressed in the next Section.

Employing the covariant derivatives \eqref{cov}, the standard form of the chiral and antichiral conditions is as follows
\bea
   {\rm (a)}\;\;  \bar{\cal D}_i \Phi =0\,,\qquad {\rm (b)} \;\;{\cal D}^i\bar{\Phi}=0\,.\label{ch1}
\eea
This implies the existence of the left and right chiral subspaces \cite{DSQM}:
\bea
    (t_L,\theta_i),\qquad  (t_R, \bar{\theta}^i)\,,
\eea
where
\bea
    t_L = t +i\,\bar{\theta}^k\theta_k -\frac{i}{2}\,m\left(\theta\right)^2\left(\bar{\theta}\,\right)^2 ,\qquad {\rm and \;\;c.c.}. \label{left}
\eea
These coordinate sets are closed under the $SU(2|1)$ transformations
\bea
    \delta \theta_{i}=\epsilon_{i} +
    2m\,\bar{\epsilon}^k\theta_k\theta_{i}\,,\qquad
    \delta t_L=2i\,\bar{\epsilon}^k\theta_k\,, \qquad {\rm and \;\;c.c.}.\label{newtr}
\eea

One can require that the complex superfield $\Phi$ with the minimal field contents $({\bf 2, 4, 2})$ possesses
a fixed overall $U(1)$ charge
\bea
    \tilde{F}\Phi = 2\kappa\, \Phi\,,\qquad\tilde{I}^i_k\,\Phi = 0\,. \label{FPhi}
\eea
The general solution of \eqref{ch1} for an arbitrary real $\kappa$ reads:
\bea
\Phi\big(t,\theta,\bar{\theta}\,\big) = \left(1+ 2m\,\bar{\theta}^k\theta_k \right)^{-\kappa}\Phi_L (t_L,\theta\,)\,,\quad
    \Phi_L (t_L,\theta\,)= z+\sqrt{2}\,\theta_i \xi^i  +(\theta)^2 B\,, \quad \overline{\left(\xi^i\right)}= \bar{\xi}_i\,.
    \label{chiral0}
\eea
The chiral superfield $\Phi$ transforms as
\bea
    \delta\Phi = 2\kappa m\left( \bar{\epsilon}^i\theta_i +\epsilon_i\bar{\theta}^i\right)\Phi\,,\quad
    \delta\Phi_L = 4\kappa m\, \bar{\epsilon}^i\theta_i\, \Phi_L \,. \label{epsilonChir}
\eea
This transformation law implies the following off-shell $SU(2|1)$ transformations of the component fields in \eqref{chiral0}
\bea
    &&\delta z  =-\sqrt{2}\, \epsilon_k\xi^k ,\quad
    \delta \xi^i =  \sqrt{2}\, \bar{\epsilon}^i \left(i\dot z-2\kappa m z\right) -\sqrt{2}\,\epsilon^i B,\nn
    &&\delta B = -\sqrt{2}\, \bar{\epsilon}_k\left[i\dot{\xi}^k -\left(2\kappa -\frac{1}{2}\right)m\,\xi^k\right].\lb{compChir1}
\eea

As in case of the multiplet $({\bf 1, 4, 3})$, for analyzing the superconformal properties of the multiplet $({\bf 2, 4, 2})$
it will be convenient to pass to the supercoset \eqref{coset1}, in which the time-translation generator is ${\cal H}\in so(2,1)$.
Imposing the constraints \eqref{ch1} with the covariant derivatives defined in \eqref{cov1} and choosing $\kappa =0$,
we come to the left chiral subspace parametrized by the same coordinates $(t_L, \theta_i)$ as before, with  the definition \eqref{left} being valid.
It is straightforward to check that this set of coordinates is closed under the superconformal transformations generated
by \eqref{Q} and \eqref{S} only for $\alpha = -1\,$. The relevant coordinate transformations read
\bea
    \delta \theta_{i}=\epsilon_{i} +
    2\mu\,\bar{\epsilon}^k\theta_k\theta_{i}+\varepsilon_{i}e^{-i\mu t_L}\,,\qquad
    \delta t_L=2i\,\bar{\epsilon}^k\theta_k + 2i\,\bar{\varepsilon}^k\theta_k e^{i \mu t_L}.\label{newtr1}
\eea
This agrees with the observation that under the action of the generators $C$, $\bar{C}$ \eqref{bos} belonging to the group $SU'(2)$
the constraints \eqref{ch1} are not covariant. Thus the chiral subspaces are closed, and, respectively, the chirality constraints are covariant,
only for the conformal supergroup $D(2,1;\alpha{=}-1)$ $= PSU(1,1|2) \rtimes U(1)_{\rm ext}$. Note that the chiral superfields in the flat ${\cal N}=4$ superspace
are also known to preserve the superconformal $D(2,1; \alpha)$ covariance only for the values $\alpha = -1, 0\,$\cite{SCM,SCM1}.

At $\alpha = -1$ the overall $U(1)$ charge operator $\tilde{F}$ drops out from the covariant derivatives \eqref{cov1}, so the only
solution of \p{ch1} in this case, i.e. the solution consistent with the superconformal covariance, corresponds to the choice $\kappa = 0$ in \p{chiral0}. As
follows from \p{epsilonChir}, the $Q^i, \bar Q_k$ transformations of $\Phi_L$ at $\kappa=0$ (i.e. those with $\epsilon_i, \bar\epsilon^i$)
do not involve any weight terms. Since in the appropriate basis $S^i(\mu) = Q^i(-\mu)\,, \bar S_i(\mu) = \bar Q_i(-\mu)$, the same should
be true for the $\varepsilon_i, \bar\varepsilon^i$ transformations, i.e.
\bea
\delta_\epsilon \Phi_L = \delta_{\varepsilon} \Phi_L = 0\,.\label{superconf0}
\eea
On the other hand, the measure of integration over the $SU(2|1)$ superspace $d\zeta$ is not superconformally invariant at any $\alpha$
(recall \p{tr-measure}), so it is impossible to construct a homogeneous superconformally invariant action out of the
superfield $\Phi_L$ transforming as in \p{superconf0}.

One way to construct the superconformal action is to pass to its  inhomogeneous version as it was done for the $\alpha=0$ case of
the multiplet $({\bf 1, 4, 3})$ in Section 4.5. This will be performed in Section 5.3. Another way which allows one to construct a more general class
of superconformal actions is to start from the embedding of $SU(2|1)$ into a central-charge extension of $PSU(1,1|2)$, i.e. the
supergroup $SU(1,1|2)$ with the superalgebra given in Appendix A, eqs.  \eqref{conf-anticomm2} -- \eqref{conf-comm2}. The corresponding $su(2|1)$ subalgebra
is specified by
the anticommutator
\bea
\lbrace Q^{i}, \bar{Q}_{j}\rbrace = 2\mu I^i_j +2\delta^i_j \left({\cal H} - \mu\,Z_1\right),  \lb{QQZ1}
\eea
where the central charge generator $Z_1$ {\it commutes} with all other generators. The natural modification of the supercoset \p{coset1}
for $\alpha =-1$ is as follows
\bea
    \frac{SU(2|1)\rtimes U(1)_{\rm ext}}{SU(2)\times U(1)_{\rm int}\times U(1)_{\rm ext}}\, \sim \,
    \frac{\{Q^{i},\bar{Q}_{j}, {\cal H}, Z_1, F, I^i_j \}}{\{I^i_j, Z_1, F\}}\, \sim \,
    \frac{\{Q^{i},\bar{Q}_{j}, {\cal H}, Z_1, I^i_j \}}{\{I^i_j, Z_1 \}} \,,\label{cosetZ}
\eea
where $SU(2|1)$ in the numerator is defined through the $Z_1$ extended anticommutation relation \p{QQZ1} and we placed $Z_1$ into the
stability subgroup. Recall that the former internal generator $F$ becomes an outer automorphism generator at $\alpha = -1$ and
is completely split from the remaining $su(1,1|2)$ generators.

An element of the coset \p{cosetZ} coincides with \eqref{element1}. However, due to the appearance of the new generator $Z_1$
in the stability subgroup and the modification of the basic $SU(2|1)$ anticommutator as in \p{QQZ1}, the covariant spinor derivatives
\eqref{cov1} at $\alpha =-1$ should be extended as
%the new generator $Z_1$:
\bea
{\cal D}^i \;\;\Rightarrow \;\; {\cal D}^i_{Z} = {\cal D}^i + \mu\,e^{-\frac{i}{2}\mu t}\bar{\theta}^i Z_1\,, \qquad
\bar{{\cal D}}_j \;\;\Rightarrow \;\;\bar{{\cal D}}_{Zj} =  \bar{{\cal D}}_j -\mu e^{\frac{i}{2}\mu t}\theta_j Z_1\,. \lb{covZ2}
\eea
Now we can require that the superfield $\Phi$ has a non-zero charge with respect to $Z_1$:
\bea
    Z_1\Phi = b \,\Phi\,. \lb{Z1charge}
\eea
Then, imposing the chirality condition \p{ch1} with the modified covariant derivative \p{covZ2}, i.e.,
\bea
\bar{{\cal D}}_{Zj}\Phi = \left(\bar{\cal D}_j -\mu e^{\frac{i}{2}\mu t}\theta_j Z_1\right)\Phi =0\,,\lb{ch1b}
\eea
one obtains the solution
\bea
\Phi\big(t,\theta,\bar{\theta}\,\big) = \left(1+ 2\mu\,\bar{\theta}^k\theta_k \right)^{-\frac{b}{2}}\Phi_L (t_L,\theta\,)\,,\quad
    \Phi_L (t_L,\theta\,)= z+\sqrt{2}\,\theta_i \xi^i e^{\frac{i}{2}\mu t_L} +(\theta)^2 B e^{i\mu t_L}\,,\label{chiral}
\eea
which looks as \eqref{chiral0}, with $b=2\kappa$ and the fields redefined as
\bea
    \xi^i(t) \rightarrow \xi^i(t) e^{\frac{i}{2}\mu t} ,\qquad B(t) \rightarrow B(t) e^{i\mu t}\,, \qquad{\rm and\;\; c.c.}.\label{B1}
\eea

To preserve this $b\neq 0$ chirality, the holomorphic chiral superfield $\Phi_L$ should have the following $\epsilon$ and $\varepsilon$ transformation laws
\bea
    \delta_{\epsilon}\Phi_L = 2b \mu\, \bar{\epsilon}^i\theta_i\, \Phi_L\, ,\qquad \delta_{\varepsilon}\Phi_L = - 2b \mu \,
    \bar{\varepsilon}^i\theta_i \,e^{i \mu t_L}\,  \Phi_L \,,
    \label{trPhi}
\eea
or, in terms of the superfield $\Phi\,$,
\bea
    \delta_{\epsilon}\Phi = b \mu\left( \bar{\epsilon}^i\theta_i +\epsilon_i\bar{\theta}^i\right)\Phi\,,\quad
    \delta_{\varepsilon}\Phi = - b \mu \left(3\bar{\varepsilon}^i\theta_i \,e^{i \mu t} -\varepsilon_i\bar{\theta}^i\,e^{-i \mu t}\right)
    \left(1 - \mu\,\bar{\theta}^k\theta_k \right)  \Phi \,.\label{trPhi1}
\eea
Under the odd transformations \eqref{newtr1}, \p{trPhi}, the component fields in \eqref{chiral} are transformed as:
\bea
    &&\delta z  =-\sqrt{2}\, \epsilon_k\xi^k e^{\frac{i}{2}\mu t}-\sqrt{2}\, \varepsilon_k\xi^k e^{-\frac{i}{2}\mu t},\nn
    &&\delta \xi^i =  \sqrt{2}\, \bar{\epsilon}^i \left(i\dot z - b \mu z\right)e^{-\frac{i}{2}\mu t}
    -\sqrt{2}\,\epsilon^i B e^{\frac{i}{2}\mu t}  +\sqrt{2}\, \bar{\varepsilon}^i \left(i\dot z + b \mu z\right)e^{\frac{i}{2}\mu t}
    -\sqrt{2}\,\varepsilon^i B e^{-\frac{i}{2}\mu t},\nn
    &&\delta B = -\sqrt{2}\, \bar{\epsilon}_k\left[i\dot{\xi}^k - \left(b -\frac{1}{2}\right)\mu\,\xi^k\right]e^{-\frac{i}{2}\mu t}
    -\sqrt{2} \,\bar{\varepsilon}_k\left[i\dot{\xi}^k + \left(b -\frac{1}{2}\right) \mu \,\xi^k\right]e^{\frac{i}{2}\mu t}.\label{offsh}
\eea

To avoid a possible confusion, let us point out that, leaving aside the issues of superconformal covariance, the $SU(2|1)$ chirality based
on the coset \p{coset} and the covariant derivatives defined in \p{cov} (eqs. \p{ch1} - \p{compChir1}) is equivalent to that based on the coset \p{cosetZ}
and the covariant derivatives \p{covZ2}.
Indeed, using the relation ${\cal H}=H -\mu F$, one can rewrite \p{QQZ1} as
$$
\lbrace Q^{i}, \bar{Q}_{j}\rbrace = 2\mu I^i_j +2\delta^i_j \left[H -\mu\,(F + Z_1)\right]\,,
$$
which has the same form as the anticommutator in \p{alg}, with $m =\mu$ and the substitution $F \rightarrow F + Z_1$. The generator $F+ Z_1$ cannot be distinguished
from $F$ since $Z_1$ commutes with anything and does not act on the superspace coordinates. Then one can start from the supercoset \p{coset},
make the shift $F \rightarrow F + Z_1$, and impose, instead of \p{FPhi}, the condition $(\tilde F + Z_1)\Phi = 2\kappa \Phi$  which can be realized
either with $\tilde F\Phi = 2\kappa \Phi, \, Z_1\Phi = 0$ or with $\tilde F\Phi = 0, \, Z_1\Phi = 2\kappa \Phi, \, b\equiv 2\kappa$.
The relevant covariant derivatives \p{cov} and \p{covZ2}, equally as the solutions \p{chiral0} and \p{chiral},
have the same form for both options. The difference between $F$ and $Z_1$ is displayed at the full superconformal level: In the basis $({\cal H}, F)$ the generator $F$
entirely splits from all other superconformal generators, while there is no way to make $Z_1$ not to appear on the right-hand sides of the relevant
anticommutators (see eqs. \eqref{conf-anticomm2} -- \eqref{conf-comm2} for the case $Z_2= Z_3= 0$).

\subsection{Superconformal Lagrangian}
The general $SU(2|1)$ invariant action of the chiral superfields is defined as
\bea
    S(\Phi)= \int dt\, L = \frac{1}{4}\int d\zeta\, f\left(\Phi , \bar{\Phi}\right),
    \label{kinterm}
\eea
where $ f\left(\Phi ,\bar{\Phi}\right)$ is a K\"ahler potential.
The corresponding component Lagrangian
reads
\bea
    {L} &=& g\dot{\bar{z}}\dot{z}
    + \frac{i }{2}\, g\left(\bar{\xi}_i\dot{\xi}^i-\dot{\bar{\xi}}_i\xi^i \right) -\frac{i}{2}\,\bar{\xi}_k\xi^k
    \left(\dot{\bar z}g_{\bar{z}}-\dot{z}g_z\right) -\frac{1}{2}\left(\xi\right)^2
    \bar{B}g_z-\frac{1}{2}\left(\bar{\xi}\,\right)^2 B g_{\bar{z}}\nn
    &&+ \,g\bar{B}B +\frac{1}{4}\left(\xi\right)^2 \left(\bar{\xi}\,\right)^2
    g_{z\bar{z}} + i b \mu \left(\dot{\bar{z}}z -\dot{z}\bar{z}\right)g-\frac{i}{2}\,\mu\left(\dot{\bar{z}}f_{\bar{z}} -\dot{z}f_{z}\right)\nn
    &&-\,\mu\,\bar{\xi}_k\xi^k\,U- \mu^2 V \,,\lb{chir1Lagr}
\eea
where
\bea
    V &=& \frac{b}{2} \left(\bar{z}\partial_{\bar{z}} +z\partial_z \right)f-\frac{b^2}{4} \left(\bar{z}\partial_{\bar{z}} +z\partial_z \right)^2 f\,, \nn
    U &=&\frac{b}{2}\left(\bar{z}\partial_{\bar{z}} +z\partial_z \right)g +\left(b -1\right)g+\frac{g}{2}\,.
\eea
Here, the lower case indices denote the differentiation
in $z,\bar{z}\,$,   $f_{z \bar{z}}=\partial_z \partial_{\bar z}f\,$, and $g :=f_{z \bar{z}}$ is the metric on a K\"ahler manifold.
Performing the redefinition \eqref{B1} in \p{chir1Lagr} and choosing $b=2\kappa$, one can see that this Lagrangian coincides with
the chiral $SU(2|1)$ Lagrangian given in \cite{DSQM} on the basis of the supercoset \p{coset},
in accord with the equivalency of two definitions of chirality, as was discussed in the end of the previous subsection.

According to \eqref{trPhi1}, in order to render the action \eqref{kinterm} superconformal,
one needs to define the K\"ahler potential as
\bea
    f_{\rm sc}^{(b)}(\Phi, \bar\Phi)=\left(\Phi\bar\Phi\right)^{\frac{1}{2b}}\,. \lb{confKalChi}
\eea
Then the Lagrangian
\bea
    {L}_{\rm sc}^{(b)} &=& \frac{\left(z \bar{z}\right)^{\frac{1}{2b}-1}}{4b^2}\left[ \dot{\bar{z}}\dot{z}
    + \frac{i }{2}\left(\bar{\xi}_i\dot{\xi}^i-\dot{\bar{\xi}}_i\xi^i \right)+ \bar{B}B\right]
    +\frac{(2b -1)^2}{64 b^4} \left(z \bar{z}\right)^{\frac{1}{2b}-2}\left(\xi\right)^2 \left(\bar{\xi}\,\right)^2
    \nn
    && +\,\frac{2b-1}{8 b^3}\left(z \bar{z}\right)^{\frac{1}{2b}-2}\left[\frac{i}{2}\,\bar{\xi}_k\xi^k
    \left(\dot{\bar z}z-\dot{z}\bar{z}\right) + \frac{1}{2}\left(\xi\right)^2
    \bar{B}\bar{z}+\frac{1}{2}\left(\bar{\xi}\,\right)^2 B z \right] \nn
    &&-\, \frac{\mu^2}{4}\left(z \bar{z}\right)^{\frac{1}{2b}}
    \label{L2}
\eea
is invariant under the superconformal transformations \eqref{offsh}.

The simplest case of \p{L2} corresponding to the choice $b=1/2$ and yielding
the free action,
\bea
    {L}_{\rm sc}^{(b= 1/2)} = \dot{\bar{z}}\dot{z}
    + \frac{i }{2}\left(\bar{\xi}_i\dot{\xi}^i-\dot{\bar{\xi}}_i\xi^i \right)+ \bar{B}B - \frac{\mu^2}{4}\,z \bar{z}\,, \lb{b12}
\eea
was previously worked out in \cite{DSQM}\footnote{These actions become identical after choosing $\kappa=1/4$ and making the redefinition \eqref{B1}
in the action of ref. \cite{DSQM}, which eliminates there the term $\sim\bar{\xi}\xi$.
Note that the $su(2|2)$ symmetry found in this problem in \cite{DSQM} appears only at the quantum level and is not related
to the superconformal symmetry $SU(1,1|2)\rtimes U(1)$ which is                                                                                                                                                                                      present already at the classical level.}.

Thus we observe that the superconformal sigma-model type action for the multiplet $({\bf 2, 4, 2})$ exists only for the non-zero central charge $Z_1$,
i.e. the relevant invariance supergroup is $SU(1,1|2)$, not its quotient $PSU(1,1|2)$. It is worth noting that the action \p{kinterm} with
the superfield Lagrangian \p{confKalChi},
at any $b\neq 0\,$, is in fact related to the free bilinear action through the field redefinition
\bea
(\Phi_L)^{\frac{1}{2b}}  = \hat{\Phi}_L\,, \quad
S_{\rm sc}^{(b)}(\Phi)  = S_{\rm sc}^{(b=1/2)}(\hat\Phi) =
\frac14 \int d\zeta \left(1+ 2\mu\,\bar{\theta}^k\theta_k \right)^{-\frac12} \hat{\Phi}_L \bar{\hat{\Phi}}_R\,.\lb{genKinChi}
\eea
In other words, without loss of generality, we can always choose $b=1/2$ and deal with the Lagrangian \p{b12}. The same equivalence
to the free actions is valid also for other types of the superconformal sigma-model term of the multiplet $({\bf 2, 4, 2})$.

\subsection{Conformal superpotential}
One can define the chiral superspace measure $d\zeta_L$ which is invariant under the superconformal transformations \eqref{newtr1}:
\bea
    d\zeta_L = dt_L\,d^2\theta\,e^{-i\mu t_L},\qquad\delta_\epsilon\left( d\zeta_L\right) = \delta_\varepsilon\left( d\zeta_L\right)=0\,.\label{measureL}
\eea
Taking into account the explicit form of the superconformal transformations with $b\neq 0$, eqs. \eqref{trPhi}, the only superpotential term
respecting superconformal invariance is:
\bea
    S_{\rm sc}^{\rm pot}(\Phi) = \nu\int d\zeta_L \, \ln{\Phi_L}+ {\rm c.c.}.\lb{supPot}
\eea
The corresponding superconformal Lagrangian reads
\bea
    { L}_{\rm sc}^{\rm pot} = \nu\left(\frac{2B}{z}+\frac{\xi_i\xi^i}{z^2}\right)+ {\rm c.c.}\,.\label{sp}
\eea
After summing it with \p{b12} and eliminating the auxiliary fields, the on-shell superconformal trigonometric Lagrangian acquires the standard conformal potential
\bea
-4\,\frac{|\nu |^2}{z \bar z}\,, \lb{chirPot}
\eea
in addition to the oscillator term $-\frac{\mu^2}{4}\,z\bar z \,$. Thus the non-trivial dynamics in the Lagrangian of
the multiplet $({\bf 2, 4, 2})$ invariant under the trigonometric realization of the superconformal group  arises solely due to
the superpotential term \p{supPot}. For the parabolic realization, the same statement can be traced back to \cite{ikrlev}.

\subsection{Inhomogeneous superconformal action at $b=0$}
As was already mentioned, at $b = 0$ (or, equivalently, at $\kappa =0$) we encounter difficulties, when trying to construct the superconformal action.
It is still possible to define the inhomogeneous superconformal action with $b = 0$ by resorting to the same procedure as in Section \ref{143-0}.
Indeed, the parameter $b$ can be identified with a central charge of $su(1,1|2)$, therefore
one can identify $-b$ with the scaling dimension $\lambda_{D}$ of the chiral multiplet \cite{HT,kuto,khto}. Making the redefinition
\bea
    z\to z+\frac{\rho}{b}\,,\qquad \bar{z}\to \bar{z}+\frac{\rho}{b}\,, \lb{bosshift1}
\eea
detaching the singular factors and, finally, sending $b \to 0$, we obtain the Lagrangian
\bea
    {L}^{(b = 0, \,\rho)}_{\rm sc} &=& e^{\frac{z+\bar{z}}{2\rho}}\left[ \dot{\bar{z}}\dot{z}
    + \frac{i}{2}\left(\bar{\xi}_i\dot{\xi}^i-\dot{\bar{\xi}}_i\xi^i \right)+ \bar{B}B\right]
    -\frac{i}{4\rho}\,\bar{\xi}^k\xi_k \left(\dot{\bar z}-\dot{z}\right)e^{\frac{z+\bar{z}}{2\rho}}\nn
    &&-\,\frac{1}{4\rho}\,\left[\left(\xi\right)^2
    \bar{B} + \left(\bar{\xi}\,\right)^2 B \right]e^{\frac{z+\bar{z}}{2\rho}}
    + \frac{1}{16\rho^2 }\,e^{\frac{z+\bar{z}}{2\rho}}\left(\xi\right)^2 \left(\bar{\xi}\,\right)^2
     - \mu^2\rho^2\,e^{\frac{z+\bar{z}}{2\rho}}.\label{L-rho}
\eea
It can be derived from the following $SU(2|1)$ superfield action
\bea
     \quad S^{(b = 0, \,\rho)}_{\rm sc}(\Phi) = \int dt\, {L}^{(b = 0, \,\rho)}_{\rm sc} =\rho^2\int d\zeta
     \left(1+ 2\mu\,\bar{\theta}^k\theta_k \right)^{-\frac{1}{2}}e^{\frac{\Phi_L+\bar{\Phi}_R}{2\rho}}.\lb{b0act}
\eea
The relevant supersymmetric transformations \eqref{offsh} with $b = 0$ should be extended by the inhomogeneous pieces
\bea
    \delta_{(\rho)}\xi^i = -\sqrt{2}\,\rho\mu \left(\bar{\epsilon}^i e^{-\frac{i}{2}\mu t}
    -\bar{\varepsilon}^i e^{\frac{i}{2}\mu t}\right),\quad \delta_{(\rho)}z = \delta_{(\rho)}B=0\,.
\eea
This is equivalent to saying that, at $b = 0$, the ``passive'' variation of the holomorphic chiral superfield $\Phi_L$
under both supersymmetries involves only the inhomogeneous parts
\bea
\delta_{(\rho)} \Phi_L = 2\rho \mu \left(\bar{\epsilon}^k\theta_k -\bar{\varepsilon}^k\theta_k e^{i\mu t_L}\right). \lb{inhom100}
\eea
It can be obtained from the transformation \eqref{trPhi}, where $\Phi_L$ is shifted as
\bea
    \Phi_L \to \Phi_L + \frac{\rho}{b}
\eea
in conjunction with the shift \p{bosshift1}. Then we can write the invariant superpotential term as
\bea
    S'{}_{\rm sc}^{\rm pot}(\Phi) = \nu\int d\zeta_L \, \Phi_L + {\rm c.c.}\quad \Rightarrow\quad {L}'{}_{\rm sc}^{\rm pot} = 2\nu B+ 2\bar{\nu}\bar{B}.\label{sp1}
\eea
The action \p{b0act}, like its $b\neq 0$ counterpart, can be reduced to the bilinear action by means of the redefinition
$$
e^{\frac{\Phi_L}{2\rho}} \sim \hat{\Phi}_L\,, \quad \Phi_L \sim \ln \hat{\Phi}_L\,.
$$
Then the full $b=0$ superconformal superfield action amounts to a sum of the free kinetic action and the logarithmic superconformal potential.

Note that the action \eqref{b0act} can be rewritten as
\bea
     \quad S^{(b = 0, \,\rho)}_{\rm sc}(\Phi) = \rho^2\int d\zeta\,e^{\frac{\Phi+\bar{\Phi}}{2\rho}},
\eea
where
\bea
    \Phi\big(t,\theta,\bar{\theta}\,\big) = \Phi_L (t_L,\theta\,)-\rho\mu\,\bar{\theta}^k\theta_k\left(1 - \mu\,\bar{\theta}^i\theta_i \right), \lb{SupPhi}
\eea
and
\bea
    \delta\Phi = \rho \mu\left( \bar{\epsilon}^i\theta_i +\epsilon_i\bar{\theta}^i\right)
    - \rho \mu \left(1-\mu\,\bar{\theta}^k\theta_k\right)\left(3\bar{\varepsilon}^k\theta_k e^{i\mu t} - \varepsilon_k \bar{\theta}^k e^{-i\mu t}\right).
\eea
The superfield \p{SupPhi} can be regarded as a solution of the chirality condition (\ref{ch1}a) with the covariant derivative  \eqref{covZ2},
in which the central charge $Z_1$ acts on $\Phi$ as the pure shift
\bea
    Z_1\Phi = \rho\, .
\eea
In this way, the parameter $\rho\neq 0$ activates a non-vanishing central charge in $su(1,1|2)\,$. Thus the superconformal sigma-model type action at $b=0$
exists only on account of a non-zero central charge in $su(1,1|2)\,$, like in the $b\neq 0$ case.

\subsection{The limit $\mu = 0$}
As an instructive example, we consider the parabolic chiral model obtained in the limit $\mu = 0$.

In this limit, the superconformally invariant action of the chiral multiplet becomes
\bea
    S_{\rm sc}^{(\mu =0)}(\Phi)=\frac{1}{4}\int dt\,d^2\theta \,d^2\bar{\theta}\, \left(\Phi \bar{\Phi}\right)^{\frac{1}{2b}}.
\eea
The chiral superfield $\Phi$ transforms under the superconformal charges as
\bea
    \delta\Phi = -4ib \, {\bar{\varepsilon}'^i}\theta_i\, \Phi\,,
\eea
while transforming as a scalar under the $d=1$ Poincar\'e supersymmetry
with the parameters ${\epsilon}'^i, \bar\epsilon'_i$.
The whole amount of superconformal transformations is derived from the trigonometric ones according to the procedure \eqref{parabolic}.
The parameter $b$ is still interpreted as the central charge of $su(1,1|2)$.
Then the superconformal component off-shell Lagrangian
\bea
    {L}_{\rm sc}^{(\mu = 0)} &=& \frac{\left(z \bar{z}\right)^{\frac{1}{2b}-1}}{4 b^2}\left[ \dot{\bar{z}}\dot{z}
    + \frac{i }{2}\left(\bar{\xi}_i\dot{\xi}^i-\dot{\bar{\xi}}_i\xi^i \right)+ \bar{B}B\right]
    +\frac{(2b -1)^2}{64 b^4} \left(z \bar{z}\right)^{\frac{1}{2b}-2}\left(\xi\right)^2 \left(\bar{\xi}\,\right)^2
    \nn
    && +\,\frac{2b-1}{8 b^3}\left(z \bar{z}\right)^{\frac{1}{2b}-2}\left[\frac{i}{2}\,\bar{\xi}_k\xi^k
    \left(\dot{\bar z}z-\dot{z}\bar{z}\right) + \frac{1}{2}\left(\xi\right)^2
    \bar{B}\bar{z}+\frac{1}{2}\left(\bar{\xi}\,\right)^2 B z \right]
\eea
is invariant under both the Poincar\'e  and the superconformal ${\cal N}=4, d=1$ transformations
\bea
    &&\delta z  =-\sqrt{2}\, {\epsilon'_k}\xi^k + \sqrt{2}\,t\, {\varepsilon'_k}\xi^k ,\nn
    &&\delta \xi^i =  \sqrt{2}\,i\, {\bar{\epsilon}'^i}\dot z
    -\sqrt{2}\,{\epsilon}'^i B  -\sqrt{2}\,i\, {\bar{\varepsilon}'^i} \left(t\dot z - 2b  z\right)
    +\sqrt{2}\,t\,{\varepsilon'^i} B,\nn
    &&\delta B = -\sqrt{2}\,i\, {\bar{\epsilon}'_k}\dot{\xi}^k
    +\sqrt{2} \,i\,{\bar{\varepsilon}'_k}\left[t\,\dot{\xi}^k -\left(2b -1\right)\xi^k\right].
\eea

The inhomogeneous superconformal Lagrangian at $b=0$ reads
\bea
    {L}^{(\mu =0, \,b = 0,\,\rho)}_{\rm sc} &=& e^{\frac{z+\bar{z}}{2\rho}}\left[ \dot{\bar{z}}\dot{z}
    + \frac{i}{2}\left(\bar{\xi}_i\dot{\xi}^i-\dot{\bar{\xi}}_i\xi^i \right)+ \bar{B}B\right]
    -\frac{i}{4\rho}\,\bar{\xi}^k\xi_k \left(\dot{\bar z}-\dot{z}\right)e^{\frac{z+\bar{z}}{2\rho}}\nn
    &&-\,\frac{1}{4\rho}\,\left[\left(\xi\right)^2
    \bar{B} + \left(\bar{\xi}\,\right)^2 B \right]e^{\frac{z+\bar{z}}{2\rho}}
    + \frac{1}{16\rho^2 }\,e^{\frac{z+\bar{z}}{2\rho}}\left(\xi\right)^2 \left(\bar{\xi}\,\right)^2,
\eea
and it can be deduced from the superfield action
\bea
     \quad S^{(\mu=0, \,b = 0, \,\rho)}_{\rm sc}(\Phi) = \int dt\, {L}^{(\mu =0, \,b = 0, \,\rho)}_{\rm sc}=\rho^2\int dt\,d^2\theta \,d^2\bar{\theta}\,e^{\frac{\Phi+\bar{\Phi}}{2\rho}}.\lb{Inhmu0}
\eea
In the inhomogeneous case, the superconformal transformation of the superfield $\Phi$ involves only the inhomogeneous piece
\bea
    \delta_{(\rho)} \Phi = -4i\rho\,{\bar{\varepsilon}'^k}\theta_k \,.
\eea

Since the superpotential terms \eqref{sp} and \eqref{sp1} do not depend on $\mu$, their form is preserved in the parabolic limit $\mu = 0$.
The only peculiarity is that the invariant chiral integration measure \eqref{measureL} turns into the flat measure $dt_L\,d^2\theta$. Obviously,
the kinetic superfield term \p{Inhmu0} is reduced to the free one after the appropriate holomorphic redefinition of $\Phi$.
\setcounter{equation}{0}
\section{Generalized chiral multiplet}\label{Sec. 6}
\subsection{Another type of chiral $SU(2|1)$ superspace}
In \cite{SKO}, there was defined a different kind of $SU(2|1)$ chiral superfields. Let us consider the general coset \eqref{cosetZ}.
The chiral condition \eqref{ch1} can be generalized as
\bea
   {\rm (a)}\;\;  \bar{\tilde{{\cal D}}}_i \varphi =0\,,\qquad {\rm (b)} \;\;\tilde{{\cal D}}^i\bar{\varphi}=0\,,\label{ch2}
\eea
where the spinor derivatives $\tilde{{\cal D}}^i, \bar{\tilde{{\cal D}}}_i$ are the following linear combinations of the covariant derivatives
defined in \eqref{cov1}:
\bea
    \bar{\tilde{{\cal D}}}_i = \cos{\lambda}\,\bar{{\cal D}}_i -\sin{\lambda}\,{\cal D}_i \,,\qquad \tilde{{\cal D}}^i
    =\cos{\lambda}\,{\cal D}^i +\sin{\lambda}\,\bar{{\cal D}}^i\,.\label{rotated}
\eea
One can treat such combinations as the result of particular rotation by an extra $SU'(2)$ group with the generators $\{C,\bar{C},F\}$.
In general, the $SU'(2)$ transformations break the covariance of the constraints \p{ch2}. The latter remain covariant only under
the special combination of the $SU'(2)$ generators,
\bea
F' = F\cos{2\lambda}+\frac{1}{2}\left(C+\bar{C}\right)\sin{2\lambda}\,.\label{U1}
\eea

Thus the constraints \eqref{ch2} are covariant under the superconformal group $D\left(2,1;\alpha\right)$ only for $\alpha=-1$, when
it is reduced to the supergroup $PSU(1,1|2)$, and under the external automorphism $U(1)$ group with the generator $F'$ \eqref{U1}.
The Hamiltonian ${\cal H}$ is identified with the whole internal $U(1)$ generator of the non-extended subalgebra $su(2|1) \subset psu(1,1|2)$ for $\alpha = -1$, $m=\mu$.

The conditions \eqref{ch2} amounts to the existence of the left and right chiral subspaces:
\bea
    (\hat{t}_L,\hat\theta_i),\qquad  \big(\hat{t}_R, \bar{\hat{\theta}}^i\big)\,,\label{leftGen}
\eea
where
\bea
\hat{t}_L = t +i\,\bar{\hat\theta}^k\hat{\theta}_k \,,\qquad
    \hat{\theta}_i = \left(\cos{\lambda}\,\theta_i e^{\frac{i}{2}\mu t}+\sin{\lambda}\,
    \bar{\theta}_i e^{-\frac{i}{2}\mu t}\right)
    \left(1 -\frac{\mu}{2}\,\bar{\theta}^k\theta_k\right).\label{subs}
\eea
As expected, the coordinate set $(\hat{t}_L,\hat\theta_i)$ is closed under the $SU(2|1)$ transformations
\bea
    &&\delta\hat{\theta}_{i}=\cos{\lambda}\left(\epsilon_{i}\,e^{\frac{i}{2}\mu \hat{t}_L}
    + \mu\,\bar{\epsilon}^k\hat{\theta}_k\hat{\theta}_{i}\, e^{-\frac{i}{2}\mu \hat{t}_L}\right)
    +\sin{\lambda}\left(\bar{\epsilon}_i\, e^{-\frac{i}{2}\mu \hat{t}_L} +\mu\,\epsilon^k\hat{\theta}_k\hat{\theta}_{i}\, e^{\frac{i}{2}\mu \hat{t}_L}\right),\nn
    && \delta \hat{t}_L=2i\cos{\lambda}\,\bar{\epsilon}^k\hat{\theta}_k\, e^{-\frac{i}{2}\mu \hat{t}_L} -2i\sin{\lambda}\,
    \epsilon^k\hat{\theta}_k \,e^{\frac{i}{2}\mu \hat{t}_L}. \label{tr1-ch}
\eea
The second $SU(2|1)$ transformations
\bea
    &&\delta\hat{\theta}_{i}=\cos{\lambda}\left(\varepsilon_{i}\,e^{-\frac{i}{2}\mu \hat{t}_L}
    - \mu\,\bar{\varepsilon}^k\hat{\theta}_k\hat{\theta}_{i}\, e^{\frac{i}{2}\mu \hat{t}_L}\right)
    + \sin{\lambda}\left(\bar{\varepsilon}_i\, e^{\frac{i}{2}\mu \hat{t}_L} - \mu\,\varepsilon^k\hat{\theta}_k\hat{\theta}_{i}\, e^{-\frac{i}{2}\mu \hat{t}_L}\right),\nn
    && \delta \hat{t}_L=2i\cos{\lambda}\,\bar{\varepsilon}^k\hat{\theta}_k\, e^{\frac{i}{2}\mu \hat{t}_L}
    -2i\sin{\lambda}\,\varepsilon^k\hat{\theta}_k \,e^{-\frac{i}{2}\mu \hat{t}_L}
    \label{tr2-ch}
\eea
are generated by \eqref{S} for $\alpha=-1$ and also leave the left chiral subspace invariant.
The chiral subspace \eqref{subs} is not closed under the $SU'(2)$ transformations generated by $\{C,\bar{C},F\}$, except
those generated by the $U(1)$ generator \p{U1}.

Since at $\alpha=-1$ the superconformal group admits the central extension, in what follows we will assume that the $\alpha=-1$ spinor
covariant derivatives in the definition \p{rotated} are replaced by the central-extended ones ${\cal D}^i_Z, \bar{\cal D}_{Zi}$ \p{covZ2},
i.e. in the chirality constraints \p{ch2} we will use
\bea
    \bar{\tilde{{\cal D}}}_i = \cos{\lambda}\,\bar{{\cal D}}_{Zi} -\sin{\lambda}\,{\cal D}_{Zi} \,,\qquad \tilde{{\cal D}}^i
    =\cos{\lambda}\,{\cal D}^i_Z +\sin{\lambda}\,\bar{{\cal D}}^i_Z\,.\label{rotated1}
\eea
Assuming that the central charge acts on the superfield as\footnote{The eigenvalue of the central charge $Z_1$ in this case is not obliged to be the same $b$
as in Section 5. We hope that denoting it also by $b$ will not give rise to any confusion.}
\bea
    Z_1\varphi = b\cos{2\lambda}\,\varphi\,,\label{Z1varphi}
\eea
the solution of \eqref{ch2} is given by
\bea
    \label{newvarphi}
    \varphi \big(t, \hat{\theta}, \bar{\hat\theta}\,\big)= e^{-b\mu\cos{2\lambda} \,\bar{\hat\theta}^k\hat{\theta}_k}
    \varphi_L\big(\hat{t}_L, \hat{\theta}\,\big),\quad
    \varphi_L \big(\hat{t}_L, \hat{\theta}\,\big)= z+\sqrt{2}\,\hat{\theta}_k\xi^k +(\hat\theta)^2 B\, .
\eea
As we will see, the parameter $|b|$ is associated with the norm of the triplet of central charges like in the previous Section,
since in the case under consideration the superalgebra $psu(1,1|2)$ turns out to be extended by three constant central charges.
This is consistent with the limit $\cos{2\lambda}=1$ in the generalized conditions \eqref{ch2}.

The transformations of the superfield $\varphi$ are given by
\bea
    \delta_{\epsilon}\varphi &=& b \mu\cos{2\lambda}\,\bar{\epsilon}^i e^{-\frac{i}{2}\mu t}\left[\cos{\lambda}\, \hat{\theta}_i
    \left(1+\frac{\mu}{2}\,\bar{\hat\theta}^k\hat{\theta}_k\right)- \sin{\lambda}\,
    \bar{\hat\theta}_i\left(1-\frac{\mu}{2}\,\bar{\hat\theta}^k\hat{\theta}_k\right)\right]\varphi \nn
    &&+\, b \mu\cos{2\lambda}\, \epsilon_i e^{\frac{i}{2}\mu t}\left[\cos{\lambda}\,\bar{\hat\theta}^i\left(1+\frac{\mu}{2}\,
    \bar{\hat\theta}^k\hat{\theta}_k\right)+ \sin{\lambda}\,
    \hat{\theta}^i\left(1-\frac{\mu}{2}\,\bar{\hat\theta}^k\hat{\theta}_k\right)\right]\varphi\, ,\nn
    \delta_{\varepsilon}\varphi &=&b \mu\cos{2\lambda}\,\bar{\varepsilon}^i e^{\frac{i}{2}\mu t}\left[\cos{\lambda}\,
    \hat{\theta}_i\left(1-\frac{\mu}{2}\,\bar{\hat\theta}^k\hat{\theta}_k\right)- \sin{\lambda}\,
    \bar{\hat\theta}_i\left(1+\frac{\mu}{2}\,\bar{\hat\theta}^k\hat{\theta}_k\right)\right]\varphi \nn
    &&+\, b \mu\cos{2\lambda}\, \varepsilon_i e^{-\frac{i}{2}\mu t}\left[\cos{\lambda}\,\bar{\hat\theta}^i
    \left(1-\frac{\mu}{2}\,\bar{\hat\theta}^k\hat{\theta}_k\right)+ \sin{\lambda}\,
    \hat{\theta}^i\left(1+\frac{\mu}{2}\,\bar{\hat\theta}^k\hat{\theta}_k\right)\right]\varphi\nn
    &&-\, 4b \mu\left[\cos{\lambda}\,
    \bar{\varepsilon}^i\hat{\theta}_i  \left(1-\frac{\mu}{2}\,\bar{\hat\theta}^k\hat{\theta}_k\right)e^{\frac{i}{2}\mu t}
    + \sin{\lambda}\,\varepsilon^i\hat{\theta}_i
    \left(1 + \frac{\mu}{2}\,\bar{\hat\theta}^k\hat{\theta}_k\right)e^{-\frac{i}{2}\mu t}\right]\varphi \,.
    \label{varphi}
\eea
The relevant ``passive'' transformations of the holomorphic superfield $\varphi_L$ are
\bea
    \delta_{\epsilon}\varphi_L &=& 2b \mu\cos{2\lambda}\left(\cos{\lambda}\,\bar{\epsilon}^i  e^{-\frac{i}{2}\mu \hat{t}_L}
    -\sin{\lambda}\,\epsilon^i e^{\frac{i}{2}\mu \hat{t}_L}\right) \hat{\theta}_i\varphi_L\, ,\nn
    \delta_{\varepsilon}\varphi_L &=& 2b \mu\cos{2\lambda}\left(\cos{\lambda}\,\bar{\varepsilon}^i  e^{\frac{i}{2}\mu \hat{t}_L}
    - \sin{\lambda}\,\varepsilon^ie^{-\frac{i}{2}\mu \hat{t}_L}\right)\hat{\theta}_i\varphi_L \nn
    &&-\, 4b \mu\left(\cos{\lambda}\,\bar{\varepsilon}^i  e^{\frac{i}{2}\mu \hat{t}_L}+ \sin{\lambda}\,\varepsilon^i
    e^{-\frac{i}{2}\mu \hat{t}_L}\right)\hat{\theta}_i\varphi_L \,.\label{varphiL}
\eea
Then the full set of the off-shell transformations of the component fields is generated by \p{varphiL} and by the coordinate transformations
\eqref{tr1-ch}, \eqref{tr2-ch}:
\bea
    \delta z  &=&-\,\sqrt{2}\cos{\lambda}\,\epsilon_k\xi^k e^{\frac{i}{2}\mu t} - \sqrt{2}\sin{\lambda}\,\bar{\epsilon}_k\xi^k e^{-\frac{i}{2}\mu t},\nn
    \delta \xi^i &=&  \sqrt{2}\, \bar{\epsilon}^i\left(i\cos{\lambda}\,\dot z -b\mu\cos{2\lambda}\,\cos{\lambda}\,z -\sin{\lambda} \,B\right)e^{-\frac{i}{2}\mu t}\nn
    &&
    -\,\sqrt{2}\,\epsilon^i\left(i\sin{\lambda}\,\dot z - b\mu\cos{2\lambda}\,\sin{\lambda}\,z+ \cos{\lambda} \,B\right) e^{\frac{i}{2}\mu t},\nn
    \delta B &=& -\,\sqrt{2}\cos{\lambda}\, \bar{\epsilon}_k\left[i\dot{\xi}^k+\frac{\mu}{2}\left(1-2b\cos{2\lambda}\right)\xi^k\right] e^{-\frac{i}{2}\mu t}\nn
    &&+\,\sqrt{2}\sin{\lambda}\,\epsilon_k\left[i\dot{\xi}^k-\frac{\mu}{2}\left(1+2b\cos{2\lambda}\right)\xi^k\right] e^{\frac{i}{2}\mu t},\label{tr1-comp2}
\eea
\bea
    \delta z &=& -\,\sqrt{2}\cos{\lambda}\,\varepsilon_k\xi^k e^{-\frac{i}{2}\mu t} - \sqrt{2}\sin{\lambda}\,\bar{\varepsilon}_k\xi^k e^{\frac{i}{2}\mu t},\nn
    \delta \xi^i &=&  \sqrt{2}\, \bar{\varepsilon}^i\left[i\cos{\lambda}\,\dot z + 2b\mu\cos{\lambda}\left(1-\frac{1}{2}\cos{2\lambda}\right)z
    -\sin{\lambda} \,B\right]e^{\frac{i}{2}\mu t}\nn
    &&-\,\sqrt{2}\,\varepsilon^i\left[i\sin{\lambda}\,\dot z - 2b\mu\sin{\lambda}\left(1+\frac{1}{2}\cos{2\lambda}\right)z+ \cos{\lambda} \,B\right]
    e^{-\frac{i}{2}\mu t}\,,\nn
    \delta B &=& -\,\sqrt{2}\cos{\lambda}\, \bar{\varepsilon}_k\left[i\dot{\xi}^k -\frac{\mu}{2}\,\xi^k + 2b\mu\left(1-\frac{1}{2}\cos{2\lambda}\right)\xi^k\right]
    e^{\frac{i}{2}\mu t}\nn
    &&+\,\sqrt{2}\sin{\lambda}\,\varepsilon_k\left[i\dot{\xi}^k +\frac{\mu}{2}\,\xi^k - 2b\mu\left(1+\frac{1}{2}\cos{2\lambda}\right)\xi^k\right]
    e^{-\frac{i}{2}\mu t}\,.\label{tr2-comp2}
\eea

The new set of the transformations \eqref{tr1-comp2}, \eqref{tr2-comp2} closes on the
centrally extended superalgebra \eqref{conf-anticomm2} -- \eqref{conf-comm2} with the central charges
\bea
    Z_1=b\cos{2\lambda}\,,\qquad Z_2=b\sin{2\lambda}\,,\qquad Z_3=-b\sin{2\lambda}\,,\qquad
    (Z_1)^2-Z_2 Z_3=b^2.
\eea
The precise realization of the central charges on the superfields $\varphi, \bar\varphi$ is given by the following transformations
\bea
\delta \varphi =2ib\mu\left(a_1\cos{2\lambda} + a_2 \sin{2\lambda}\right)\varphi,
\eea
where $a_1$, $a_2$ are infinitesimal parameters associated with $Z_1$ and $Z_2=-Z_3\,$.
\subsection{The superconformal Lagrangian}
The most general sigma-model part of the $SU(2|1)$ invariant action of the generalized chiral superfields
$\varphi\big(t, \hat\theta,\bar{\hat\theta}\,\big)$ is specified
by an arbitrary K\"ahler potential $ f(\varphi , \bar{\varphi})$:
\bea
    S (\varphi) = \int dt\, \tilde{L} = \frac{1}{4}\int d\hat\zeta \,f(\varphi , \bar{\varphi})\,,
\eea
where the $SU(2|1)$ invariant measure is
\bea
    d\hat\zeta = dt\, d^2\hat\theta\, d^2\bar{\hat{\theta}}\left[1+ \mu\cos{2\lambda}\,\bar{\hat{\theta}}^k\hat{\theta}_k
    -\frac{\mu}{2}\sin{2\lambda}\,\big(\bar{\hat{\theta}}\,\big)^2 -\frac{\mu}{2}\sin{2\lambda} \,(\hat{\theta})^2\right].\label{measure1}
\eea
This measure is not invariant under the second-type  $SU(2|1)$ transformations (with $\mu \rightarrow -\mu$).

The transformations of $d\hat\zeta$ can be canceled, using
the inhomogeneity of the chiral superfield $\varphi$ transformation \eqref{varphi} for $b\neq 0$.
One can check that the superconformal action is uniquely specified by the following K\"ahler potential
\bea
    f_{\rm sc}^{(b)}(\varphi, \bar\varphi)=\left(\varphi \bar{\varphi}\right)^{\frac{1}{2b}}.
\eea
The corresponding full superconformally invariant off-shell component Lagrangian
reads
\bea
    \tilde{L}_{\rm sc}^{(b)} &=& \frac{\left(z \bar{z}\right)^{\frac{1}{2b}-1}}{4b^2}\left[ \dot{\bar{z}}\dot{z}
    + \frac{i}{2}\left(\bar{\xi}_i\dot{\xi}^i-\dot{\bar{\xi}}_i\xi^i \right)+ \bar{B}B \right]
    +\frac{(2b -1)^2}{64 b^4} \left(z \bar{z}\right)^{\frac{1}{2b}-2}\left(\xi\right)^2 \left(\bar{\xi}\,\right)^2\nn
    && +\,\frac{2b-1}{8b^3}\left(z \bar{z}\right)^{\frac{1}{2b}-2}\left[\frac{i}{2}\,\bar{\xi}_k\xi^k
    \left(\dot{\bar z}z-\dot{z}\bar{z}\right) + \frac{1}{2}\left(\xi\right)^2
    \bar{B}\bar{z}+\frac{1}{2}\left(\bar{\xi}\,\right)^2 B z \right]\nn
    &&-\,\frac{2b-1}{16b^2}\left(z \bar{z}\right)^{\frac{1}{2b}-2}
    \mu\sin{2\lambda}\left[\bar{z}^2\left(\xi\right)^2 +z^2\left(\bar{\xi}\,\right)^2\right],\nn
    && -\,\frac{\left(z \bar{z}\right)^{\frac{1}{2b}-1}}{2b}\left[
    \frac{\mu}{2}\sin{2\lambda}\left(\bar{B}z+B\bar{z}\right)+\frac{b \mu^2}{2}\cos^2{2\lambda}\, z \bar{z}\right].\label{L3}
\eea
In the particular case $\cos\lambda = 1$, one comes back to the Lagrangian \eqref{L2}.

\subsection{Remark}\label{remark}
Let us make the following redefinition in \p{L3}
\bea
    B= \tilde{B}+b\mu\sin{2\lambda}\,z,\quad{\rm and\;\; c.c.}.\label{B2}
\eea
The redefined superconformal Lagrangian \eqref{L3} exactly coincides  with the previously constructed
superconformal Lagrangian \eqref{L2} (with $B \rightarrow \tilde{B}$). However, it is invariant under the following modified transformations
\bea
    \delta z &=& -\,\sqrt{2}\left(\cos{\lambda}\,\epsilon_k + \sin{\lambda}\,\bar{\varepsilon}_k\right)\xi^k e^{\frac{i}{2}\mu t}
    - \sqrt{2}\left(\sin{\lambda}\,\bar{\epsilon}_k + \cos{\lambda}\,\varepsilon_k \right)\xi^k e^{-\frac{i}{2}\mu t},\nn
    \delta \xi^i &=&  \sqrt{2}\, \left(\cos{\lambda}\,\bar{\epsilon}^i - \sin{\lambda}\,\varepsilon^i\right)\left(i\dot z -  b \mu z\right)e^{-\frac{i}{2}\mu t}
    -\sqrt{2}\left( \sin{\lambda} \,\bar{\epsilon}^i+ \cos{\lambda} \,\varepsilon^i\right)\tilde{B} e^{-\frac{i}{2}\mu t}\nn
    &&-\,\sqrt{2}\, \left(\sin{\lambda}\,\epsilon^i - \cos{\lambda}\,\bar{\varepsilon}^i\right)\left(i\dot z + b \mu z\right)e^{\frac{i}{2}\mu t}
    -\sqrt{2}\left(\cos{\lambda}\, \epsilon^i + \sin{\lambda} \,\bar{\varepsilon}^i\right)\tilde{B} e^{\frac{i}{2}\mu t},\nn
    \delta \tilde{B} &=& -\,\sqrt{2}\left(\cos{\lambda}\, \bar{\epsilon}_k -\sin{\lambda}\,\varepsilon_k\right)\left[i\dot{\xi}^k  -
    \left(b - \frac{1}{2}\right) \mu\,\xi^k\right] e^{-\frac{i}{2}\mu t}\nn
    &&+\,\sqrt{2}\left(\sin{\lambda}\,\epsilon_k - \cos{\lambda}\, \bar{\varepsilon}_k\right)\left[i\dot{\xi}^k +
    \left(b - \frac{1}{2}\right)\mu\,\xi^k\right] e^{\frac{i}{2}\mu t}\,,\label{offsh1}
\eea
which are just \p{tr1-comp2}, \p{tr2-comp2} rewritten in terms of $\tilde{B}$  defined in \eqref{B2}. These transformations are induced by
\eqref{tr1-ch}, \eqref{tr2-ch} and the superfield transformations
\bea
    \delta\tilde\varphi_L= 2b \mu\, \left[\left(\cos{\lambda}\,\bar{\epsilon}^i - \sin{\lambda}\,\varepsilon^i\right)\hat{\theta}_i\,e^{-\frac{i}{2} \mu \hat{t}_L}
    - \left(\cos{\lambda}\,\bar{\varepsilon}^i - \sin{\lambda}\,\epsilon^i \right)\hat{\theta}_i \,e^{\frac{i}{2} \mu \hat{t}_L}\right]  \tilde\varphi_L \,.\label{varphi-tr}
\eea
The newly defined  chiral superfield $\tilde\varphi_L$ encompasses the field set $(z, \xi^k, \tilde{B})$ and is related to \eqref{newvarphi} as
\bea
    \varphi_L\big(\hat{t}_L,\hat\theta\,\big) = \left[1+b\mu\sin{2\lambda}\,(\hat{\theta})^2\right]\tilde\varphi_L\big(\hat{t}_L,\hat\theta\,\big),\qquad
    \tilde\varphi_L\big(\hat{t}_L,\hat\theta\,\big)=z+\sqrt{2}\,\hat{\theta}_k\xi^k +(\hat\theta)^2 \tilde{B}\,. \label{B3}
\eea
Note that the $\varepsilon_i, \bar\varepsilon^k$ transformations in \p{offsh1}, \p{varphi-tr} are obtained from the $\epsilon_i, \bar\epsilon^k$ ones
just by the replacement $\mu \rightarrow -\mu$ in the latter, in the agreement with the general statement of Section 3.

After passing to the new independent linear combinations of the infinitesimal parameters $\{\epsilon, \bar{\epsilon}, \varepsilon, \bar{\varepsilon}\}$ as
\bea
    \tilde\epsilon_k = \cos{\lambda}\,\epsilon_k + \sin{\lambda}\,\bar{\varepsilon}_k\,,\qquad
    \tilde\varepsilon_k = \cos{\lambda}\,\varepsilon_k + \sin{\lambda}\,\bar{\epsilon}_k\,, \qquad {\rm and\;\; c.c.},\label{epsilon}
\eea
the above transformations take just the form of \eqref{offsh}. These new combinations of the parameters correspond to the following redefinition of the
$D(2,1;\alpha{=}-1)$ supercharges
\bea
    \tilde{Q}^i = \cos{\lambda}\,Q^i -\sin{\lambda}\,\bar{S}^i ,\qquad \tilde{S}^i= \cos{\lambda}\,S^i -\sin{\lambda}\,\bar{Q}^i ,\qquad {\rm and\;\; c.c.}.
\eea
The redefined  supercharges close on the superalgebra \eqref{conf-anticomm2} -- \eqref{conf-comm2} with the {\it single} central charge $Z_1=b$, i.e.,
the superconformal models of the generalized chiral multiplet prove to be equivalent to the superconformal models associated with the standard chiral multiplet.
One can check that the generator \eqref{U1} is the  $U(1)$ automorphism generator of the $su(1,1|2)$ superalgebra with the supercharges
$\tilde{Q}^i$, $\tilde{S}^i$. Thus, as far as the superconformal $SU(2|1)$ mechanics is concerned, the generalized $SU(2|1)$ chiral
multiplet does not give rise to new models compared to the ``standard'' chiral multiplet.

More details on connection between the standard and generalized $SU(2|1)$ chiralities from the superspace point of view are given in Appendix \ref{App-B}.

\setcounter{equation}{0}
\section{The ``mirror'' multiplet $({\bf 2, 4, 2})$}\label{Sec. 7}
The $\alpha =0$ version of the chirality conditions \p{ch1} or \p{ch2} is not covariant under the full second $SU'(2) \propto \{F, C, \bar C\}$ and,
hence, under the superconformal group $D(2,1;\alpha{=}0)$ which necessarily contains $SU'(2)$ as a subgroup.

However, one can define the ``mirror'' chiral multiplet $({\bf 2, 4, 2})$ which respects the covariance under the $\alpha = 0$ superconformal group
realized in the coset \eqref{coset2}.
Using the $\alpha =0$ covariant derivatives \eqref{cov2}, we may impose the relevant chiral conditions as
\bea
    \bar{\cal D}_1 \tilde\Phi ={\cal D}^2 \tilde\Phi = 0\,.\label{ch3}
\eea
It is straightforward to show that at $\alpha = 0$ these conditions  are covariant  with respect to the superconformal symmetry $PSU(1,1|2)\rtimes U(1)_{\rm ext}$,
with the internal $SU(2)$ group generated by $\{F, C, \bar C\}$ and ${\cal H}$ as the Hamiltonian.
The generator $I_1^1=-I_2^2$ plays the role of an external automorphism $U(1)_{\rm ext}$ generator, while the generators $I_1^2$, $I_2^1$ violate the covariance
of \eqref{ch3} and so should be thrown away. Since the $SU'(2)$ generators $\{F, C, \bar{C}\}$ form a subalgebra of $psu(1,1|2)$,
allowing the chiral superfield to have
an external $U(1)$ charge with respect to $\tilde{F}$ would entail the necessity to attach the whole $SU'(2)$ index to $\tilde\Phi$. This
would result in extension of the field contents of $\tilde\Phi$. In order to deal with the chiral multiplet possessing the minimal field contents $({\bf 2, 4, 2})$,
we are so led to require that
\bea
    \tilde{F}\tilde{\Phi}=0\,.\label{Fzero}
\eea

The conditions \eqref{ch3} amount to the existence of the chiral subspace $(t_L,\theta_1,\bar{\theta}^2)$, where
\bea
    &&t_L = t +i\,\bar{\theta}^1\theta_1 - i\,\bar{\theta}^2\theta_2\, .
\eea
It is closed under the superconformal transformations
\bea
    &&\delta t_L = 2i\left(\bar{\epsilon}^1\theta_1 + {\epsilon}_2\bar{\theta}^2+\bar{\varepsilon}^1\theta_1
    e^{i \mu t_L}+{\varepsilon}_2\bar{\theta}^2  e^{-i \mu t_L}\right),\nn
    &&\delta \theta_{1}=\epsilon_{1}+\varepsilon_{1}e^{-i\mu t_L} + 2\mu\, \varepsilon_{2}\bar{\theta}^2\theta_1 e^{-i\mu t_L},\nn
    &&\delta \bar{\theta}^2 = \bar{\epsilon}^2 + \bar{\varepsilon}^2 e^{i\mu t_L} - 2\mu\, \bar{\varepsilon}^{1}\theta_1 \bar{\theta}^2 e^{i\mu t_L}.
\eea

As in case of the $\alpha=-1$ chiral multiplets, we extend the algebra \eqref{alpha0} by the central charge generator:
\bea
    &&\lbrace Q^{i}, \bar{Q}_{j}\rbrace = 2\delta^i_j\left({\cal H} + \mu\,F\right) + 2\mu\,(\sigma_3)^i_j V,\nn
    &&\left[F, \bar{Q}_{l}\right]=-\frac{1}{2}\,\bar{Q}_{l}\,,\qquad \left[F, Q^{k}\right]=\frac{1}{2}\,Q^{k},\nn
    &&\left[{\cal H}, \bar{Q}_{l}\right]= \frac{\mu}{2}\,\bar{Q}_{l}\,,\qquad \left[{\cal H}, Q^{k}\right]= -\frac{\mu}{2}\, Q^{k}.\lb{ext319}
\eea
The superfield $\tilde{\Phi}$ can have a non-zero charge under $V$:
\bea
    V\tilde\Phi = a \tilde \Phi .
\eea
Then the extended algebra \p{ext319} is embedded in the $\alpha=0$ counterpart of \eqref{conf-anticomm2} -- \eqref{conf-comm2}.
Like in the coset \eqref{cosetZ}, we place the central charge in the stability subgroup
\bea
    \frac{\{Q^{i},\bar{Q}_{j}, {\cal H}, F, V\}}{\{F, V\}}\,.\label{cosetV}
\eea
The modified covariant derivatives are as follows
\bea
    &&{\cal D}^1 = e^{-\frac{i}{2}\mu t}\left(\frac{\partial}{\partial\theta_1}
    -i\bar{\theta}^1\partial_t -\mu\,\bar{\theta}^1 \tilde{F} -\mu\,\bar{\theta}^1 V\right),\nn
    &&\bar{{\cal D}}_2 = e^{\frac{i}{2}\mu t}\left(-\frac{\partial}{\partial\bar{\theta}^2}
    +i\theta_2\partial_t + \mu\,\theta_2\tilde{F}-\mu\,\theta_2 V \right).
\eea
Keeping in mind the condition \p{Fzero}, the solution of \eqref{ch3} can be  written as
\bea
    \tilde\Phi\big(t,\theta_k,\bar{\theta}^k\,\big) = e^{-a\mu\,\bar{\theta}^k\theta_k}
    \left[z(t_L)+\sqrt{2}\,\theta_1 \eta^1 (t_L)\,e^{\frac{i}{2}\mu t_L}+\sqrt{2}\,\bar{\theta}^2\bar{\eta}_2(t_L) \,
    e^{-\frac{i}{2}\mu t_L} - 2\theta_1\bar{\theta}^2 B(t_L)\right].
    \label{Phi1}
\eea
Thus the number $a$ is an analog of the charge $b$ and it can be identified with the central charge of
the conformal superalgebra $su(1,1|2)$ of the $\alpha=0$ case.

The $\alpha=0$ chirality-preserving odd transformations of $\tilde\Phi$ read
\bea
    \delta_{\epsilon}\tilde\Phi &=& a\mu\left(\bar{\epsilon}^1\theta_1 -\epsilon_2\bar{\theta}^2\right) \tilde\Phi
    - a\mu\left(\bar{\epsilon}^2\theta_2 -\epsilon_1\bar{\theta}^1\right) \tilde\Phi ,\nn
    \delta_{\varepsilon}\tilde\Phi &=&
     -\,3 a\mu \left(\bar{\varepsilon}^1\theta_1 e^{i\mu t}-\varepsilon_2\bar{\theta}^2 e^{-i\mu t}\right)\left(1-\frac{\mu}{3}\,\bar{\theta}^k\theta_k\right) \tilde\Phi \nn
     &&-\, a\mu \left(\bar{\varepsilon}^2\theta_2 e^{i\mu t}-\varepsilon_1\bar{\theta}^1 e^{-i\mu t}\right)\left(1-3\mu\,\bar{\theta}^k\theta_k\right) \tilde\Phi .
\eea
They generate the off-shell transformations of the component fields
\bea
    \delta z &=& -\sqrt{2}\, \epsilon_1\eta^1 \,e^{\frac{i}{2}\mu t}-\sqrt{2}\, \bar{\epsilon}^2\bar{\eta}_2\, e^{-\frac{i}{2}\mu t}-\sqrt{2}\,
    \varepsilon_1\eta^1 \,e^{-\frac{i}{2}\mu t}
    -\sqrt{2}\, \bar{\varepsilon}^2\bar{\eta}_2\, e^{\frac{i}{2}\mu t},\nn
    \delta \eta^1 &=& \sqrt{2}\, \bar{\epsilon}^1 \left(i\dot z - a\mu z\right) e^{-\frac{i}{2}\mu t}
    +\sqrt{2}\,\bar{\epsilon}^2 B e^{-\frac{i}{2}\mu t}  +\sqrt{2}\, \bar{\varepsilon}^1 \left(i\dot z + a\mu z\right) e^{\frac{i}{2}\mu t}
    +\sqrt{2}\,\bar{\varepsilon}^2  B e^{\frac{i}{2}\mu t},\nn
    \delta \bar{\eta}_2 &=& \sqrt{2}\, \epsilon_2 \left(i\dot z + a\mu z\right) e^{\frac{i}{2}\mu t}
    -\sqrt{2}\,\epsilon_1 B e^{\frac{i}{2}\mu t}  +\sqrt{2}\, \varepsilon_2  \left(i\dot z - a\mu z\right) e^{-\frac{i}{2}\mu t}
    -\sqrt{2}\,\varepsilon_1  B e^{-\frac{i}{2}\mu t} ,\nn
    \delta B &=& -\,\sqrt{2}\, \epsilon_2\left[i\dot{\eta}^1 + \left(a -\frac{1}{2}\right)\mu\,\eta^1\right]e^{\frac{i}{2}\mu t}
    -\sqrt{2} \,\varepsilon_2\left[i\dot{\eta}^1 - \left(a -\frac{1}{2}\right)\mu\,\eta^1\right]e^{-\frac{i}{2}\mu t}\nn
    &&+\,\sqrt{2}\, \bar{\epsilon}^1\left[i\dot{\bar{\eta}}_2 - \left(a -\frac{1}{2}\right)\mu\,\bar{\eta}_2\right]e^{-\frac{i}{2}\mu t}
    +\sqrt{2} \,\bar{\varepsilon}^1\left[i\dot{\bar{\eta}}_2 + \left(a -\frac{1}{2}\right)\mu\,\bar{\eta}_2\right]e^{\frac{i}{2}\mu t}.
\eea
The superconformally invariant superfield action
\bea
     \quad S^{(\alpha = 0, \,a)}_{\rm sc}(\tilde{\Phi}) = \int dt\, {L}^{(\alpha = 0, \,a)}_{\rm sc}
     =-\frac{1}{4}\int dt\,d^2\theta \,d^2\bar{\theta}\, \left(\tilde\Phi \bar{\tilde\Phi}\right)^{\frac{1}{2a}}
\eea
yields the following component superconformal Lagrangian:
\bea
    {L}^{(\alpha = 0, \,a)}_{\rm sc} &=& \frac{\left(z \bar{z}\right)^{\frac{1}{2a}-1}}{4 a^2}\left[ \dot{\bar{z}}\dot{z}
    + \frac{i }{2}\left(\bar{\eta}_i\dot{\eta}^i-\dot{\bar{\eta}}_i\eta^i \right)+ \bar{B}B\right]
    -\frac{(2a-1)^2}{64 a^4} \left(z \bar{z}\right)^{\frac{1}{2a}-2}\left(\eta\right)^2 \left(\bar{\eta}\,\right)^2
    \nn
    && +\,\frac{2a-1}{8 a^3}\left(z \bar{z}\right)^{\frac{1}{2a}-2}\left[\frac{i}{2}\left(\bar{\eta}_1\eta^1
    -\bar{\eta}_2\eta^2\right)
    \left(\dot{\bar z}z-\dot{z}\bar{z}\right) + \bar{\eta}_2\eta^1
    \bar{B}\bar{z}+\bar{\eta}_1\eta^2 B z \right] \nn
    && - \, \frac{\mu^2}{4}\left(z \bar{z}\right)^{\frac{1}{2a}}.
\eea
One can cast it into the form of the Lagrangian \eqref{L2} by passing to the fermions $\xi^{i^\prime}$ with the {\it primed} doublet indices as
\bea
    &&\eta^1 = \xi^{1^\prime},\qquad \bar{\eta}_1 = \bar{\xi}_{1^\prime},\qquad
    \eta^2 = \bar{\xi}_{2^\prime}, \qquad \bar{\eta}_2 = \xi^{2^\prime},\\
    && \overline{\left(\xi^{i^\prime}\right)}= \bar{\xi}_{i^\prime}\,,\qquad \overline{\left(\eta^i\right)}= \bar{\eta}_i\,.
\eea
This redefinition makes manifest the property that the fermionic fields are transformed as doublets
of the $SU'(2)$ group with the generators $\{F,C,\bar{C}\}$.

As in the case of $\alpha = -1$, we can add the superconformal superpotential term
\bea
    S^{{\rm pot}(\alpha = 0)}_{\rm sc}(\tilde{\Phi}) = s\int dt_L \,d\theta_1\,d\bar{\theta}^2\, \ln{\tilde\Phi_L}+ {\rm c.c.}\quad\Rightarrow\quad
    {L}^{{\rm pot}(\alpha = 0)}_{\rm sc}= 2s\left(\frac{B}{z}+\frac{\bar{\eta}_2\eta^1}{z^2}\right)+ {\rm c.c.},
\eea
which yields on shell the standard conformal mechanics potential in addition to the oscillator-type term $\sim \mu^2$ coming from the superconformal
superfield kinetic term. The latter can be reduced to the free one as in the previous cases.

Thus the superconformal action at $\alpha = 0$ can be constructed using the superfield approach associated with the $\alpha=0$ supercoset \eqref{coset2},
while the $\alpha=-1$ action \eqref{L2} was based on the $SU(2|1)$ supercoset. In the parabolic limit $\mu=0$, both supercosets
are reduced to the standard flat ${\cal N}=4, d=1$ superspace.
\setcounter{equation}{0}

\section{$D$-module representation approach}\label{Sec. 8}
Here we sketch a different approach to the $d=1$ superconformal actions based solely on the component field considerations \cite{kuto,khto,HT}.

\subsection{The ${\cal N} = 4$ linear supermultiplets}
As a preamble, it is instructive, following ref. \cite{HT},  to give a concise account of the general superconformal properties of the set of linear ${\cal N} = 4$
supermultiplets $({\bf k, 4, 4-k})$ for $k=0,1,2,3,4$, despite the fact that in the present paper we deal  with the cases $k=1,2$ only.

The linear supermultiplets $({\bf k, 4, 4-k})$ for $k=0,1,2,3,4$ exist in the parabolic and hyperbolic/trigonometric variants \cite{HT}.
The parabolic variant leads to actions which are both superconformally invariant and show up the manifest Poincar\'e supersymmetry.
The hyperbolic/trigonometric variants lead to superconformally invariant actions in which the $d=1$ Poincar\'e supersymmetry is implicit
(the corresponding supercharges are not a ``square root'' of the canonical Hamiltonian as the time-translation generator) so
they look  non-supersymmetric or weakly supersymmetric. The potentials are bounded from below in the trigonometric version (i.e., they are well-behaved).
They are unbounded (bad-behaved) in the hyperbolic version. In the parabolic case the Hamiltonian is a Cartan generator of the conformal $so(2,1)$ subalgebra.
In the hyperbolic/trigonometric case the canonical Hamiltonian is a root generator of $so(2,1)$.

The connection of these ${\cal N} = 4$ linear supermultiplets with the ${\cal N} = 4$ superconformal algebras and
the corresponding scaling dimensions $\lambda_{D}$ is as follows \cite{kuto, khto, HT}:
\begin{itemize}
\item $({\bf 0, 4, 4})$: $D\left(2,1;\alpha{=} 2\lambda_{D}\right)$;
\item $({\bf 1, 4, 3})$: $D\left(2,1;\alpha{=} \lambda_{D}\right)$. At $\alpha = -1$ and $\alpha = 0$\,, the extra
inhomogeneous constant parameter $c$ is allowed;
\item $({\bf 2, 4, 2})$: $su(1,1|2)$. The scaling dimension $\lambda_{D}$ is associated with a central charge of $su(1,1|2)$ as $\lambda_{D}=-b$\,;
\item $({\bf 3, 4, 1})$: $D\left(2,1;\alpha{=} -\lambda_{D}\right)$;
\item $({\bf 4, 4, 0})$: $D\left(2,1;\alpha{=} -2\lambda_{D}\right)$.
\end{itemize}
For all multiplets except $({\bf 2, 4, 2})$, the superconformal algebra at $\alpha = -1,0$ can be reduced to the superalgebra $psu(1,1|2)$.

Another type of inhomogeneous linear transformation \cite{HT} is only present at $\lambda_{D} = 0$. The inhomogeneous parameter is $\rho$.
The supermultiplets $({\bf k, 4, 4-k})_{\rho}$  carry a representation of $psu(1,1|2)$ for $k=0,1,3,4$ and its central-extended version $su(1,1|2)$ for $k=2$.
One should note that the superconformal actions based on $({\bf k, 4, 4-k})$ at a given $\lambda_{D}$ are not defined at $\lambda_{D} = 0$.
On the other hand, the superconformal actions based on $({\bf k, 4, 4-k})_{\rho}$ are well-defined.

\subsection{Superconformally invariant $({\bf 2, 4, 2})$ actions from the $D$-module approach}
In \cite{HT}, all hyperbolic $D$-module representations for the ${\cal N} = 4$ linear multiplets $({\bf k, 4, 4-k})$ were obtained and
the trigonometric $D$-module representations can be easily derived from the hyperbolic representations.
Then one can construct the hyperbolic/trigonometric superconformal actions proceeding from the $D$-module representations. The method of construction is described in \cite{kuto}.
Some superconformal actions of the supermultiplet $({\bf 1,4,3})$ were found in this way in \cite{HT}.
Here we present the realization of the ${\cal N}=4$ superconformal algebras and perform the construction of the superconformal actions
for the supermultiplet $({\bf 2,4,2})$ in this alternative approach.

We use the same notation and definitions for the component fields and superconformal generators as in the previous Sections.
The action of generators of the conformal algebra are given below:
\bea
    &&{\cal H}z = i\dot{z},\qquad {\cal H}\xi^i = i\dot{\xi}^i,\qquad {\cal H}B = i\dot{B},\nn
    &&T z = e^{-i\mu t}\left(i\dot{z} - b\mu z\right) ,\; T\xi^i = e^{-i\mu t}\left[i\dot{\xi}^i - \left(b-\frac{1}{2}\right)\mu \xi^i\right] ,\;
    T B = e^{-i\mu t}\left[i\dot{B} - \left(b-1\right)\mu B\right],\nn
    &&\bar{T} z = e^{i\mu t}\left(i\dot{z} + b\mu z\right) ,\; \bar{T}\xi^i = e^{i\mu t}\left[i\dot{\xi}^i + \left(b-\frac{1}{2}\right)\mu \xi^i\right] ,\;
    \bar{T} B = e^{i\mu t}\left[i\dot{B} + \left(b-1\right)\mu B\right].%,\nn
%   &&I_k^j\,z=0\,,\qquad I_k^j \, \xi^i = \frac{\delta_k^j}{2} \, \xi^i - \delta_k^i \, \xi^j\,,\qquad I_k^jB=0\,.
\eea
The fermionic generators are specified by
\bea
    &&Q^i z  =-\sqrt{2}\, \xi^i e^{\frac{i}{2}\mu t},\qquad Q^i \xi^k  =\sqrt{2}\, \varepsilon^{ik}B e^{\frac{i}{2}\mu t},\qquad Q^i B = 0\,,\nn
    &&\bar{Q}_i z  =0\,,\quad \bar{Q}_i \xi^k  = -\sqrt{2}\,\delta_i^k \left(i\dot z - b \mu z\right)e^{-\frac{i}{2}\mu t},\quad
    \bar{Q}_i B = -\sqrt{2}\,\varepsilon_{ik}\left[i\dot{\xi}^k - \left(b -\frac{1}{2}\right)\mu\,\xi^k\right]e^{-\frac{i}{2}\mu t},\nn
    &&S^i z  =-\sqrt{2}\, \xi^i e^{-\frac{i}{2}\mu t},\qquad S^i \xi^k  =\sqrt{2}\, \varepsilon^{ik}B e^{-\frac{i}{2}\mu t},\qquad S^i B = 0\,,\nn
    &&\bar{S}_i z  =0\,,\quad \bar{S}_i \xi^k  = -\sqrt{2}\,\delta_i^k \left(i\dot z + b \mu z\right)e^{\frac{i}{2}\mu t},\quad
    \bar{S}_i B = -\sqrt{2}\,\varepsilon_{ik}\left[i\dot{\xi}^k + \left(b -\frac{1}{2}\right)\mu\,\xi^k\right]e^{\frac{i}{2}\mu t}.
\eea
Since all $({\bf 2,4,2})$ Lagrangians can be reduced to the free Lagrangian \eqref{b12}, it is enough to consider the free case $b=1/2$.
Then the superconformally invariant action is generated from the {\it prepotential} $f(z,\bar{z})$ by acting with the supercharges $Q_i$
on the propagating bosons $z,\bar{z}$ as
\bea
    {L}^{(b=1/2)}_{\rm sc} = \frac{1}{16}\,Q_i Q^i \bar{Q}^k\bar{Q}_k\, f(z,\bar{z})\,.\label{D-reps-act}
\eea
The prepotential $f(z,\bar{z})$ can be found from the constraint that the action of conformal generators on the Lagrangian
produces a total time-derivative:
\bea
    T\,{L}^{(b=1/2)}_{\rm sc}=\frac{d}{dt}\,M,\qquad   \bar{T}\, {L}^{(b=1/2)}_{\rm sc} =\frac{d}{dt}\,\bar{M}\,, \label{D-reps-constr}
\eea
where the explicit form of $M$ is of no interest for our purposes. Solving these constraints, we obtain the {\it prepotential}
\bea
    f(z,\bar{z})=z\bar{z}\,.
\eea
The corresponding superconformal action \eqref{D-reps-act} generated from the $D$-module representations can be shown to coincide with the
superconformal action \eqref{b12} derived from the $SU(2|1)$ superspace approach.

The superpotential term \eqref{sp} can also be equivalently constructed using the $D$-module approach. We define
\bea
    {L}_{\rm sc}^{\rm pot} =\frac{1}{2}\, Q_i Q^i\, {h}(z) +\frac{1}{2}\,\bar{Q}^i\bar{Q}_i\,\bar{h}(\bar{z}) \label{D-reps-sp}
\eea
and impose the conformal constraints in the same way as for \eqref{D-reps-constr}. As their solution we uniquely obtain
\bea
    {h}(z) = -\nu \ln{z}\,,\qquad \bar{h}(\bar{z})=-\bar{\nu}\ln\bar{z}\,.
\eea
It is direct to check that \eqref{D-reps-sp} for such ${h}(z)$ coincides  with \eqref{sp}.

Note that the superfield and $D$-module approaches can be regarded as complementary to each other. The second method directly
yields the component off-shell Lagrangians. On the other hand, the superfield techniques bring to light some properties
which are hidden in the component formulations. For instance, the reducibility of the general sigma-model type action of the multiplet
$({\bf 2, 4,2})$ to the free one is immediately seen, when using the chiral $SU(2|1)$ superfield language, as in Sections 5 - 7.

\setcounter{equation}{0}
\section{Summary and outlook}\label{Sec. 9}
In this paper, we presented the superspace realization of the trigonometric-type ${\cal N}=4, d=1$ superconformal symmetry.
This realization can be given in terms of the $SU(2|1)$ superspace  at $\alpha\neq 0$ or in terms of the $U(1)$ deformed flat ${\cal N}=4, d=1$ superspace
 at $\alpha = 0$. In the contraction limit $\mu=0$, the relevant superconformal models are reduced to the standard models
 of the parabolic superconformal mechanics, with the superconformal Lagrangians constructed out of the standard ${\cal N}=4$, $d=1$ superfields.
 The main advantage of the $SU(2|1)$ superfield approach (or its
degenerate $\alpha=0$ version) is that it automatically yields the trigonometric-type realization of the superconformal symmetry, with the correct-sign harmonic
oscillator term $\sim \mu^2$ in the component actions.

Our construction is based on the new observation that the most general ${\cal N}=4, d=1$ superconformal algebra $D(2,1;\alpha)$ at $\alpha\neq 0$
in the $SU(2|1)$ superspace realizations can be represented as a closure of its two $su(2|1)$ subalgebras, one of which defines the superisometry
of the underlying $SU(2|1)$ superspace
while the other is obtained from the first one by the reflection of
the contraction parameter as $\mu \rightarrow -\mu$. This suggests the simple selection rule for singling out the superconformally invariant
actions in the general set of the $SU(2|1)$ invariant actions constructed in \cite{DSQM,SKO}. The superconformal $SU(2|1)$ actions are those which are even functions of $\mu$.
The superalgebra $D(2,1;\alpha{=}0)\sim psu(1,1|2) \oplus su(2)$ (and its central extensions) admit a similar closure structure, this time in terms
of two $\mu$-dependent $U(1)$ deformed flat ${\cal N}=4, d=1$ superalgebras.

We gave an off-shell superfield formulation of the trigonometric superconformal actions of the multiplet $({\bf 1, 4, 3})$ some of which were constructed
earlier at the component level in \cite{HT}, and presented new trigonometric superconformal actions for the chiral multiplet $({\bf 2, 4, 2})$.
For the latter multiplet the superconformal actions exists only for $\alpha =-1$ and $\alpha =0$, and they are always reduced to
a sum of the free kinetic (sigma-model type) $SU(2|1)$ superfield action and the superconformal superfield potential, yielding, in the bosonic component sector,
a sum of the standard conformal mechanics potential $\sim \frac{1}{|z|^2}$ and the oscillator term $\sim \mu^2 |z|^2$. The $SU(2|1)$ superfield approach
provides a simple proof of this notable property.
Another feature easily revealed in the $SU(2|1)$ superfield approach is that the superconformal $\alpha = -1$ models corresponding to the generalized
$({\bf 2, 4, 2})$ chirality \cite{SKO} proved to be equivalent to the superconformal models associated with the standard chiral $SU(2|1)$ multiplet.
The common property of all superconformal sigma-model type  $({\bf 2, 4, 2})$ actions (at $\alpha=-1$ and $\alpha=0$) is that they exist only on account
of non-zero central charge in the corresponding superconformal algebras $su(1,1|2)$. We also presented an alternative way of deriving the component superconformal
$({\bf 2, 4, 2})$ actions, based on the $D$-module representation approach developed in \cite{HT,kuto,khto}, and found the nice agreement with the superfield
considerations.

It would be interesting to use the $SU(2|1)$ superspace approach to construct analogous models with the trigonometric realization of superconformal symmetry
for other off-shell $SU(2|1)$ supermultiplets, with the field contents $({\bf 3, 4, 1})$ and $({\bf 4, 4, 0})$, as well as the multi-particle
generalizations of all such models (including those studied in the present paper). Also, it seems important to better understand the relationship between
the $SU(2|1)$ superfield approach and the component approach based on the $D$-module representations of the superconformal symmetries, including
$D(2,1;\alpha)$. Finding out the possible links with the superconformal structures in the higher-dimensional theories based on curved analogs of
flat rigid supersymmetries (see, e.g., \cite{KuSor}) is also an urgent subject for the future study.

\section*{Acknowledgements}
We are particularly grateful to N. L. Holanda for valuable comments.
We thank S. Fedoruk, J. Lukierski and A. Nersessian for interest in this work. E.I. and S.S. acknowledge partial support from the RFBR grants No 12-02-00517,
No 13-02-90430, No 13-02-91330, and a grant of the Heisenberg-Landau program. F.T. is grateful to JINR - Dubna for hospitality during part of the work.
His research was supported by CNPq under PQ Grant 306333/2013-9.

\appendix
\section{Central extension of superconformal algebra}\label{App-A}
At $\alpha=-1$ (or $\alpha=0$) it is possible to extend the superalgebra $D\left(2,1;\alpha\right)$ by additional central charges.
In this particular case, the (anti)commutators \eqref{basD}, \eqref{D12} can be cast in the form
\begin{eqnarray}
    &&\{ {Q}_{\alpha i i^\prime},  {Q}_{\beta j j^\prime}\}=
2\,\Big(\epsilon_{ij}\epsilon_{i^\prime j^\prime} {T}_{\alpha\beta} - \epsilon_{\alpha\beta}\epsilon_{i^\prime j^\prime} {J}_{ij}-
    \epsilon_{\alpha\beta}\epsilon_{ij} C_{i^\prime j^\prime}\Big)\,,\nn
    &&\left[{T}_{\alpha\beta}, {Q}_{\gamma i i^\prime}\right] =
-i\,\epsilon_{\gamma(\alpha}{Q}_{\beta) i i^\prime}\,,\qquad \left[{T}_{\alpha\beta}, {T}_{\gamma\delta}\right] =
i\left(\epsilon_{\alpha\gamma}{T}_{\beta\delta} +\epsilon_{\beta\delta}{T}_{\alpha\gamma}\right),\nn
    &&\left[{J}_{ij}, {Q}_{\alpha k i^\prime}\right] =
-i\,\epsilon_{k(i}{Q}_{\alpha j) i^\prime}\,,\qquad\left[{J}_{ij}, {J}_{kl}\right] =
i\left(\epsilon_{ik}{J}_{jl} +\epsilon_{jl}{J}_{ik}\right),\label{D21-1}
\end{eqnarray}
where the central charges ${C}_{i^\prime j^\prime}$ commute with all other generators. They form a vector with respect to the automorphism $SU'(2)_{\rm ext}$
transformations acting on the indices $i^\prime, j^\prime$.
The norm of the vector ${C}_{i^\prime j^\prime}$ of central charges,
\begin{equation}
    |C|^2:=\frac{1}{2}\,{C}^{i^\prime k^\prime}{C}_{i^\prime k^\prime}\,,
\end{equation}
is an invariant of these $SU'(2)_{\rm ext}$ transformations. Hence, in the case of constant central charges,
we can choose the $SU'(2)_{\rm ext}$ frame in such a way that only one non-vanishing central charge remains, e.g., its third component:
\begin{equation}
{C}_{1^\prime 2^\prime}\neq 0\,,\qquad {C}_{1^\prime 1^\prime}={C}_{2^\prime 2^\prime}=0\,.
\end{equation}
Simultaneously, $SU'(2)_{\rm ext}$ is reduced to the automorphism $U(1)_{\rm ext}$.

One can equivalently rewrite the superalgebra \p{D21-1} as  the appropriate extension of \eqref{conf-anticomm} -- \eqref{conf-alg} at $\alpha=-1$:
\bea
    &&\lbrace Q^{i}, \bar{Q}_{j}\rbrace = 2\mu I^i_j +2\delta^i_j \left({\cal H} - \mu\,Z_1\right)\, ,\qquad
    \{S^i, \bar{S}_j\}=-2\mu I^i_j + 2\delta^i_j \left({\cal H} + \mu\,Z_1\right)\,  ,\nn
    &&\{S^i, \bar{Q}_j\}=2\delta^i_j T,\qquad \{Q^i, \bar{S}_j\}=2\delta^i_j \bar{T},\nn
    &&\{Q^i,S^k\}=2\mu\varepsilon^{ik}Z_2\,,\qquad\{\bar{Q}_j, \bar{S}_k\}=2\mu\varepsilon_{jk}Z_3\,,\label{conf-anticomm2}
\eea
\bea
    &&\left[I^i_j,  I^k_l\right] = \delta^k_j I^i_l - \delta^i_l I^k_j\,,\nn
    &&\left[I^i_j, \bar{Q}_{l}\right] = \frac{1}{2}\,\delta^i_j\bar{Q}_{l}-\delta^i_l\bar{Q}_{j}\, ,\qquad \left[I^i_j, Q^{k}\right]
    = \delta^k_j Q^{i} - \frac{1}{2}\,\delta^i_j Q^{k},\nn
    &&\left[I^i_j, \bar{S}_{l}\right] = \frac{1}{2}\,\delta^i_j\bar{S}_{l}-\delta^i_l\bar{S}_{j}\, ,\qquad \left[I^i_j, S^{k}\right]
    = \delta^k_j S^{i} - \frac{1}{2}\,\delta^i_j S^{k},
\eea
\bea
    &&\left[T , \bar{T}\right] = - 2 \mu {\cal H} , \qquad \left[{\cal H}, T\right]= \mu T,\qquad \left[{\cal H}, \bar{T}\right]=-\mu\bar{T},\nn
    &&\left[{\cal H}, \bar{S}_{l}\right]= -\frac{\mu}{2}\,\bar{S}_{l}\,,\quad \left[{\cal H}, S^{k}\right]= \frac{\mu}{2}\,S^{k},\quad
    \left[{\cal H}, \bar{Q}_{l}\right]= \frac{\mu}{2}\,\bar{Q}_{l}\,,\quad \left[{\cal H}, Q^{k}\right]= - \frac{\mu}{2}\,Q^{k},\nn
    &&\left[T , Q^i\right]=-\mu S^i,\quad \left[T , \bar{S}_j\right]= -\mu \bar{Q}_j\,,\quad
    \left[\bar{T} , \bar{Q}_j\right]= \mu\bar{S}_j\,,\quad\left[\bar{T} , S^i\right]=\mu Q^i.\label{conf-comm2}
\eea
According to \eqref{sca}, the central charges appearing here are related to the central charges defined in \eqref{D21-1} as
\begin{equation}
{C}_{1^\prime 2^\prime}={C}_{2^\prime 1^\prime}=iZ_1\,,\qquad {C}_{1^\prime 1^\prime}=iZ_2\,,\qquad {C}_{2^\prime 2^\prime}=iZ_3\,,
\qquad |C|^2 = (Z_1)^2-Z_2 Z_3\, .
\end{equation}
\section{More on $SU(2|1)$ chiralities}\label{App-B}
As was demonstrated, the superconformal $SU(2|1)$ models of the $({\bf 2, 4, 2})$ superfield defined by the generalized (central-charge extended) chirality
condition \eqref{ch2} are in fact equivalent to those constructed on the basis of the superfield subjected to the ``standard'' chirality condition \eqref{ch1}
(or its central-charge extended version \p{ch1b}). So in the superconformal case the parameter $\lambda$ entering \eqref{ch2}, \eqref{rotated}, \eqref{rotated1}
is unessential. This is in contrast with the pure $SU(2|1)$ invariant models
in which $\lambda$ is a physical parameter specifying a new class of such models \cite{SKO}.

Let us discuss the interplay between two types of the $SU(2|1)$ chirality in more detail, based upon the superspace considerations.
It will be useful to pass to the coordinates  $\{t,\tilde{\theta}_j,\bar{\tilde\theta}^i\}$ defined by the relations \p{altpar}.
Being specialized to $\alpha = -1$, these relations read:
\bea
    &&\tilde{\theta}_j = e^{\frac{i}{2}\mu t}\theta_j\left(1-\frac{\mu}{2}\,\bar{\theta}^k\theta_k\right)=e^{\frac{i}{2}\mu t_L}\theta_j\,, \qquad
    t_L = t +i\,\bar{\tilde{\theta}}^i\tilde{\theta}_i\,,\nn
    &&\bar{\tilde\theta}^i = e^{-\frac{i}{2}\mu t}\bar{\theta}^i\left(1-\frac{\mu}{2}\,\bar{\theta}^k\theta_k\right)
= e^{-\frac{i}{2}\mu t_R}\bar\theta^i\,,\qquad
    t_R = t - i\,\bar{\tilde{\theta}}^i\tilde{\theta}_i\,.
\eea
The $SU(2|1)$ supercharges \eqref{Q2} are rewritten as
\bea
    &&Q^i = e^{\frac{i}{2}\mu t}\bigg\{\left[1 - \frac{\mu}{2}\,\bar{\tilde\theta}^k\tilde{\theta}_k
    - \frac{\mu^2}{16}\,(\tilde{\theta})^2 \big(\bar{\tilde\theta}\,\big)^2 \right]\frac{\partial}{\partial\tilde{\theta}_i}
    -\mu\,\bar{\tilde\theta}^i\bar{\tilde\theta}^k\frac{\partial}{\partial\bar{\tilde\theta}^k}
    +\, i\bar{\tilde\theta}^i \left(1 + \frac{\mu}{2}\,\bar{\tilde\theta}^k\tilde{\theta}_k\right)\partial_t\bigg\},\nn
    &&\bar{Q}_j = e^{-\frac{i}{2}\mu t} \bigg\{\left[1 -\frac{\mu}{2}\,\bar{\tilde\theta}^k\tilde{\theta}_k
    - \frac{\mu^2}{16}\,(\tilde{\theta})^2 \big(\bar{\tilde\theta}\,\big)^2 \right] \frac{\partial}{\partial\bar{\tilde\theta}^j}
     +\mu\,\tilde{\theta}_j\tilde{\theta}_k\frac{\partial}{\partial\tilde{\theta}_k}
     + i\tilde{\theta}_j\left(1 + \frac{\mu}{2}\,\bar{\tilde\theta}^k\tilde{\theta}_k\right) \partial_t\bigg\}.
     \label{Q3}
\eea
The extra generators $S_i$ completing $SU(2|1)$ to $D(2,1;\alpha {=} -1)$ are represented in this basis as $S(\mu) = Q(-\mu)$.
The covariant derivatives \eqref{covZ2} take the form
\bea
    &&{\cal D}^i_Z = \left[1+\frac{\mu}{2}\,\bar{\tilde\theta}^k\tilde{\theta}_k-\frac{\mu^2}{16}\,(\tilde{\theta})^2 \big(\bar{\tilde\theta}\,\big)^2\right]
    \left(\frac{\partial}{\partial\tilde{\theta}_i}-i\bar{\tilde\theta}^i\partial_{t}+\mu\,\bar{\tilde\theta}^i Z_1\right),\nn
    &&\bar{{\cal D}}_{Zj} = \left[1+\frac{\mu}{2}\,\bar{\tilde\theta}^k\tilde{\theta}_k-\frac{\mu^2}{16}\,(\tilde{\theta})^2 \big(\bar{\tilde\theta}\,\big)^2\right]
    \left(-\frac{\partial}{\partial\bar{\tilde{\theta}}^j}+i\tilde{\theta}_j\partial_{t}-\mu\,\tilde{\theta}_j Z_1\right).\label{D}
\eea
We ignore the matrix $SU(2)$ generators $\tilde{I}^l_j$ in ${\cal D}^i, \bar{{\cal D}}_j$, because the generalized chiral superfields defined by \p{ch2}
cannot carry external $SU(2)$ indices owing to the compatibility relation
$$
\{\bar{\tilde{{\cal D}}}_k ,\bar{\tilde{{\cal D}}}_j \} = -2 \mu\sin{2\lambda} \,\tilde{I}_{ij}\,, \qquad \mbox{and \;c.c.}.
$$

Using the explicit expressions \p{D}, the generalized chirality condition \eqref{ch2} with $\bar{\tilde{{\cal D}}}_j$ defined according to
\p{rotated1} can be rewritten in the basis $\{t,\tilde{\theta}_j,\bar{\tilde\theta}^i\}$  as
\bea
    \bar{\tilde{{\cal D}}}_j \varphi &=& \Big[1+\frac{\mu}{2}\,\bar{\tilde\theta}^k\tilde{\theta}_k
    -\frac{\mu^2}{16}\,(\tilde{\theta})^2 \big(\bar{\tilde\theta}\,\big)^2\Big]
    \Big[\cos{\lambda}\left(-\frac{\partial}{\partial\bar{\tilde{\theta}}^j}+i\tilde{\theta}_j\partial_{t}-\mu\,\tilde{\theta}_j Z_1\right)\nn
    && -\, \varepsilon_{ji}\sin{\lambda}\left(\frac{\partial}{\partial\tilde{\theta}_i}-i\bar{\tilde\theta}^i\partial_{t}
    +\mu\,\bar{\tilde\theta}^i Z_1\right) \Big]\varphi = 0\,.\label{ch4}
\eea
It is easy to check that the coordinates $\hat{\theta}_i$ defined in \eqref{subs} and parametrizing the left chiral superspace \p{leftGen}
can be represented, for generic $\lambda$, as a particular $SU(2)$ rotation of the coordinates $\tilde{\theta}_j$, $\bar{\tilde\theta}^i$:
\bea
    \hat{\theta}_i = \cos{\lambda}\,\tilde{\theta}_i + \sin{\lambda}\,
    \bar{\tilde\theta}_i \,,\qquad
    \bar{\hat{\theta}}^i = \cos{\lambda}\,\bar{\tilde\theta}^i - \sin{\lambda}\,
    \tilde{\theta}^i.
\eea
In the basis $\{t,\hat{\theta}_j,\bar{\hat\theta}^i\}$ the condition \eqref{ch4} becomes
\bea
    \bar{\tilde{{\cal D}}}_j \varphi &=&\left[1+\frac{\mu}{2}\cos{2\lambda}\,\bar{\hat\theta}^k\hat{\theta}_k
    - \frac{\mu}{4}\sin{2\lambda}\left(\bar{\hat\theta}^k\bar{\hat\theta}_k+\hat{\theta}_k\hat{\theta}^k\right)
    - \frac{ \mu^2}{16}\,(\hat{\theta})^2 \big(\bar{\hat\theta}\,\big)^2\right]\nn
    &&\times\left(-\frac{\partial}{\partial\bar{\hat{\theta}}^j}+i\hat{\theta}_j\partial_{t}-\mu\,\hat{\theta}_j Z_1\right)\varphi = 0\,.\lb{ch44}
\eea
Comparing \p{ch44} with the ``standard'' chirality constraint \p{ch1b} written through $\bar{{\cal D}}_{Zj}$ from \eqref{D}, we see
that they have the same form, up to an unessential non-singular scalar factor and the change of Grassmann coordinates as $\tilde{\theta}\leftrightarrow\hat{\theta}$.

One can define the new supercharges
\bea
    \tilde{Q}^i = \cos{\lambda}\,Q^i -\sin{\lambda}\,\bar{S}^i ,\qquad  {\rm and \;c.c.},\label{Q4}
\eea
and check that they coincide with the generators \eqref{Q3} in which the same substitution $(\tilde{\theta}, \bar{\tilde{\theta}}) \rightarrow (\hat{\theta},
\bar{\hat\theta})$ has been performed. The same applies to the $\tilde{S}^i$ supercharges
\bea
    \tilde{S}^i= \cos{\lambda}\,S^i -\sin{\lambda}\,\bar{Q}^i ,\qquad {\rm and \;c.c.},\label{S4}
\eea
and the corresponding conformal subgroup generators. We also observe that the $U(1)$ generator \eqref{U1} takes the form
\bea
    &&F' = F\cos{2\lambda}+\frac{1}{2}\left(C+\bar{C}\right)\sin{2\lambda}=\frac{1}{2}\left(\bar{\hat\theta}^k\frac{\partial}{\partial\bar{\hat\theta}^k}
    -\hat{\theta}_k\frac{\partial}{\partial\hat{\theta}_k}\right),
\eea
which, up to the coordinate change just mentioned, coincides with the definition \eqref{bosU2} of $F\,$.

As was shown in Section \ref{remark}, the transformations \eqref{offsh1} of the component fields
under the supercharges \eqref{Q4}, \eqref{S4} with the parameters $\tilde{\epsilon}_i, \tilde{\varepsilon}_i$ defined in \p{epsilon}
have the same form as the original $(Q, S)$ transformations \p{offsh} with the parameters ${\epsilon}_i, {\varepsilon}_i\,$.
Accordingly, the  superfield $(Q, S)$ transformations \eqref{trPhi} of $\Phi_L$ can be given the same form as
the transformations \eqref{varphi-tr} of the superfield $\tilde\varphi_L(\hat{t}_L, \hat{\theta})$  under the supercharges \eqref{Q4}, \eqref{S4}
by rewriting \eqref{trPhi} through the coordinates $\big(t_L,\tilde\theta\,\big)$:
\bea
    \delta\Phi_L\big(t_L,\tilde\theta\,\big) = 2b \mu\left( \bar{\epsilon}^i\tilde{\theta}_i\,e^{-\frac{i}{2} \mu t_L} -
    \bar{\varepsilon}^i\tilde{\theta}_i \,e^{\frac{i}{2} \mu t_L}\right)  \Phi_L \big(t_L,\tilde\theta\,\big).
\eea
Thus we observe the full similarity between $\Phi_L$ and $\tilde\varphi_L$ modulo the change $(t_L,\tilde{\theta})\leftrightarrow(\hat{t}_L,\hat\theta)\,$.

This phenomenon can be summarized as follows. In the basis $\{t,\hat{\theta}_j,\bar{\hat\theta}^i\}$ the rotated superconformal generators
\p{Q4}, \p{S4} have the same form as the original supercharges $Q^i, S^i$ in the basis $\{t,\tilde{\theta}_j,\bar{\tilde\theta}^i\}$.
The superconformal subclass of the actions of the generalized multiplet $({\bf 2, 4, 2})$ is invariant under both $Q$ and $S$ supersymmetries,
hence it is invariant under their $\tilde{Q}$ and $\tilde{S}$ realizations as well. The generalized chiral $SU(2|1)$ superfield defined for the $Q, S$ realization
of the superconformal group looks just as the standard chiral $SU(2|1)$ superfield with respect to the equivalent $\tilde{Q}, \tilde{S}$ realization.
So the superconformal $({\bf 2, 4, 2})$ actions actually cannot distinguish on which kind of the chiral $SU(2|1)$ superfield they are built and, respectively,
cannot involve any dependence on the parameter $\lambda\,$.

To make the latter property manifest, let us proceed from the superconformal action of  generalized
chiral superfield $\varphi(t, \hat{\theta}, \bar{\hat{\theta}})$ as the solution \p{newvarphi} of \eqref{ch44}
\bea
    S_{\rm sc}^{(b)}(\varphi)= \frac{1}{4}\int  d\hat{\zeta}\left(\varphi \bar{\varphi}\right)^{\frac{1}{2b}}, \label{ActioN}
\eea
where the integration measure $d\hat{\zeta}$ in the $SU(2|1)$ superspace basis $\{t, \hat{\theta}, \bar{\hat{\theta}} \}$ is defined in \p{measure1}.
Since the component action \eqref{L3} has no dependence on $\lambda$, the $\lambda$ dependence of the  superfield action \eqref{ActioN} is also expected to be fake.
Using the relations \p{newvarphi} and \eqref{measure1}, we rewrite \eqref{ActioN} through the (anti)holomorphic superfields $\varphi_L$, $\bar{\varphi}_R$ as
\bea
    S_{\rm sc}^{(b)}(\varphi) = \frac{1}{4}\int  dt\, d^2\hat\theta\, d^2\bar{\hat{\theta}}\big[1-\frac{\mu^2}{4}(\bar{\hat{\theta}})^2(\hat{\theta})^2 \big]
    \big[1-\frac{\mu}{2}\sin{2\lambda} (\hat{\theta})^2 \big]\big[1-\frac{\mu}{2}\sin{2\lambda}(\bar{\hat{\theta}})^2 \big]
    (\varphi_L \bar{\varphi}_R)^{\frac{1}{2b}}.\lb{Action2}
\eea
One can absorb the (anti)holomorphic factors in this action into the redefinition of $\bar\varphi_R$, ${\varphi}_L$ as in \eqref{B3} and cast \p{Action2} in the
following final form
\bea
    S_{\rm sc}^{(b)}(\varphi) = \frac{1}{4}\int  dt\, d^2\hat\theta\, d^2\bar{\hat{\theta}} \left(1+ \mu\,\bar{\hat{\theta}}^k\hat{\theta}_k\right)
    \left(\tilde{\varphi} \bar{\tilde{\varphi}}\right)^{\frac{1}{2b}}. \lb{Action3}
\eea
Here,  the newly introduced superfield $\tilde{\varphi}$ is a solution of \p{ch44} with $Z_1\tilde\varphi_L = b\,\tilde\varphi_L$:
\bea
    \bar{\tilde{{\cal D}}}_j \tilde{\varphi} = 0\,,\quad \Rightarrow\quad
    \tilde{\varphi} \big(t,\hat\theta, \bar{\hat\theta}\,\big)= e^{-b\mu\,\bar{\hat\theta}^k\hat{\theta}_k}\tilde{\varphi}_L\big(\hat{t}_L,\hat\theta\,\big)\,,
\eea
and it does not display any $\lambda$ dependence, equally as the action \p{Action3}. Comparing it with the superconformal action \p{kinterm}, \p{confKalChi} rewritten
in the basis $\{t, \tilde{\theta}_i, \bar{\tilde{\theta}}^k \}$,
\bea
    S_{\rm sc}^{(b)}(\Phi) = \frac{1}{4}\int  dt\, d^2\tilde\theta\, d^2\bar{\tilde{\theta}} \left(1+ \mu\,\bar{\tilde{\theta}}^k\tilde{\theta}_k\right)
    \left(\Phi \bar{\Phi}\right)^{\frac{1}{2b}}, \quad
    \Phi\big(t, \tilde{\theta}, \bar{\tilde{\theta}}\,\big) = e^{-b\mu\bar{\tilde{\theta}}^i \tilde{\theta}_i}\Phi_L\big(t_L, \tilde{\theta}\,\big)\,, \lb{Action4}
\eea
where the expression \p{tilde-measure} for the $d\tilde{\zeta}$ integration measure was used, we observe its identity with \p{Action3}, up to the
interchange $\tilde{\theta} \leftrightarrow \hat{\theta}$, as was anticipated above. Note that the integration measure in \p{Action3},
\bea
    dt\, d^2\hat\theta\, d^2\bar{\hat{\theta}}\left(1+ \mu\,\bar{\hat{\theta}}^k\hat{\theta}_k\right),\label{tildeQ-measure}
\eea
is invariant with respect to $SU(2|1)$ generated by the rotated supercharges \eqref{Q4}.

The non-conformal $SU(2|1)$ invariant chiral actions are invariant under the transformations generated by $Q_i$ and $\bar Q^i$ but not under
the $\tilde{Q}_i, \bar{\tilde{Q}}^i$ transformations since the definition of the latter involve the superconformal generators $S_i$ and $\bar S^i$.
Hence they differ for the standard and generalized chiral $({\bf 2, 4, 2})$ multiplets and depend on $\lambda$ as an essential parameter. It
labels  non-equivalent $SU(2|1)$ actions and the corresponding SQM models \cite{SKO}.

\section{Hyperbolic superconformal mechanics}\label{App-C}
The hyperbolic superconformal mechanics can be obtained by substituting the deformation parameter in the trigonometric models as $\mu \rightarrow i\mu$\,.
One can see that the superconformal generators defined in Section \ref{subsect. 3.2} go over to the new generators
\bea
    && Q^i \longrightarrow \Pi^i ,\qquad \bar{Q}_k \longrightarrow \bar{\Theta}_k\,,\qquad S^i \longrightarrow \Theta^i ,\qquad \bar{S}_k \longrightarrow \bar{\Pi}_k\,,\nn
    && T \longrightarrow T_2\,,\qquad \bar{T} \longrightarrow T_1\,,\qquad {\cal H}\longrightarrow {\cal H}_{\rm h}\,,
\eea
which behave under the Hermitian conjugation as
\bea
    \left(\Pi^{k}\right)^{\dagger} = \bar{\Pi}_{k}\,,\qquad \left(\Theta^{k}\right)^{\dagger} = \bar{\Theta}_{k}\,\quad\Rightarrow\quad
    \left(T_2\right)^{\dagger} = T_2\,,\qquad\left(T_1\right)^{\dagger} = T_1\,,
    \qquad\left({\cal H}_{\rm h}\right)^{\dagger}={\cal H}_{\rm h}\,.\label{QS-conj}
\eea
In this basis, the basic anticommutation relations of $D\left(2,1;\alpha\right)$ can be rewritten as
\bea
    &&\lbrace \Pi^{i}, \bar{\Theta}_{j}\rbrace = -2i\alpha\mu\, I^i_j +2\delta^i_j\big[{\cal H}_{\rm h} + i\left(1+\alpha\right)\mu\,F\big] ,\nn
    &&\{\Theta^i, \bar{\Pi}_j\}=2i\alpha\mu\, I^i_j + 2\delta^i_j \big[{\cal H}_{\rm h} -i\left(1+\alpha\right) \mu\,F\big] ,\nn
    &&\{\Theta^i, \bar{\Theta}_j\}=2\delta^i_j T_2\,,\qquad \{\Pi^i, \bar{\Pi}_j\}=2\delta^i_j T_1\,,\nn
    &&\{\Pi^i , \Theta^k\} = -2i\left(1+\alpha\right)\mu\,\varepsilon^{ik}C,\qquad\{\bar{\Theta}_j, \bar{\Pi}_k\}=2i\left(1+\alpha\right)\mu\, \varepsilon_{jk}\bar{C}.
\eea

The bosonic truncation of the corresponding conformal group generators \eqref{trig} yields their hyperbolic realization:
\bea
    {\cal H}_{\rm h}=i\partial_t\,,\qquad T_1 = i e^{-\mu t}\partial_t \,,\qquad
    T_2 = i e^{\mu t}\partial_t\,.\label{Hyper}
\eea
The corresponding hyperbolic realization of \eqref{trigonso21} now reads
\bea
    \hat{H}=\frac{i}{2}\left(1 + \cosh{\mu t}\,\right)\partial_t\,,\quad
    \hat{K}=-\frac{2i}{\mu^2}\left(1 - \cosh{\mu t}\,\right)\partial_t\,,\quad
    \hat{D}=\frac{i}{\mu}\sinh{\mu t}\,\partial_t\,,\qquad \mu \neq 0\,. \label{hyperso21}
\eea
In contrast to the trigonometric case, the time-translation generator ${\cal H}_{\rm h}$ is now
\bea
    {\cal H}_{\rm h}= \hat{H} - \frac{\mu^2}{4}\,\hat{K}.\lb{Hhyp}
\eea
Due to the minus sign before $\mu^2\hat{K}$, we face the quantum mechanical problem in which the potentials accompanying the kinetic terms are not bounded from below, like
in the parabolic case \cite{DFF}. This difficulty could of course be cured in a similar way by passing to
\bea
    {\cal H}_{\rm trig}= \hat{H} + \frac{\mu^2}{4}\,\hat{K}=\cosh{\mu t}\,\partial_t
\eea
as the correct time-evolution operator. The discrete energy spectrum with the canonical Hamiltonian
can be obtained only in the trigonometric models of superconformal mechanics. Note that $D\left(2,1;\alpha\right)$ contains no any {\it self-conjugated} subalgebra with four real
supercharges, in which ${\cal H}_{\rm h}$ would appear on the r.h.s. of the basic anticommutator, in contrast to the parabolic and trigonometric cases.
\subsection{Example}
As an instructive example, we consider the simplest free case $b=1/2$ of the multiplet ${\bf (2,4,2)}$
\bea
    {L}^{(b=1/2)}_{\rm sc} = \dot{\bar{z}}\dot{z}
    + \frac{i }{2}\left(\bar{\xi}_i\dot{\xi}^i-\dot{\bar{\xi}}_i\xi^i \right)+ \bar{B}B \lb{AA}
    - \frac{\mu^2}{4}\,z \bar{z}\,,\label{L-trig}
\eea
and the relevant superconformal transformations
\bea
    &&\delta z  =-\sqrt{2}\, \epsilon_k\xi^k e^{\frac{i}{2}\mu t}-\sqrt{2}\, \varepsilon_k\xi^k e^{-\frac{i}{2}\mu t},\nn
    &&\delta \xi^i =  \sqrt{2}\, \bar{\epsilon}^i \left(i\dot z - \frac{\mu}{2}\, z\right)e^{-\frac{i}{2}\mu t}
    -\sqrt{2}\,\epsilon^i B e^{\frac{i}{2}\mu t}  +\sqrt{2}\, \bar{\varepsilon}^i \left(i\dot z + \frac{\mu}{2}\, z\right)e^{\frac{i}{2}\mu t}
    -\sqrt{2}\,\varepsilon^i B e^{-\frac{i}{2}\mu t},\nn
    &&\delta B = -\sqrt{2}\,i\, \bar{\epsilon}_k\dot{\xi}^k e^{-\frac{i}{2}\mu t}
    -\sqrt{2}\,i \,\bar{\varepsilon}_k\dot{\xi}^k e^{\frac{i}{2}\mu t}.\lb{BB}
\eea
These transformations correspond to the superalgebra \eqref{conf-anticomm2} -- \eqref{conf-comm2} with $Z_1=1/2$\,.

After the change $\mu \rightarrow i\mu$ in \p{AA}, \p{BB}, we obtain the hyperbolic mechanics Lagrangian as
\bea
    {L}^{(b=1/2)}_{\rm sc\, (h)} = \dot{\bar{z}}\dot{z}
    + \frac{i }{2}\left(\bar{\xi}_i\dot{\xi}^i-\dot{\bar{\xi}}_i\xi^i \right)+ \bar{B}B
    + \frac{\mu^2}{4}\,z \bar{z}\,,
\eea
and the superconformal transformations as
\bea
    &&\delta z  =-\sqrt{2}\,\upsilon_k\xi^k e^{-\frac{1}{2}\mu t}-\sqrt{2}\, \varsigma_k\xi^k e^{\frac{1}{2}\mu t},\nn
    &&\delta \xi^i =  \sqrt{2}\,i\, \bar{\upsilon}^i \left(\dot z + \frac{\mu}{2}\, z\right)e^{-\frac{1}{2}\mu t}
    -\sqrt{2}\,\upsilon^i B e^{-\frac{1}{2}\mu t}  +\sqrt{2}\,i\, \bar{\varsigma}^i \left(\dot z - \frac{\mu}{2}\, z\right)e^{\frac{1}{2}\mu t}
    -\sqrt{2}\,\varsigma^i B e^{\frac{1}{2}\mu t},\nn
    &&\delta B = -\sqrt{2}\,i \,\bar{\upsilon}_k\dot{\xi}^k e^{-\frac{1}{2}\mu t}-\sqrt{2}\,i\, \bar{\varsigma}_k\dot{\xi}^k e^{\frac{1}{2}\mu t}.\lb{C12}
\eea
The parameters $\upsilon, \bar{\upsilon}$ and $\varsigma, \bar{\varsigma}$ correspond to
the supercharges $\Pi , \bar{\Pi}$ and $\Theta , \bar{\Theta}\,$, respectively. Note that the original $SU(2|1)$ transformations are embedded in \p{C12}
as
\bea
&&\delta z  =- \epsilon_k\xi^k \left(e^{-\frac{1}{2}\mu t}+ ie^{\frac{1}{2}\mu t}\right),\quad
\delta B  =- i\bar\epsilon_k\dot\xi^k \left(e^{-\frac{1}{2}\mu t}- ie^{\frac{1}{2}\mu t}\right),\nn
&&\delta \xi^i =  i\, \bar{\epsilon}^i \left[\dot z\left(e^{-\frac{1}{2}\mu t}- ie^{\frac{1}{2}\mu t}\right) + \frac{\mu}{2}z
\left(e^{-\frac{1}{2}\mu t}+ ie^{\frac{1}{2}\mu t}\right)\right] - \epsilon^i B\left(e^{-\frac{1}{2}\mu t}+ ie^{\frac{1}{2}\mu t}\right),
\eea
where $\epsilon_k := \frac{1}{\sqrt{2}} (\upsilon_k - i \varsigma_k)\,.$


\begin{thebibliography}{99}

\bibitem{FS}
G.~Festuccia, N.~Seiberg, {\it Rigid Supersymmetric Theories in Curved Superspace}, JHEP {\bf 1106} (2011) 114, {\tt arXiv:1105.0689 [hep-th]}.

\bibitem{DFS}
T.T.~Dumitrescu, G.~Festuccia, N.~Seiberg, {\it Exploring Curved Superspace},
JHEP {\bf 1208} (2012) 141, {\tt arXiv:1205.1115 [hep-th]}.

\bibitem{SamSor}
I.B.~Samsonov, D.~Sorokin, {\it Superfield theories on $S^3$ and their localization}, JHEP {\bf 1404} (2014) 102, {\tt arXiv:1401.7952 [hep-th]}; \\
I.B.~Samsonov, D.~Sorokin, {\it Gauge and matter superfield theories on $S^2$}, JHEP {\bf 1409} (2014) 097, {\tt arXiv:1407.6270 [hep-th]}.

\bibitem{DSQM}
E.~Ivanov, S.~Sidorov, {\it Deformed Supersymmetric Mechanics},  Class. Quant. Grav. {\bf 31} (2014) 075013,
 {\tt arXiv:1307.7690 [hep-th]}.

\bibitem{SKO}
E.~Ivanov, S.~Sidorov, {\it Super K\"ahler oscillator from $SU(2|1)$ superspace}, J. Phys. A {\bf 47} (2014) 292002, {\tt arXiv:1312.6821 [hep-th]}.


\bibitem{superc}
S.~Fedoruk, E.~Ivanov, O.~Lechtenfeld, {\it Superconformal Mechanics}, J. Phys. A {\bf 45} (2012) 173001,
{\tt arXiv:1112.1947 [hep-th]}.

\bibitem{WS}
A.V.~Smilga, {\it Weak supersymmetry}, Phys. Lett. B {\bf 585} (2004) 173,  {\tt arXiv:hep-th/0311023}.

\bibitem{BN3}
S.~Bellucci, A.~Nersessian, {\it (Super)Oscillator on CP(N) and Constant Magnetic Field},
Phys. Rev. D {\bf 67} (2003) 065013,
{\tt arXiv:hep-th/0211070}.

\bibitem{BN4}
S.~Bellucci, A.~Nersessian, {\it Supersymmetric K\"ahler oscillator in a constant magnetic field},
{\tt arXiv:hep-th/0401232}.

\bibitem{PP}
G.~Papadopoulos, {\it New potentials for conformal mechanics}, Class. Quant. Grav. {\bf 30} (2013) 075018,
 {\tt arXiv:1210.1719 [hep-th]}.

\bibitem{DFF}
V.~de Alfaro, S.~Fubini, G.~Furlan, {\it Conformal Invariance in Quantum Mechanics}, Nuovo Cim. A {\bf 34} (1976) 569.


\bibitem{SCM}
E.~Ivanov, S.~Krivonos, O.~Lechtenfeld, {\it New variant of $N=4$ superconformal mechanics},
JHEP {\bf 0303} (2003) 014, {\tt  arXiv:hep-th/0212303}.

\bibitem{SCM1}
E.~Ivanov, S.~Krivonos, O.~Lechtenfeld, {\it $N=4$, $d=1$ supermultiplets from nonlinear realizations of $D(2,1;\alpha)$}, Class. Quant. Grav. {\bf 21} (2004) 1031,
{\tt arXiv:hep-th/0310299}.



\bibitem{HT}
N.~L.~Holanda, F.~Toppan, {\it Four types of (super)conformal mechanics: D-module reps and invariant actions},
 J. Math. Phys. {\bf 55} (2014) 061703, {\tt arXiv:1402.7298 [hep-th]}.

\bibitem{Pash}
V.P.~Akulov, S.~Catto, O.~Cebecioglu, A.~Pashnev, {\it On the time dependent oscillator and the nonlinear realizations of the Virasoro group},
Phys. Lett. B {\bf 575} (2003) 137, {\tt arXiv:hep-th/0303134}.

\bibitem{DI}
F.~Delduc, E.~Ivanov, {\it Gauging ${\cal N}=4$ Supersymmetric Mechanics}, Nucl. Phys. B {\bf 753} (2006) 211, {\tt arXiv:hep-th/0605211}.

\bibitem{BK}
S.~Bellucci, S.~Krivonos, {\it Potentials in $N=4$ superconformal mechanics}, Phys. Rev. D {\bf 80} (2009) 065022,
{\tt arXiv:0905.4633 [hep-th]}.

\bibitem{Sorba}
L.~Frappat, A.~Sciarrino, P.~Sorba,
{\it Dictionary on Lie algebras and superalgebras},
Academic Press, 2000, {\tt arXiv:hep-th/9607161}.

\bibitem{IvSo}
E.~Ivanov, A.~Sorin, {\it The Structure of representations of conformal supergroup in the $OSp(1,4)$ basis}, Theor. Math. Phys. {\bf 45} (1980) 862
[Teor. Mat. Fiz. {\bf 45} (1980) 30].


\bibitem{ikrpash}
E.A.~Ivanov, S.O.~Krivonos, A.I.~Pashnev, {\it Partial supersymmetry breaking in N=4 supersymmetric quantum mechanics},
Class. Quant. Grav. {\bf 8} (1991) 19.

\bibitem{ikrlev}
E.A.~Ivanov, S.O.~Krivonos, V.M.~Leviant, {\it Geometric superfield approach to superconformal mechanics},
J. Phys. A {\bf 22} (1989) 4201.

\bibitem{kuto} Z. Kuznetsova and F. Toppan, {\it D-module Representations of N=2,4,8 Superconformal Algebras and Their Superconformal Mechanics}
 J. Math. Phys. {\bf 53} (2012) 043513, {\tt arXiv:1112.0995 [hep-th]}.

\bibitem{khto} S. Khodaee and F. Toppan, {\it Critical scaling dimension of D-module representations of N=4,7,8 Superconformal Algebras
and constraints on Superconformal Mechanics},
 J. Math. Phys. {\bf 53} (2012) 103518, {\tt arXiv:1208.3612 [hep-th]}.

\bibitem{KuSor}
S.M.~Kuzenko, D.~Sorokin, {\it Superconformal structures on the three-sphere}, JHEP {\bf 1410} (2014) 80, {\tt arXiv:1406.7090 [hep-th]}.


\end{thebibliography}
\end{document}